\documentclass[twocolumn,preprintnumbers,amsmath,amssymb,superscriptaddress,longbibliography]{revtex4-2}
\usepackage{amsmath}
\usepackage{float}
\usepackage{amsthm}
\usepackage{comment}
\usepackage{amssymb}
\usepackage{subfiles}
\usepackage{bm}
\usepackage[colorlinks, linkcolor=blue, citecolor=blue, urlcolor=blue, breaklinks=true]{hyperref}
\usepackage[normalem]{ulem}
\usepackage[T1]{fontenc}
\usepackage{graphicx}
\usepackage{dcolumn}
\usepackage{amssymb,amsmath,color}
\usepackage{eqnarray}
\usepackage{booktabs}
\usepackage{mathtools}
\usepackage{amsthm}
\usepackage{dsfont}
\usepackage{cases}
\usepackage{physics}
\usepackage{subfiles}

\usepackage{subcaption}
\usepackage{caption}
\usepackage{ragged2e}
\usepackage{braket}
\usepackage[dvipsnames]{xcolor}
\usepackage{ulem} 
\usepackage[normalem]{ulem}
\usepackage[utf8]{inputenc}
\usepackage{amsmath}
\newcommand{\commentold}[1]{}
\DeclareMathSymbol{:}{\mathpunct}{operators}{"3A}
\usepackage{orcidlink}

\def\be{\begin{equation}}
\def\ee{\end{equation}}
\def\bea{\begin{eqnarray}}
\def\eea{\end{eqnarray}}

\def\be{\begin{equation}}
\def\ee{\end{equation}}
\def\bal{\begin{align*}}
\def\eal{\end{align*}}
\def\bea{\begin{eqnarray}}
\def\eea{\end{eqnarray}}

\begin{document}
\date{\today}
%
\title{Two-time weak measurement protocol for ergotropy protection in open quantum batteries}
\author{André H. A. Malavazi\orcidlink{0000-0002-0280-0621}}
\email{andrehamalavazi@gmail.com}
\address{International Centre for Theory of Quantum Technologies, University of Gdańsk, Jana Bażyńskiego 1A, 80-309 Gdańsk, Poland}

\author{Rishav Sagar\orcidlink{https://orcid.org/0000-0002-6389-1897}}
\address{International Centre for Theory of Quantum Technologies, University of Gdańsk, Jana Bażyńskiego 1A, 80-309 Gdańsk, Poland}

\author{Borhan Ahmadi\orcidlink{0000-0002-2787-9321}}
\address{International Centre for Theory of Quantum Technologies, University of Gdańsk, Jana Bażyńskiego 1A, 80-309 Gdańsk, Poland}

\author{Pedro R. Dieguez\orcidlink{0000-0002-8286-2645}}
\email{dieguez.pr@gmail.com}
\address{International Centre for Theory of Quantum Technologies, University of Gdańsk, Jana Bażyńskiego 1A, 80-309 Gdańsk, Poland}
\begin{abstract}
Quantum batteries are emerging as highly efficient energy storage devices that can exceed classical performance limits. Although there have been significant advancements in controlling these systems, challenges remain in stabilizing stored energy and minimizing losses due to inevitable environmental interaction. In this paper, we propose a protocol that employs selective weak measurements to protect quantum states from such influence and mitigate battery discharging, that is feasible in state-of-the-art technologies. We establish thermodynamic constraints that allow this method to be implemented without disrupting the overall energy and ergotropy balance of the system, i.e., with no extra net recharging. Our findings demonstrate that appropriately chosen measurement intensity can reduce unwanted discharging effects, thereby preserving ergotropy and improving the stability of quantum batteries. We illustrate the protocol with single and two-qubit systems and establish the generalization for $N$-cell batteries. Additionally, we explore how weak measurements influence the coherent and incoherent components of ergotropy, providing new insights into the practical application of quantum coherence in energy storage technologies.
\end{abstract}

\maketitle
\section{Introduction}
In recent years, quantum batteries have emerged as promising devices for energy storage. These systems, composed of quantum states as energy carriers, aim to employ genuine quantum properties, such as coherence and entanglement, to surpass classical limits~\cite{PhysRevE.87.042123,binder2015quantacell,RevModPhys.96.031001,quach2022superabsorption,Andolina2018,PhysRevB.99.205437,gyhm2024beneficial}. In this sense, quantum batteries were shown to exhibit unique features such as coherence-driven and collective enhancements in ergotropy and charging processes~\cite{Andolina2018,PhysRevLett.124.130601,binder2015quantacell,PhysRevE.99.052106,PhysRevE.102.042111,PhysRevLett.120.117702,PhysRevLett.118.150601,Dou2021,PhysRevA.106.032212,PhysRevB.105.115405,PhysRevLett.128.140501,deMoraes_2024,10.3389/fphy.2022.1097564}.
Significant progress has been made in understanding and manipulating these devices, with studies exploring diverse models~\cite{PhysRevA.109.032201,PhysRevA.97.022106,PhysRevResearch.4.013172,Dou_2020,PhysRevB.109.235432,PhysRevA.106.032212,PhysRevA.110.052404,hadipour2025nonequilibrium} and approaches like collisional models~\cite{PhysRevLett.127.100601,PhysRevResearch.5.013155}, super- and sub-radiance~\cite{Glicenstein2022}, reservoir engineering~\cite{PhysRevLett.132.210402,PhysRevApplied.23.024010}, phase transitions~\cite{barra2022quantum}, micro-maser
quantum batteries~\cite{Shaghaghi2022,Shaghaghi2023}, topological features~\cite{lu2024topological,Dennis2002}, catalysis~\cite{PhysRevA.107.042419}, non-Markovian effects~\cite{kamin2020non}, wireless charging~\cite{PhysRevLett.132.090401,PhysRevE.110.064107,hu2025efficient}, reinforcement learning~\cite{sun2024cavity,PhysRevLett.133.243602}, charging by measurement~\cite{PhysRevApplied.19.064069}, and optimal control strategies~\cite{Rodríguez_2024,PhysRevA.107.032218,Hu_2022}. 

Various physical platforms have been proposed as viable candidates for quantum energy storage. Under cryogenic conditions, technologies such as superconducting circuits~\cite{RevModPhys.93.025005,PhysRevA.109.012204,batteries8050043,Hu_2022, PhysRevA.107.023725, PhysRevA.109.062432,Zheng_2022}, semiconductor quantum dots~\cite{PhysRevLett.131.260401} and ultracold-atoms~\cite{PhysRevA.110.032205} provide promising options, enabling quantum coherence and collective behavior necessary for efficient charging. On the other hand, room-temperature approaches, including organic micro-cavities~\cite{quach2022superabsorption}, optical setups~\cite{YANG2024102300}, nitrogen-vacancy (NV) centers in diamond~\cite{RevModPhys.92.015004} and nuclear spin systems~\cite{PhysRevA.106.042601,Cruz_2022}, showcase the potential for robust quantum energy storage through mechanisms that minimize decoherence, marking key advancements toward practical quantum batteries across a range of operating conditions.

Nevertheless, despite ongoing efforts, significant challenges persist, particularly in more realistic scenarios where quantum coherence is susceptible to decoherence effects~\cite{breuer2002theory, PhysRevA.100.043833, PhysRevE.103.042118,PhysRevA.109.052206,PhysRevE.100.032107,Mitchison2021chargingquantum,Liu2019,selfdis, stable1,Carrega_2020, PhysRevLett.122.210601,PhysRevB.99.035421,PhysRevLett.132.090401,PhysRevA.102.060201,PhysRevA.103.033715}.
In this context, a charged quantum battery exposed to a dissipative environment renders not only the irreversible leakage of energy from the battery into the environment but also the loss of its charge, commonly quantified by the ergotropy, i.e., the extractable work from a quantum state~\cite{A.E.Allahverdyan_2004}. For instance,
the weak contact with a thermal reservoir leads the battery to a Gibbs state, which is completely passive~\cite{Pusz1978}.
Such effects are particularly detrimental to the performance of quantum technologies, as they limit the ability to harness ergotropy for practical applications. Therefore, characterizing and mitigating the influence of thermal reservoirs is essential for the reliable operation of quantum devices in general~\cite{RevModPhys.95.045005,Galve2017,PhysRevA.105.012432,Molitor2024,aziz2025revival}.
As proposed by~\cite{PhysRevResearch.2.013095}, a possible way to preserve a quantum system from decaying is given by the continuous observation of the quantum system by a sequence of frequent projective measurements (a feature of the Zeno effect~\cite{PhysRevA.41.2295}), which can hinder the discharging process, effectively freezing the energy stored in a quantum battery. However, this method may not be experimentally friendly due to the required frequent state observation.

Furthermore, measurements are also being incorporated in quantum thermodynamics, e.g., as additional strokes in thermal cycles, with measurement-powered engines exploiting the fact that the measurement action generally disturbs the observed system, thereby altering its internal energy~\cite{yi2017single,elouard2017role,brandner2015coherence}. For instance, by employing non-selective measurements, a single temperature heat engine without feedback control to extract work was introduced~\cite{yi2017single}, 
and several other measurement-based protocols have been explored ~\cite{campisi2017feedback,chand2017single,chand2017measurement,mohammady2017quantum,elouard2017extracting,chand2018critical,ding2018measurement,elouard2018efficient,buffoni2019quantum,solfanelli2019maximal,jordan2020quantum,behzadi2020quantum,seah2020maxwell,bresque2021two,chand2021finite,lin2021suppressing,PhysRevLett.124.110604, PhysRevLett.121.030604, PRXQuantum.3.020329,
PhysRevLett.117.240502, anka2021measurement,alam2022two,manikandan2022efficiently,myers2022quantum,lisboa2022experimental,dieguez2023thermal}.
In particular, since generalized measurements can be implemented as a sequence of weak measurements~\cite{oreshkov2005weak,dieguez2018information}, their intensity can be continuously adjusted to transition between weak and strong measurement regimes~\cite{pan2020weak,mancino2018entropic}.
Thus, weakly measured systems were shown to be crucial for consistently observing work and heat contributions in an externally driven quantum stochastic evolution~\cite{alonso2016thermodynamics} and were employed to develop a deterministic protocol based on non-selective variable strength measurements to perform a heat engine cycle~\cite{behzadi2020quantum,lisboa2022experimental,VIEIRA2023100105}. Considering bipartite quantum systems, local strong (projective) measurements were shown to improve the quantum battery capacity~\cite{PhysRevA.109.042424}, while selective weak measurements can offer more work extraction than the one described by the daemonic ergotropy ~\cite{balkanlu2024selective}, i.e., the amount of extractable work that arises from information gained in a strong measurement~\cite{Francica2017}.
In this sense, weak measurements have been widely adopted across diverse quantum platforms, with successful implementations in nuclear magnetic resonance (NMR)~\cite{Lu_2014,lisboa2022experimental, protect5}, NV centers~\cite{Cujia2019}, trapped ion systems~\cite{pan2020weak}, photonic systems~\cite{Kim09, kim2012protecting, protect1}, and superconducting qubits~\cite{doi:10.1126/science.1126475, PhysRevLett.101.200401, Zhong2014,Weber2014,PhysRevLett.126.100403}, demonstrating their versatility for contemporary quantum applications. The thermodynamic cost of implementing weak measurements fundamentally depends on the underlying protocol and the physical nature of the probe, necessitating an analysis of interconnected factors such as measurement strength, energy consumption, information extraction, and thermodynamic irreversibility~\cite{latune2025thermodynamically,ferraz2025weak}. Furthermore, renormalization procedures used in post-measurement state analysis introduce additional energetic considerations, which are essential to characterize the impact of measurements on the performance of such thermodynamic protocols.
\begin{figure}
    \center
    \includegraphics[width=1\columnwidth]{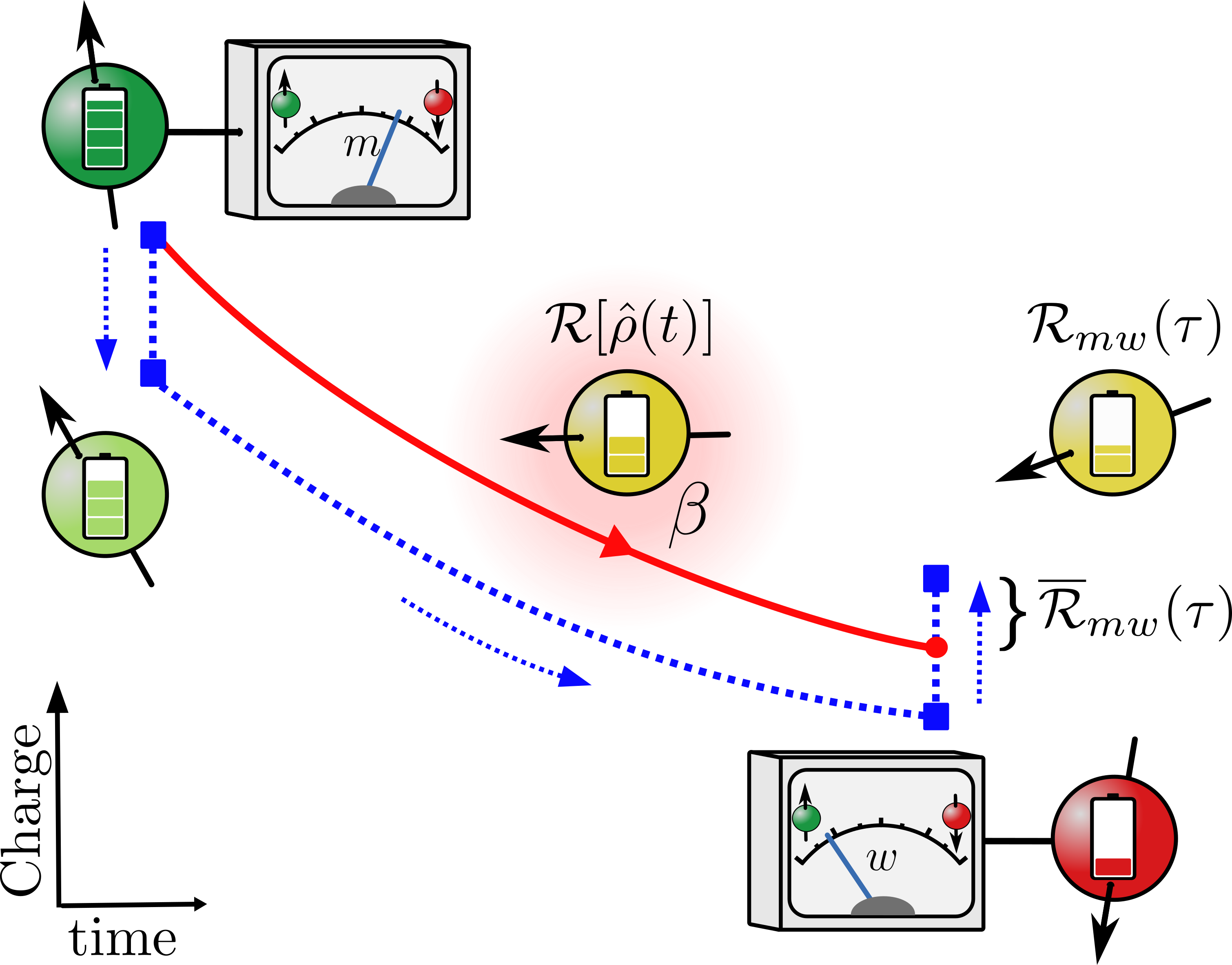}
    \caption{ \textbf{Two-time Weak Measurement (TWM) protocol}: We consider a quantum battery initially charged, which then undergoes a discharging process. Before this, a weak measurement with strength $m$ is applied in the first step, changing the system's energy and ergotropy (charge), represented by the dashed blue line. In the intermediate step, during the discharging phase, a second reversal weak measurement with strength $w$ is performed at instant $\tau$, compensating for the previous changes. By carefully tuning the measurement intensities, the protocol can offer an advantage despite null net recharging, over a purely discharging process undergone by $\mathcal{R}[\hat{\rho}(t)]$ (continuous red line), which is quantified by the ergotropy gain $\overline{\mathcal{R}}_{mw}(\tau)\coloneqq \mathcal{R}_{mw}(\tau)-\mathcal{R}[\hat{\rho}(\tau)]$. \justifying}
    \label{Diagram}
\end{figure}

Here, we investigate the potentiality of weak measurements in attenuating environment-induced losses, aiming to harness quantum mechanical subtleties for more efficient energy storage and extraction in two-level systems-based quantum batteries.
The two-time weak measurement (TWM) protocol builds on the structure of previous theoretical and experimental studies on protecting quantum states from decoherence using weak measurements and measurement reversal operations~\cite{kim2012protecting, Xiao2013,protect1,protect2,protect3, protect4, Lee:11, Lalita_2024,e27040350}. The key distinction lies in our incorporation of thermodynamic-based constraints applied on the target system while considering the measurements as resources, enabling these methods to be applied without altering the overall energetics of the protocol.  
Specifically, as shown in Fig.~\ref{Diagram}, before the discharging process, an initial weak measurement is performed, modifying the system energy and ergotropy. Subsequently, the charged system undergoes dissipative dynamics, during which a second weak measurement is carried out to compensate for the previously induced energetic and ergotropic changes. Despite the net exchange of these quantities being constrained to be null, a positive gain can be observed at the end of the protocol by tuning the relevant parameters. The feasibility of such a procedure is a direct consequence of the symmetry of weak measurement and its reversal, and the different discharging rates resembling the ergotropic counterpart of the Mpemba effect~\cite{mpemba1969cool, medina2024anomalous}. We separated the analysis to consider both the isolated incoherent and coherent components of ergotropy. Notably, without affecting the protocol's overall probability of success, quantum coherence can be employed to enhance the measurement-based battery recovery process from discharging effects, as suggested in this work. The protocol is investigated for batteries composed of single and two-qubit systems, and the generalization for $N$-cells batteries is introduced.

The remainder of this article is structured as follows. In Sec.~\ref{IntroErgotropy} we introduce the physical context and model to explain the dissipation loss mitigation protocol. In Sec.~\ref{Sec3} we present our main results by analyzing the protocol performance for the incoherent and coherent part of the ergotropy, as well as the total ergotropy, and identify parameter regions where the protocol yields promising outcomes. In Sec.~\ref{MultiQubit} we describe how the protocol is extended to multi-qubit quantum batteries, and illustrate the procedure for a two-qubit system with quantum correlations. Finally, in Sec.~\ref{DiscussionSection}, we summarize our main conclusions and discuss potential perspectives for future work.  
%
%
%
%
%
%
%
%
%
%
%
%
%
%
%
%
%
%
%
\section{Ergotropy loss mitigation: The Two-time Weak Measurement (TWM) protocol}\label{IntroErgotropy}
Consider a $d$-dimensional quantum battery represented by the density matrix $\hat{\rho} = \sum_{j}^{d} p_{j} |j\rangle \langle j|$ and governed by the Hamiltonian $\hat{H} = \sum_{k}^{d} E_{k} |E_{k}\rangle \langle E_{k}|$. We assume that the probability distribution and energy spectrum are ordered as $p_{1} > p_{2} > \dots > p_{d}$ and $E_{1} < E_{2} < \dots < E_{d}$, respectively. The maximum amount of energy extractable via a unitary process that preserves cyclic Hamiltonian control is known as ergotropy, which is defined as~\cite{A.E.Allahverdyan_2004}
\begin{equation}\label{ErgotropyDef}
    \mathcal{R}[\hat{\rho}] \coloneqq \textrm{Tr} \left[ \left( \hat{\rho} - \hat{\rho}^{p} \right) \hat{H} \right] = \sum_{k,j}^{d} E_{k} p_{j} \left( |\langle E_{k} | j \rangle|^{2} - \delta_{k,j} \right),
\end{equation}
where $\hat{\rho}^{p} = \sum_{i}^{d} p_{i} |E_{i}\rangle \langle E_{i}|$ is the passive state, which is unitarily related to the initial state $\hat{\rho}$. Generally, the battery state may exhibit coherences in the $\hat{H}$ basis, which is a unique quantum resource \cite{RevModPhys.89.041003}. Identifying the contribution of coherence to the battery ergotropy is therefore insightful. To capture this, the coherent ergotropy within $\hat{\rho}$ is defined as \cite{PhysRevLett.125.180603,guha2022activation,niu2024experimental}
\begin{equation}\label{CoherentErgoDef}
    \mathcal{R}^{coh}[\hat{\rho}]\coloneqq\mathcal{R}[\hat{\rho}]-\mathcal{R}^{inc}[\hat{\rho}],
\end{equation}
where $\mathcal{R}^{inc}[\hat{\rho}]$ represents the incoherent component of the ergotropy, such that $\mathcal{R}^{inc}[\hat{\rho}] = \mathcal{R}[\hat{\rho}^{diag}]$, where $\hat{\rho}^{diag} = \sum_{k}^{d} \langle E_{k} | \hat{\rho} | E_{k} \rangle |E_{k}\rangle \langle E_{k}|$ denotes the diagonal (incoherent) state of $\hat{\rho}$, obtained by removing its off-diagonal elements in the energy basis. Note that, by definition, $\mathcal{R}^{coh}[\hat{\rho}]$ is zero for incoherent states.
\subsection{Setup and the TWM protocol description}

For closed system dynamics, the ergotropy is preserved. Since the populations $\{p_{j}\}$ remain invariant under unitary dynamics, $\mathcal{R}[\hat{\rho}]$ remains unchanged during the free evolution of the battery. In contrast, if the battery is an open quantum system interacting with an external environment \cite{breuer2002theory}, its dynamics are no longer unitary, typically resulting in irreversible behavior that can lead to dissipation and discharging. Therefore, it is crucial to develop strategies to either stabilize its ergotropy or mitigate its inevitable leakage. To achieve this, we propose a protocol, depicted in Fig.~\ref{Diagram}, based on weak measurements performed before and after the battery discharging process.

%
%
%
%
%
%
%
%
%
%
%
%
%
%
%
%
%
%
%
More specifically, we consider a quantum battery composed of a single qubit weakly coupled to a thermal reservoir with inverse temperature $\beta$. The state of the battery is represented by $\hat{\rho}$, and its Hamiltonian is given by $\hat{H} = E_g |g\rangle \langle g| + E_e |e\rangle \langle e|$, where $E_e - E_g = \omega$ defines the qubit energy gap, with $|g\rangle$ ($|e\rangle$) corresponding to the ground (excited) state. Without loss of generality, we assume $E_g = 0$ for the rest of the paper. Notice that the maximum possible ergotropy for this system is given by $\omega$, characterized by the state $|e\rangle \langle e|$. The time evolution of $\hat{\rho}$ is governed by the GKLS master equation \cite{breuer2002theory} ($\hbar = k_B = 1$):
\begin{equation}
\label{MasterEquation} \frac{d}{dt} \hat{\rho}(t) = -i \left[\hat{H}, \hat{\rho}(t)\right] + \mathcal{D}_{\downarrow}[\hat{\rho}(t)] + \mathcal{D}_{\uparrow}[\hat{\rho}(t)],
\end{equation}
where the dissipative terms are given by
\begin{align} 
\mathcal{D}_{\downarrow}[\boldsymbol{\cdot}]&=\gamma(1-f)\left(\hat{\sigma}_{+}\boldsymbol{\cdot}\hat{\sigma}_{-}-\frac{1}{2}\left\{ \hat{\sigma}_{-}\hat{\sigma}_{+},\boldsymbol{\cdot}\right\} \right),\nonumber\\\mathcal{D}_{\uparrow}[\boldsymbol{\cdot}]&=\gamma f\left(\hat{\sigma}_{-}\boldsymbol{\cdot}\hat{\sigma}_{+}-\frac{1}{2}\left\{ \hat{\sigma}_{+}\hat{\sigma}_{-},\boldsymbol{\cdot}\right\} \right), 
\end{align}
where $\gamma$ denotes the decay rate, $\hat{\sigma}_{-} = \hat{\sigma}_{+}^{\dagger} = |e\rangle \langle g|$ are the jump operators, and the thermal population is defined as $f = e^{-\beta \omega} (1 + e^{-\beta \omega})^{-1} \in [0, 1/2)$. The characteristic timescale of this dynamics is $\tau_{\gamma} = \gamma^{-1}$. Due to the form of the dissipators, the dynamical fixed point of Eq.~\eqref{MasterEquation} is the thermal (Gibbs) state $\hat{\tau}_{\textrm{th}} = (1-f)|g\rangle \langle g| + f|e\rangle \langle e|$, determined by the reservoir temperature. Thus, regardless of the initial state, we have $\lim_{t\rightarrow\infty} \hat{\rho}(t) = \hat{\tau}_{\textrm{th}}$.
This time evolution results in an inevitable decay of ergotropy $\mathcal{R}[\hat{\rho}(t)]$ to zero, indicating the discharging of the battery as thermalization proceeds. Further details can be found in Appendix~\ref{DissAndBatteryDischar}.
%
%
%
%
%
%
%
%
%
%
%
%

The proposed protocol for mitigating ergotropy loss consists of the following steps: (i) Assume the battery is initially charged, meaning it has been prepared in an active state, such that $\hat{\rho}_{0} = \left(1 - P_{0}\right)|g\rangle \langle g| + P_{0} |e\rangle \langle e| + \left(Q_{0} |g\rangle \langle e| + h.c.\right)$, where $|Q_{0}|^{2} \leq P_{0} (1 - P_{0})$ and $h.c.$ denotes the Hermitian conjugate; (ii) Immediately after charging, a weak measurement is performed of the form $\hat{M}_{m} \coloneqq |g\rangle \langle g| + \sqrt{1 - m} |e\rangle \langle e|$, with measurement strength $0 \leq m \leq 1$. This process results in the post-measurement state
\begin{equation}
    \hat{\rho}_{m}(0)=\frac{\hat{M}_{m}\hat{\rho}_{0}\hat{M}_{m}^{\dagger}}{\textrm{N}_{m}},
\end{equation}
with probability
\begin{equation}\label{ProbStep2}
    \textrm{N}_{m}\coloneqq\textrm{Tr}[\hat{M}_{m}\hat{\rho}_{0}\hat{M}_{m}^{\dagger}]=1-mP_{0}.
\end{equation}
Notice that $\rho_{m}(0) = \rho_0$ $\forall m$ for a fully charged battery, i.e., $P_{0} = 1$ and $Q_{0} = 0$. In this particular case, the action of $\hat{M}_{m}$ has no effect. However, due to the non-unitary nature of thermalization dynamics, one can wait for $P_{0} < 1$ and proceed with the protocol;
(iii) During a dissipation time $t \in [0, \tau]$, the battery is assumed to interact with the external environment. In this interval, its dynamics are governed by Eq.~\eqref{MasterEquation}. Consequently, the state of the battery evolves as
\begin{equation}
    \hat{\rho}_{m}(t)=\left(\begin{array}{cc}
1-P_{m}(t) & Q_{m}(t)\\
Q_{m}^{*}(t) & P_{m}(t)
\end{array}\right),
\end{equation}
with
\begin{equation}
    \begin{split}
        P_{m}(t)&=\frac{P(t)-m\left[\left(1-f\right)e^{-\gamma t}+f\right]P_{0}}{\textrm{N}_{m}},\\
        Q_{m}(t)&=\frac{\sqrt{1-m}}{\textrm{N}_{m}}Q(t),
    \end{split}
\end{equation}
where $P(t) = \left(P_{0} - f\right) e^{-\gamma t} + f$ and $Q(t) = Q_{0} e^{-\frac{\gamma t}{2} + i t \omega}$ are the population and coherence, respectively, when no measurement is performed (i.e., $m = 0$) (see Appendix~\ref{DissAndBatteryDischar});
(iv) Finally, at instant $t = \tau$, a reversal measurement $\hat{W}_{w} \coloneqq \sqrt{1 - w} |g\rangle \langle g| + |e\rangle \langle e|$, with strength $0 \leq w \leq 1$, is performed. The resulting final state is then

\begin{equation}\label{FinalState}
    \hat{\rho}_{mw}(\tau)=\frac{\hat{W}_{w}\hat{\rho}_{m}(\tau)\hat{W}_{w}^{\dagger}}{\textrm{N}_{mw}(\tau)}=\left(\begin{array}{cc}
1-P_{mw}(\tau) & Q_{mw}(\tau)\\
Q_{mw}^{*}(\tau) & P_{mw}(\tau)
\end{array}\right),
\end{equation}
with probability
\begin{align}\label{ProbStep4}
        \textrm{N}_{mw}(\tau)&\coloneqq\textrm{Tr}[\hat{W}_{w}\hat{\rho}_{m}(\tau)\hat{W}_{w}^{\dagger}]=1-w+w\frac{P(\tau)}{\textrm{N}_{m}}\nonumber\\&-w\frac{\left(1-\textrm{N}_{m}\right)}{\textrm{N}_{m}}\left[\left(1-f\right)e^{-\gamma\tau}+f\right],
\end{align}
in which
\begin{equation}
    \begin{split}
        P_{mw}(\tau)&=\textrm{N}_{mw}^{-1}(\tau)P_{m}(\tau),\\Q_{mw}(\tau)&=\textrm{N}_{mw}^{-1}(\tau)\sqrt{(1-w)}Q_{m}(\tau).
    \end{split}
\end{equation}

The probability of successfully performing the protocol is defined as the product of the probabilities relative to the desired measurement outcomes as
\begin{equation}\label{TotalProbability}
    \Pi_{mw}(\tau)\coloneqq\textrm{N}_{m}\textrm{N}_{mw}(\tau).
\end{equation}
It is important to emphasize that the measurements are assumed to be instantaneous. While measurements inherently require finite time, our previous assumption holds valid when the system's characteristic dynamic timescale exceeds the measurement duration. Experimental evidence from both NMR~\cite{lisboa2022experimental} and superconducting qubits~\cite{doi:10.1126/science.1126475, PhysRevLett.101.200401, Zhong2014,PhysRevLett.126.100403} corroborates that weak measurements can be performed in time-frames significantly shorter than the relevant relaxation and decoherence times. Also, note that the diagonal states and probabilities at each step are independent of the initial coherence $Q_{0}$. This independence arises from the commutation relations $[\hat{M}_{m}, \hat{H}] = [\hat{W}_{w}, \hat{H}] = 0$ and from Eq.~\eqref{MasterEquation}, implying that the population changes induced by the successive actions of $\hat{M}_{m}$, thermalization dynamics, and $\hat{W}_{w}$ do not couple them with the off-diagonal elements. The opposite, however, does not hold; that is, the coherence is coupled with the population during the measurements.
%
%
%
%
%
%
%
%
%

As previously mentioned, the ergotropy can be decomposed into coherent and incoherent parts, allowing us to quantify the final ergotropy of the battery as $\mathcal{R}_{mw}(\tau) \coloneqq \mathcal{R}[\hat{\rho}_{mw}(\tau)] = \mathcal{R}_{mw}^{{coh}}(\tau) + \mathcal{R}_{mw}^{{inc}}(\tau)$. For two-level systems simple analytical expressions for both components can be derived~\cite{PhysRevLett.125.180603,guha2022activation}. The incoherent component of the final ergotropy can be expressed as
\begin{equation}\label{FinalIncErgo}
    \mathcal{R}_{mw}^{inc}(\tau)=\omega\left(2P_{mw}(\tau)-1\right)\left[1-\Theta\left(w^{\prime}-w\right)\right],
\end{equation}
where $\Theta(x)$ is the Heaviside function, such that $\Theta(x)=1$ if $x\geq0$ and null otherwise, and
\begin{equation}\label{wBar}
    w^{\prime}\coloneqq \frac{\left(1-2P_{m}(\tau)\right)\textrm{N}_{m}}{\textrm{N}_{m}+\left(1-\textrm{N}_{m}\right)\left[\left(1-f\right)e^{-\gamma\tau}+f\right]-P(\tau)}
\end{equation}
is the threshold strength for obtaining a finite final incoherent ergotropy, i.e., one has to guarantee $w>w^{\prime}$ to have an active post-reversal measurement diagonal state.
In general, the coherent component for two-level systems with Hamiltonian $\hat{H} = \omega|e\rangle \langle e|$ can be written as
\begin{equation}\label{CoherentErgoFormula}
    \mathcal{R}^{coh}[\hat{\rho}]=\frac{\omega}{2}\left(\psi-\sqrt{\psi^{2}-4\left|Q\right|^{2}}\right),
\end{equation}
where $\psi\coloneqq\sqrt{2\mu[\hat{\rho}]-1}$, with $\mu[\hat{\rho}]\coloneqq\textrm{Tr}\left[\hat{\rho}^{2}\right]$ being the purity of the state $\hat{\rho}$, and $Q$ is the off-diagonal element.
Thus, at the end of the protocol, one has $\mathcal{R}_{mw}^{coh}(\tau)= \mathcal{R}^{coh}[\hat{\rho}_{mw}(\tau)]$.

After successfully applying the described sequence of steps, a higher ergotropy in the final state is expected, compared to the amount one would have obtained without executing the protocol. In this sense, to quantify the performance of the procedure, we define the \textit{ergotropy gain} as
\begin{equation}\label{ErgoDiff}
\overline{\mathcal{R}}_{mw}(\tau)\coloneqq \mathcal{R}_{mw}(\tau)-\mathcal{R}[\hat{\rho}(\tau)],
\end{equation}
with $\hat{\rho}(\tau)$ being the state subjected only through the purely dissipative dynamics given by Eq.~\eqref{MasterEquation}.
This quantity can be also decomposed into coherent and incoherent parts, i.e.,
\begin{equation}
    \overline{\mathcal{R}}_{mw}(\tau)=\overline{\mathcal{R}}_{mw}^{coh}(\tau)+\overline{\mathcal{R}}_{mw}^{inc}(\tau),
\end{equation}
where
\begin{equation}
    \overline{\mathcal{R}}_{mw}^{coh(inc)}(\tau) \coloneqq \mathcal{R}_{mw}^{coh(inc)}(\tau)-\mathcal{R}^{coh(inc)}[\hat{\rho}(\tau)].
\end{equation}
In such a way the role of the initial coherence $Q_{0}$ in the protocol performance is fully captured by $\overline{\mathcal{R}}_{mw}^{coh}(\tau)$, i.e., given an initial state $\hat{\rho}_{0}$, this term quantifies the effect of $Q_{0}$ relative to the incoherent state $\hat{\rho}_{0}^{{diag}}$.
%
%
%
%
%
%
%
%
%
%
%
%
%
%
%
%
%
%
%
%
%
%
%
%
%
%
%
%
%
%
%
%
%
%
%
%
%
%
%
%
%
%
%
%

\subsection{Protocol working regime}
Note that the weak measurement and its reversal generally disturb the battery state; specifically, the former brings the battery closer to the ground state, while the latter pushes it toward the excited state. As a result, these protocol steps inevitably induce energy changes, potentially charging (or discharging) the battery. We therefore define $\varepsilon_{mw}(\tau)$ as the energy shift of the battery, quantifying the total (net) energy change induced by the measurements, i.e., as
\begin{equation}\label{EqEnergyCostAppendix}
   \begin{split}
       \varepsilon_{mw}(\tau)&\coloneqq\Delta E_{m}+\Delta E_{mw}(\tau)\\&=\omega\left[\left(\frac{(1-m)}{\textrm{N}_{m}}-1\right)P_{0}+P_{mw}(\tau)-P_{m}(\tau)\right],
   \end{split}
\end{equation}
where
\begin{equation}\label{EnergyChanges}
   \begin{split}
       \Delta E_{m}&\coloneqq\textrm{Tr}\left[\hat{H}\left(\hat{\rho}_{m}(0)-\hat{\rho}_{0}\right)\right],\\
       \Delta E_{mw}(\tau)&\coloneqq\textrm{Tr}\left[\hat{H}\left(\hat{\rho}_{mw}(\tau)-\hat{\rho}_{m}(\tau)\right)\right],
   \end{split}
\end{equation}
are the energy changes induced by the weak and reversal measurements, respectively.
It is worth highlighting that this quantity is independent of the initial coherence of the battery $Q_{0}$. In fact, $\varepsilon_{mw}(\tau)$ can be defined and computed only in terms of the diagonal (incoherent) component of the states.

These energetic changes are expected to be accompanied by ergotropic counterparts. Thus, to account for the total measurement-induced ergotropy difference, we define the ergotropy shift $\mathcal{W}_{mw}(\tau)$ as 
\begin{equation}
    \mathcal{W}_{mw}(\tau)\coloneqq\Delta\mathcal{R}_{m}+\Delta\mathcal{R}_{mw}(\tau),
\end{equation}
where
\begin{equation}
   \begin{split}
       \Delta\mathcal{R}_{m}&\coloneqq\mathcal{R}[\hat{\rho}_{m}(0)]-\mathcal{R}[\hat{\rho}_{0}],
       \\\Delta\mathcal{R}_{mw}(\tau)&\coloneqq\mathcal{R}[\hat{\rho}_{mw}(\tau)]-\mathcal{R}[\hat{\rho}_{m}(\tau)],
   \end{split}
\end{equation}
are the individual ergotropy contributions for each measurement. Alternatively, from the ergotropy definition given by Eq.~\eqref{ErgotropyDef}, one can rewrite the ergotropy shift as $\mathcal{W}_{mw}(\tau)=\varepsilon_{mw}(\tau)-\varepsilon_{mw}^{p}(\tau)$, where $\varepsilon_{mw}^{p}(\tau)$ represents the total energetic changes relative to the passive states of each step, i.e. $\varepsilon_{mw}^{p}(\tau)=\textrm{Tr}\left[\hat{H}\left(\hat{\rho}_{m}^{p}-\hat{\rho}_{0}^{p}\right)\right]+\textrm{Tr}\left[\hat{H}\left(\hat{\rho}_{mw}^{p}(\tau)-\hat{\rho}_{m}^{p}(\tau)\right)\right]$.

Observe that if at the end of the protocol one has induced net positive (negative) ergotropy, part of the recovered (loss) charge is due to the measurements themselves, which means that the procedure would be directly (dis)charging the battery. Hence, to ensure that the TWM protocol does not actively provide a net external resource to the battery and effectively (dis)charge it, one should specify the measurement strengths $m$, $w$, and the dissipation time $\tau$ such that
\begin{equation}
\begin{split}
    \varepsilon_{mw}(\tau)&=0,\\\qquad\mathcal{W}_{mw}(\tau)&=-\varepsilon_{mw}^{p}(\tau)=0.
    \end{split}
\end{equation}
The simultaneous validation of both requirements restricts the parameter space to valid working regions for our proposed protocol.
In fact, given Eq.~\eqref{EnergyChanges} for the energetic shift, and fixed values of $\gamma$, $f$, $P_{0}$, weak measurement strength $m$, and dissipation time $\tau$, one can find the expression of the reversal measurement strength $\tilde{w}$ needed to restore the energy lost due to the weak measurement, i.e., from the constraint for $\varepsilon_{m\tilde{w}}(\tau)=0$, one can invert the relation and show that
\begin{widetext}
\begin{equation}\label{NullCostW}
    \tilde{w}=\textrm{N}_{m}\frac{1-P_{0}-\textrm{N}_{m}\left(1-f\right)+e^{\gamma\tau}\left[\textrm{N}_{m}\left(P_{m}(\tau)+P_{0}-1-f\right)+1-P_{0}\right]}{\left[\textrm{N}_{m}(P_{m}(\tau)+P_{0}-1)+1-P_{0}\right]\left[\textrm{N}_{m}(1-f)\left(e^{\gamma\tau}-1\right)+1-P_{0}\right]}
\end{equation}
\end{widetext}
ensures a null energetic shift. In addition, we should limit ourselves to physical regions, characterized by strengths that satisfy $\tilde{w}\in[0,1]$. Hence, given Eq. ~\eqref{NullCostW}, one can also find the subspace with null ergotropic shift, satisfying $\mathcal{W}_{m\tilde{w}}(\tau)=0$ (see Appendix \ref{EnergeticCostAppendix} for further discussion).

Along these lines, it is instructive to consider the particular case of equal measurement strengths, such that $m = w = \eta \in [0,1]$. In such a scenario, it is possible to show that Eq.~\eqref{NullCostW} reduces to the following null energetic shift strength curves
\begin{equation}\label{NullCostCurves}
    \begin{cases}
\eta_{1}\coloneqq0,\\
\eta_{2}\coloneqq\frac{f-P_{0}}{(f-1)P_{0}},\\
\eta_{3}\coloneqq\frac{e^{\gamma\tau}(f+P_{0}-1)-f+P_{0}}{P_{0}\left(e^{\gamma\tau}(f+P_{0}-1)-f+1\right)},
\end{cases}
\end{equation}
where $\eta_{1}$ is analogous to the trivial case of no measurements. The $\eta_{2}$-curve is constant for fixed values of $f$ and initial population $P_{0}$, and it corresponds to protocols such that 
\begin{equation}
    \hat{\rho}_{\eta_{2}}(0)=\hat{\tau}_{\textrm{th}}+f\sqrt{\frac{f-1}{f(P_{0}-1)P_{0}}}\left(\begin{array}{cc}
0 & Q_{0}\\
Q_{0}^{*} & 0
\end{array}\right)
\end{equation}
and
\begin{equation}
   \hat{\rho}_{\eta_{2}\eta_{2}}(\tau)=\hat{\rho}_{0}^{diag}+e^{-\frac{1}{2}\tau\gamma}\left(\begin{array}{cc}
0 & Q_{0}e^{i\omega\tau}\\
Q_{0}^{*}e^{-i\omega\tau} & 0
\end{array}\right)
\end{equation}
\begin{figure*}
    \centering
    \includegraphics[width=0.9\textwidth]{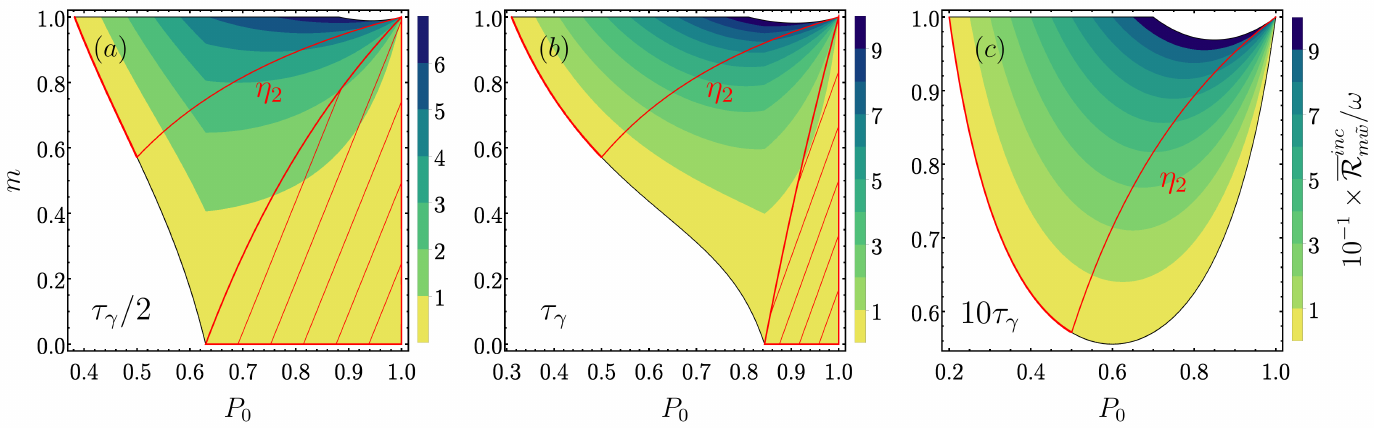}
    \caption{\textbf{Incoherent ergotropy gain} as a function of the weak measurement strength $m$ and initial population $P_{0}$ along the null energetic shift reversal measurement strength $\tilde{w}\in [0,1]$ for distinct dissipation times (a) $\tau=\tau_{\gamma}/2$, (b) $\tau=\tau_{\gamma}$, and (c) $\tau=10\tau_{\gamma}$. The red-hatched region and red curves indicate that $\mathcal{W}_{m\tilde{w}}(\tau)=0$. The $\eta_{2}$ lines are highlighted. 
    \justifying}
    \label{TauGammaWtilde}
\end{figure*}
for the post-measurement states. Note that after applying $\hat{M}_{\eta_{2}}$, the battery diagonal state, $\hat{\rho}^{diag}_{\eta_{2}}(0)$, is thermal with the inverse of temperature $\beta$, which remains invariant under the dissipative dynamics described by Eq.~\eqref{MasterEquation} (see Appendix~\ref{DissAndBatteryDischar}). This implies that while coherence decreases, no energy is lost due to environmental interaction. Then, after the application of $\hat{W}_{\eta_{2}}$, the diagonal state returns to its initial form, $\hat{\rho}_{0}^{diag}$, regardless of the elapsed time. Thus, the energy shifts induced by both measurements satisfy $\Delta E_{\eta_{2}} = -\Delta E_{\eta_{2}\eta_{2}}(\tau) = \omega (f - P_{0})$ for all $\tau$, ensuring $\varepsilon_{\eta_{2}\eta_{2}}(\tau) = 0$.
For incoherent states, where $Q_{0} = 0$, this results in a cyclic transformation $\hat{\rho}_{0} \rightarrow \hat{\tau}_{\textrm{th}} \rightarrow \hat{\rho}_{0}$, so that $\Delta \mathcal{R}_{\eta_{2}} = -\Delta \mathcal{R}_{\eta_{2}\eta_{2}}(\tau)$ and $\mathcal{W}_{\eta_{2}\eta_{2}}(\tau) = 0$.
On the other hand, the $\eta_{3}$-curve explicitly depends on the dissipation time $\tau$, leading to a non-trivial interpretation. Nonetheless, one can show that along this curve, it holds that $\Delta E_{\eta_{3}} = -\Delta E_{\eta_{3}\eta_{3}}(\tau) = \omega \left(f - P_{0} - e^{\gamma \tau}(f + P_{0} - 1)\right)$.

In short, a successful and suitable application of the described TWM protocol is characterized both by (i) a positive ergotropy gain, $\overline{\mathcal{R}}_{mw}(\tau) > 0$, and (ii) zero energy and ergotropy shifts, $\varepsilon_{mw}(\tau) = 0$ and $\mathcal{W}_{mw}(\tau) = 0$. While the former means the TWM protocol provided some advantage, the latter guarantees this gain is not due to extra net recharging.
In the following, we present our main results, demonstrating that this protocol yields an ergotropy gain, effectively mitigating battery discharging when the two measurement intensities and dissipation time are appropriately selected. Additionally, we identify the impact of both the coherent and incoherent components of the battery ergotropy. Without loss of generality and unless stated otherwise, it will be assumed $f=0.3$ and $\gamma=10^{-2}$ for the following analyzes.
%
%
%
%
%
%
%
%
%
%
%
\section{Ergotropy gain: The role of incoherent and coherent parts}
\label{Sec3}
\subsection{Incoherent ergotropy}\label{AppendixIncoherent}

We begin our analysis by characterizing the protocol steps for the incoherent component of ergotropy, given by $\mathcal{R}^{{inc}}[\hat{\rho}] = \mathcal{R}[\hat{\rho}^{{diag}}]$. Note that, since the state populations are independent of the initial coherence at every step, this is equivalent to describing an incoherently prepared battery. Thus, for the initial state $\hat{\rho}_{0}$, the corresponding incoherent ergotropy can be written as
\begin{equation}   \mathcal{R}^{inc}_{i}=\omega\left(2P_{0}-1\right)\left[1-\Theta\left(1/2-P_{0}\right)\right],
\end{equation}
where $P_{0} > 1/2$ ensures that the state is active. During step (ii), the weak measurement slightly projects the battery toward the ground state, which inevitably decreases the battery ergotropy, i.e., $\mathcal{R}^{{inc}}_{ii} - \mathcal{R}^{{inc}}_{i} \leq 0$. Note that the post-weak measurement state $\hat{\rho}_{m}(0)$ will remain active as long as $m < 2 - P_{0}^{-1}$ is satisfied, meaning that if the measurement strength is high enough, the battery state will become passive. Consequently, the ergotropy can be expressed as
\begin{equation}
    \mathcal{R}^{inc}_{ii}=\omega\left(1-2\frac{1-P_{0}}{\textrm{N}_{m}}\right)\left[1-\Theta\left(m-2+P_{0}^{-1}\right)\right].
\end{equation}
\begin{figure}
    \center
    \includegraphics[width=0.70\columnwidth]{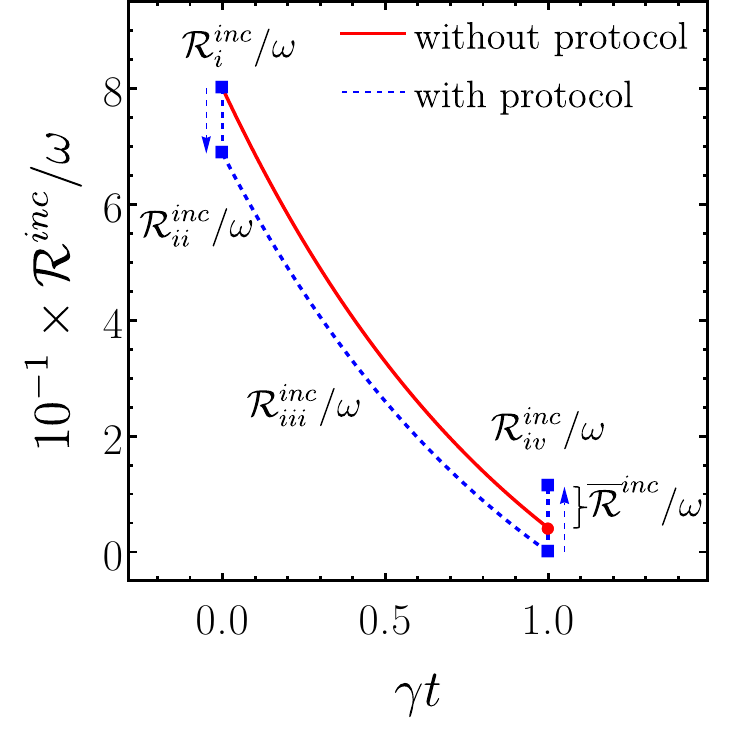}
    \caption{ \textbf{Incoherent ergotropy comparison.} Incoherent ergotropy changes with (dashed blue) and without (continuous red) the execution of the protocol for $P_{0}=0.9$, dissipation time $\tau=\tau_{\gamma}$, and strengths $m=0.4$ and $w=\tilde{w}\approx 0.2$. The blue arrows highlight the incoherent ergotropy jumps due to the measurements.
    \justifying}
    \label{fig:ErgoIncSteps}
\end{figure}

As previously mentioned, the unavoidable coupling with an external environment is detrimental to the stored ergotropy, which is asymptotically leaked, leading to $\mathcal{R}^{inc}_{iii}-\mathcal{R}^{inc}_{ii}\leq0$ (see Appendix~\ref{DissAndBatteryDischar} for more details). During this process, the ergotropy is given by
\begin{align}
    \mathcal{R}^{inc}_{iii}&=2\omega\left[\left(\frac{(1-m)P_{0}}{\textrm{N}_{m}}-f\right)e^{-\gamma t}+f-\frac{1}{2}\right]\nonumber\\&\times\left[1-\Theta\left(m-2+P_{0}^{-1}\right)\right]\left[1-\Theta\left(t-\tau_{1/2}\right)\right],
\end{align}
where
\begin{equation}\label{tauHalf}
    \tau_{1/2}=-\frac{1}{\gamma}\textrm{ln}\left[\left(\frac{1}{2}-f\right)\left(\frac{(1-m)P_{0}}{\textrm{N}_{m}}-f\right)^{-1}\right]>0
\end{equation}
is the time required for the diagonal state to become passive ($P_{m}(\tau_{1/2})=1/2$), i.e., completely discharged.
Finally, after some dissipation time $\tau$, the reversal measurement is performed, restoring part of the ergotropy, such that $\mathcal{R}^{inc}_{iv}>\mathcal{R}^{inc}_{iii}$, where $\mathcal{R}^{inc}_{iv}=\mathcal{R}^{inc}_{mw}(\tau)$ is given by Eq.~\eqref{FinalIncErgo} for the incoherent component of the final ergotropy.

Along with null energy and ergotropy shifts, the protocol performance is characterized by the ergotropy gain, quantifying the amount of ergotropy saved, relative to a quantum battery subjected to a purely discharging process.
In this context, Fig.~\ref{TauGammaWtilde} illustrates the behavior of incoherent ergotropy gain as a function of the weak measurement strength $m$ and the initial population $P_0$, computed along $\tilde{w} \in [0, 1]$ at various dissipation times: (a) $0.5 \tau_{\gamma}$, (b) $\tau_{\gamma}$, and (c) $10 \tau_{\gamma}$. For clarity, the region where $\overline{\mathcal{R}}_{m\tilde{w}}^{inc}(\tau) \leq 0$—indicating no gain following protocol execution—has been omitted. The red-hatched areas and red curves denote operational regions where $\mathcal{W}_{m\tilde{w}}(\tau) = 0$. The red lines $\eta_2$ show points characterized by the equal-strength curve, as defined in Eq.~\eqref{NullCostCurves}.
Notice that, in general, stronger measurements $m$ yield higher gains, being a direct consequence of the smaller discharging rate for less charged states. It is also evident that not all parameter space corresponds to valid operational regions. The size of these working regions depends strongly on the dissipation time $\tau$, gradually decreasing over time and eventually saturating for long dissipation times, characterized by the reversal measurement strength $\lim_{\tau \rightarrow \infty} \tilde{w}$.
In this regime, the only remaining valid execution of the protocol with a finite gain is characterized by $\eta_{2}$.
\begin{figure}[htb]
    \center
    \includegraphics[width=0.85\columnwidth]{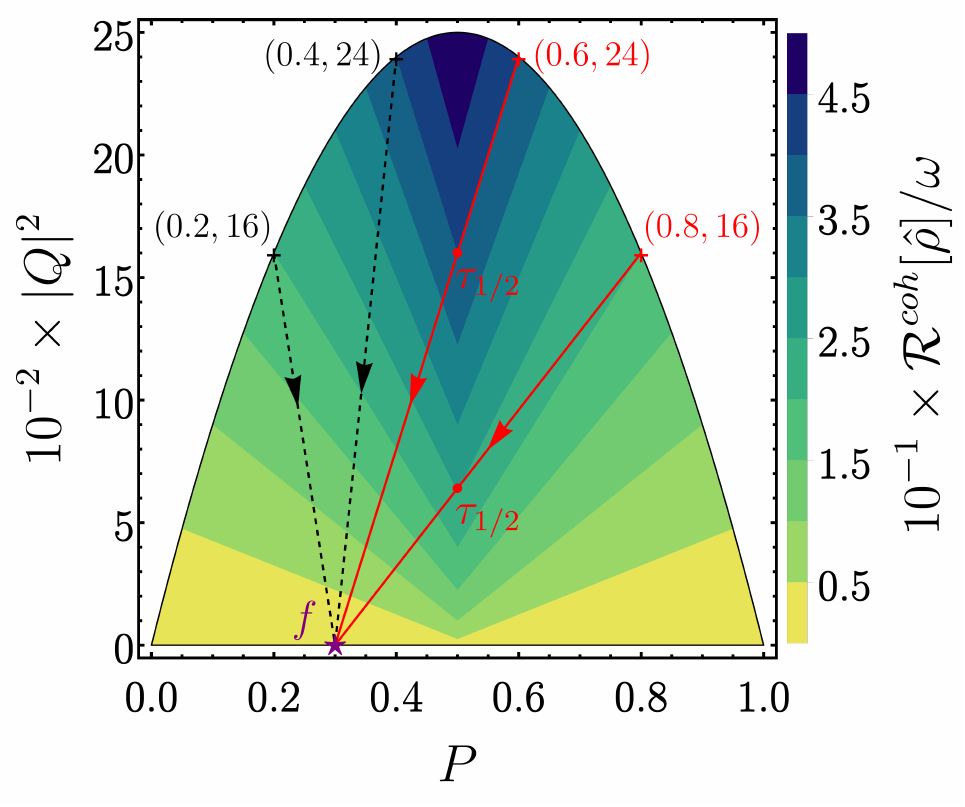}
    \caption{\textbf{Coherent ergotropy profile} in terms of $P$ and $|Q|^2\in[0,P(1-P)]$. The black (red) curves represent the dissipative dynamics of Eq.~\eqref{MasterEquation} for initial states with $P(0)<1/2$ ($P(0)>1/2$). The purple star highlights the thermal state, while the red dots indicate the instant $\tau_{1/2}$ when $P(\tau_{1/2})=1/2$.
    \justifying}
    \label{fig:CoheErgoPROFILE}
\end{figure}

To better illustrate how the protocol functions, Fig.~\ref{fig:ErgoIncSteps} shows the variation of the incoherent ergotropy during each step (dashed-blue line), assuming $P_{0} = 0.9$, $m=0.4$,
dissipation time $\tau = \tau_{\gamma}$, and $w=\tilde{w}\approx 0.2$, which ensure $\varepsilon_{m\tilde{w}}(\tau_{\gamma})=0$. This is compared to the purely discharging process (continuous-red line). The blue arrows indicate the discontinuous (instantaneous) ergotropy jumps caused by the weak and reversal measurements. According to Fig.~\ref{TauGammaWtilde}b, the chosen parameters lie within the operational region, with $\mathcal{W}_{m\tilde{w}}(\tau_{\gamma}) = 0$. Note that while $\hat{M}_{m}$ decreases the initial battery ergotropy, the execution of the final step, $\hat{W}_{\tilde{w}}$, increases it. Furthermore, the battery becomes nearly fully discharged due to environmental interaction at step (iii), since $\tau_{\gamma} \approx \tau_{1/2}$. As a result, at the end of the protocol, the final incoherent ergotropy $\mathcal{R}^{inc}_{iv}$ exceeds the red curve, yielding $\overline{\mathcal{R}}_{m\tilde{w}}^{inc}(\tau_{\gamma}) \approx 0.71 \omega \times 10^{-1}$, which represents a saving of $100\times\overline{\mathcal{R}}_{m\tilde{w}}^{inc}(\tau_{\gamma})/\mathcal{R}_{i}\approx8.89\%$ of the initial charge. It is important to emphasize that this gain is achieved without actively introducing any additional (net) external resources, which is ensured by an appropriate choice of parameters that results in null energetic and ergotropic shifts.
%
%
%
%
%
%
%
%
%
%
%
%
%
%
%
%
%
%
%
%
%
%
%
\subsection{Coherent ergotropy}\label{AppendixCoherent}
\begin{figure*}
    \centering
    \includegraphics[width=\textwidth]{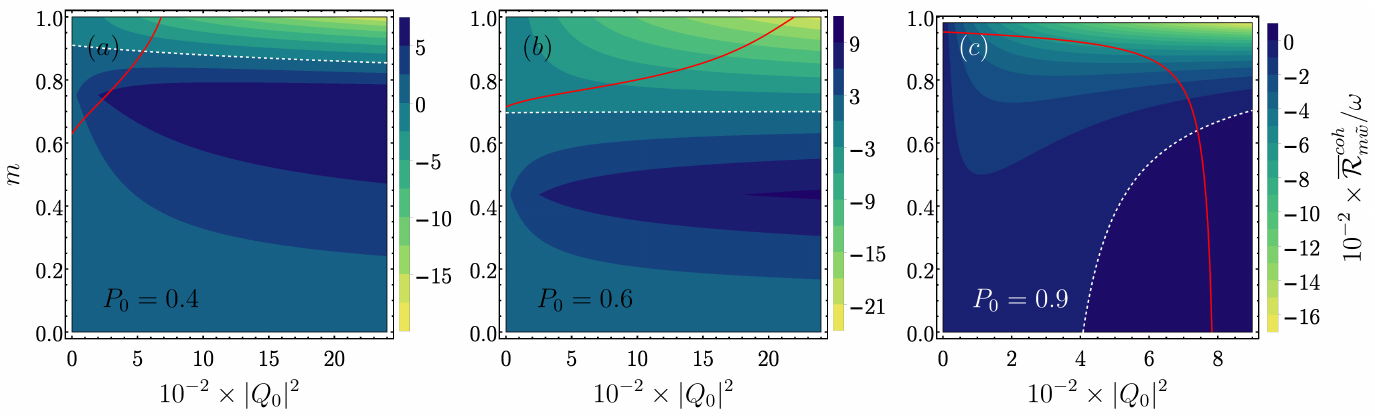}
    \caption{\textbf{Coherent ergotropy gain} as a function of $m$ and initial coherence $|Q_{0}|^{2}\in[0,P_{0}(1-P_{0})]$ computed along $\tilde{w}\in [0,1]$ and $\tau_{\gamma}$ for (a) $P_{0}=0.4$, (b) $P_{0}=0.6$, (c) $P_{0}=0.9$. The white line represents the points where $\overline{\mathcal{R}}^{coh}_{m\tilde{w}}(\tau_{\gamma})=0$. The red lines highlight the region where $\mathcal{W}_{m\tilde{w}}(\tau)=0$.
    \justifying}
    \label{CoherentErgoFig}
\end{figure*}
\begin{figure}[htb]
    \center
    \includegraphics[width=0.9\columnwidth]{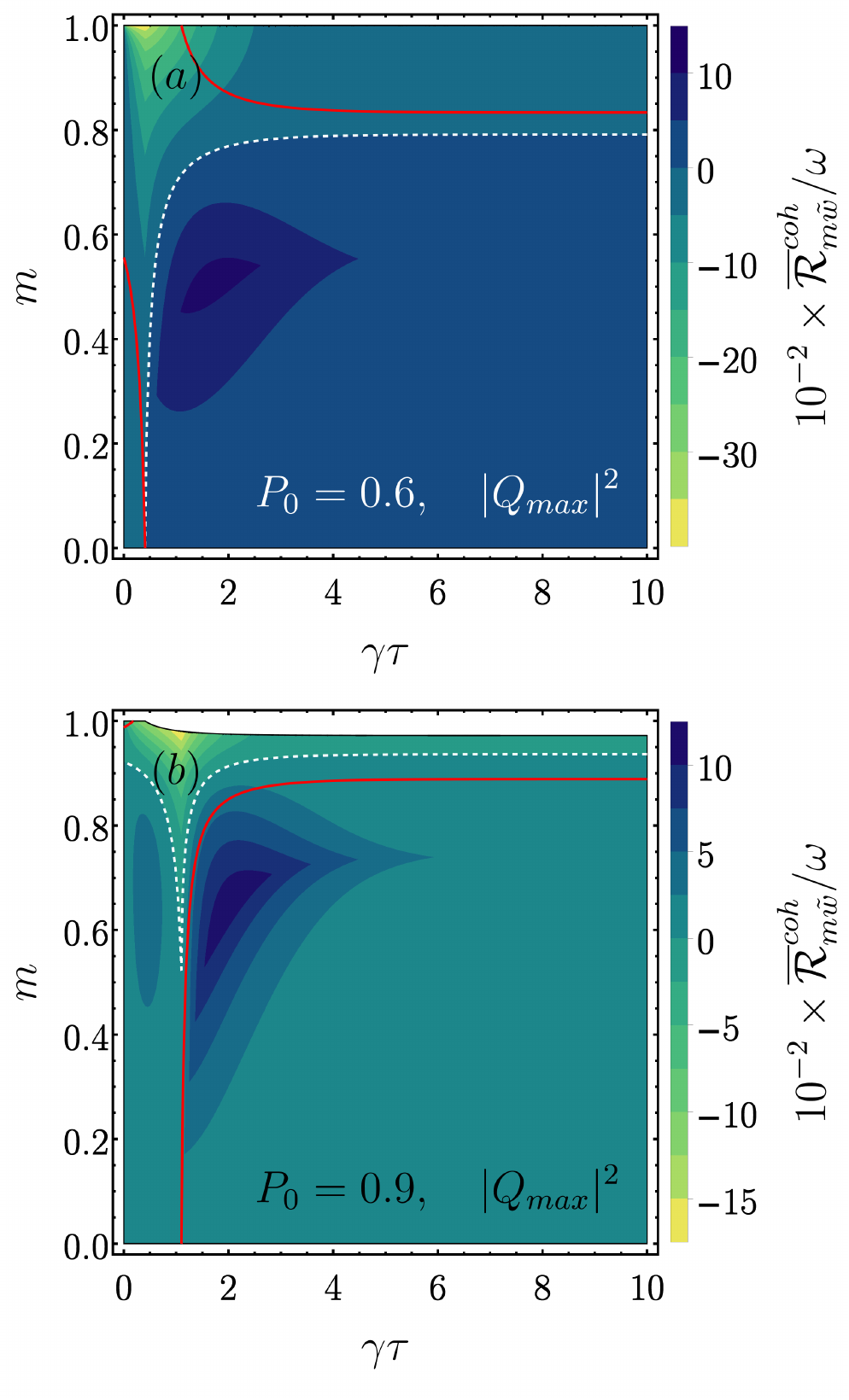}
    \caption{ \textbf{Coherent ergotropy gain} in terms of dissipation time and measurement strength $m$ for (a) $P_{0}=0.6$ and (b) $P_{0}=0.9$, assuming $|Q_{0}|^{2}=|Q_{max}|^{2}$ and $\tilde{w}\in [0,1]$. The white dashed lines represent the points with null coherent ergotropy gain. The red curves indicate the operational points, where $\mathcal{W}_{m\tilde{w}}(\tau)=0$.
    \justifying}
    \label{fig:CoherenceVStime}
\end{figure}
The effect of quantum coherence in the protocol can be analyzed by employing the coherent ergotropy, defined in Eq.~\eqref{CoherentErgoFormula}. 
Before analyzing each step, it is useful to highlight some general properties that hold regardless of the protocol.
Consider a quantum state of the form $\hat{\rho} = \left(1 - P\right)|g\rangle \langle g| + P |e\rangle \langle e| + \left(Q |g\rangle \langle e| + h.c.\right)$. The purity of this state is given by $\mu[\hat{\rho}]=P^{2}+(1-P)^{2}+2|Q|^{2}\leq1$, which implies that the coherent term is constrained by $\left|Q\right|^{2}\leq P\left(1-P\right)$. From this, the coherent ergotropy is expressed as
\begin{equation}\label{CoheErgo}
     \mathcal{R}^{coh}[\hat{\rho}]=\frac{\omega}{2} \left(-| 1-2 P|+\sqrt{(1-2 P)^2+4 |Q|^2} \right).
\end{equation}
By construction, $\mathcal{R}^{coh}[\hat{\rho}]$ is always non-negative and vanishes whenever coherence is absent ($Q=0$), including at the limits $P=0$ and $P=1$. Additionally, this function is symmetric with respect to population inversion and coherence conjugation, i.e., $\mathcal{R}^{coh}[\hat{\rho}(P, Q)]=\mathcal{R}^{coh}[\hat{\rho}(1-P, Q)]$ and $\mathcal{R}^{coh}[\hat{\rho}(P, Q)]=\mathcal{R}^{coh}[\hat{\rho}(P, Q^{*})]$, with maximum value $\mathcal{R}^{coh}[\hat{\rho}]=\omega|Q|$ which occurs at $P=1/2$, leading to an increasing behavior in terms of $P$ for $P<1/2$ and a decreasing behavior for $P>1/2$. Furthermore, the convexity condition $\partial^{2}\mathcal{R}^{coh}[\hat{\rho}]/\partial P^{2}>0$ holds for all values of $P$ and $|Q|$.

Figure~\ref{fig:CoheErgoPROFILE} illustrates the coherent ergotropy as a function of population $P$ and coherence $|Q|^2$, highlighting these key properties. Any dynamical evolution follows a trajectory within this landscape. Specifically, under the evolution governed by Eq.~\eqref{MasterEquation}, the behavior of $\mathcal{R}^{coh}[\hat{\rho}(t)]$ depends on both the initial state $(P(0), |Q(0)|^{2})$ and the stationary thermal state characterized by a fixed population $f \in [0,1/2)$. During thermalization, $|Q(t)|^2$ always decays exponentially, while $P(t)$ asymptotically approaches $f$, with $dP(t)/dt>0$ for $0 \leq P(0)<f$ and $dP(t)/dt<0$ for $f < P(0)\leq 1$. 
In Figure~\ref{fig:CoheErgoPROFILE}, black (red) curves represent trajectories for initial states with $P(0) < 1/2$ ($P(0) > 1/2$). Regardless of the initial condition, the system reaches the stationary state given by $f$ (purple star).
When $P(0) < 1/2$, $\mathcal{R}^{coh}$ decreases monotonically. However, for $P(0) > 1/2$, we observe a more complex behavior: the coherent ergotropy initially increases until the population reaches $P = 1/2$ at time $\tau_{1/2}$, after which it begins to decrease (asymptotically reaching $f$). This non-monotonic behavior means that before the inevitable discharge, systems with $P(0)>1/2$ experience a temporary enhancement of their coherent ergotropy during the interval $t\in[0,\tau_{1/2}]$.

Considering the TWM protocol, since the measurement operations are incoherent, the coherent ergotropy is non-null iff $Q_{0}\neq0$. Hence, given $\hat{\rho}_{0}$ and Eq.~\eqref{CoheErgo}, the initial coherent component of the ergotropy is written as
\begin{equation}
    \mathcal{R}_{i}^{coh}=\frac{\omega}{2}\left(-|1-2P_{0}|+\sqrt{(1-2P_{0})^{2}+4|Q_{0}|^{2}}\right).
\end{equation}
After performing the weak measurement at step (ii), in terms of the relevant parameters, the ergotropy jumps to
\begin{equation}
    \begin{split}
        \mathcal{R}_{ii}^{coh} = &-\omega\frac{|2-2P_{0}-\textrm{N}_{m}|}{2\textrm{N}_{m}}\\&+\omega\frac{\sqrt{(2-2P_{0}-\textrm{N}_{m})^{2}-4|Q_{0}|^{2}\frac{(1-P_{0}-\textrm{N}_{m})}{P_{0}}}}{2\textrm{N}_{m}}.
    \end{split}
\end{equation}
After contact with the thermal reservoir, during step (iii), one has   
\begin{equation}
    \begin{split}
        \mathcal{R}_{iii}^{coh}= &-\frac{\omega}{2}|1-2P_{m}(t)|\\
        &+\frac{\omega}{2}\sqrt{(1-2P_{m}(t))^{2}+4|Q_{m}(t)|^{2}}.
    \end{split}
\end{equation}
As mentioned above, in contrast to the incoherent counterpart, the transient dynamical behavior of the coherent ergotropy during the thermalization process is considerably less trivial. Thus, depending on the initial state of the dissipation, given by the post-weak measurement state and population $P_m(0)$, $\mathcal{R}_{iii}^{coh}$ may exhibit non-monotonic behavior with a temporary increase.
Finally, after the reversal measurement at step (iv), the coherent ergotropy becomes
\begin{equation}
    \begin{split}
        \mathcal{R}_{iv}^{coh}&=-\omega\frac{|\textrm{N}_{mw}(\tau)-2P_{m}(\tau)|}{2\textrm{N}_{mw}(\tau)}\\&+\omega\frac{\sqrt{(\textrm{N}_{mw}(\tau)-2P_{m}(\tau))^{2}+4|Q_{m}(\tau)|^{2}\frac{(P_{m}(\tau)-\textrm{N}_{mw}(\tau))}{\left(P_{m}(\tau)-1\right)}}}{2\textrm{N}_{mw}(\tau)}.
    \end{split}
\end{equation}
To provide a broader characterization of the role of coherence, Fig.~\ref{CoherentErgoFig} illustrates the landscape of coherent ergotropy gain in terms of $m$ and initial coherence $|Q_{0}|^{2} \in [0, P_{0}(1-P_{0})]$. This is computed along $\tilde{w}\in [0,1]$ at $\tau_{\gamma}$ for fixed initial populations: (a) $P_{0}=0.4$, (b) $P_{0}=0.6$, and (c) $P_{0}=0.9$. As before, the red lines indicate the working points, where $\mathcal{W}_{m\tilde{w}}(\tau)=0$. 
It is evident that, depending on the parameters, coherence can offer an advantage to the protocol and contribute positively to the ergotropy gain.
Nevertheless, at $P_{0}=0.6$, no combination of $Q_{0}$ and $m$ yields a $\overline{\mathcal{R}}^{coh}_{m\tilde{w}}(\tau_{\gamma})>0$ at operational points. Still, it is noteworthy that its scale is one order of magnitude lower than the incoherent part, suggesting that the overall protocol may still be beneficial despite small negative contributions. Also, notice that, given the monotonic decay of the coherence independently of the protocol, no gain is anticipated for long dissipation times.
In this sense, Fig.~\ref{fig:CoherenceVStime} complements Fig.~\ref{CoherentErgoFig}b and Fig.~\ref{CoherentErgoFig}c by showing how the coherent ergotropy gain behaves in terms of $\tau$ for different measurement strengths $m$ and $\tilde{w}\in [0,1]$, assuming maximum coherence allowed for each $P_{0}$, given by $|Q_{0}|^{2}=|Q_{max}|^{2}\coloneqq P_{0}(1-P_{0})$ for (a) $P_{0}=0.6$, (b) $P_{0}=0.9$. It is possible to see that, for small $\tau$, the operational values of $m$ change. After that, it asymptotically approaches a stationary value.
\begin{figure}
    \center
    \includegraphics[width=0.70\columnwidth]{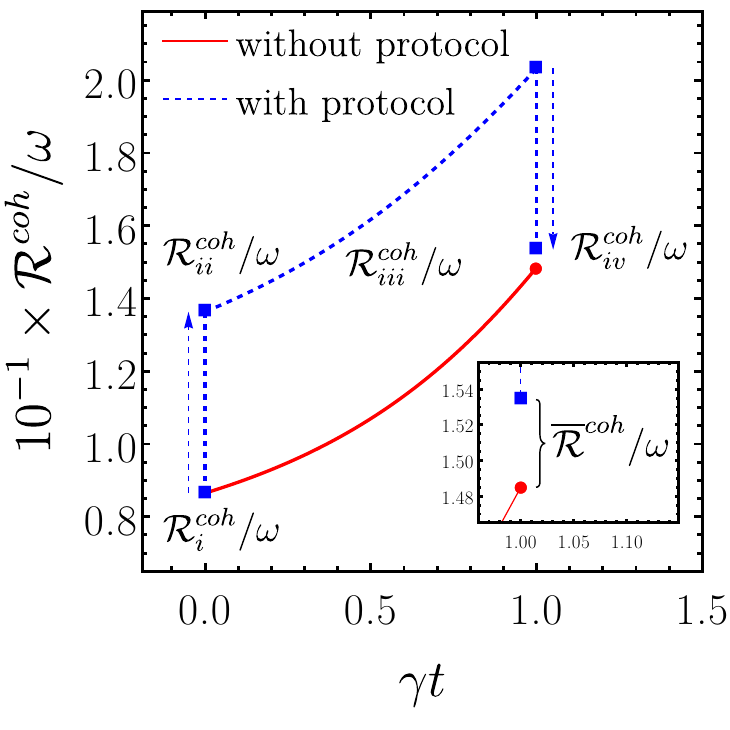}
    \caption{ \textbf{Coherent ergotropy comparison.} Coherent ergotropy changes with (dashed blue) and without (continuous red) the execution of the protocol for the same parameters used in Fig.~\ref{fig:ErgoIncSteps} and $|Q_{0}|^{2} \approx 7.67\times10^{-2}$. The blue arrows highlight the coherent ergotropy jumps due to the measurements.\justifying}
    \label{fig:ErgoCohSteps}
\end{figure}

Finally, as an illustrative example, Fig.~\ref{fig:ErgoCohSteps} displays how the coherent ergotropy changes at each step of the protocol, assuming the same set of parameters used in Fig.~\ref{fig:ErgoIncSteps} with the additional initial coherence satisfying $|Q_{0}|^{2} \approx 7.67\times10^{-2}$ to achieve an operational point (see Fig.~\ref{CoherentErgoFig}c). 
As anticipated, since $P_{m}(0)=P_{0}(1-m)/\textrm{N}_{m}>1/2$, the coherent ergotropy exhibits non-trivial behavior during the environmental interaction.
Namely, $ \mathcal{R}_{iii}^{coh}$ increases over time, reaching a turning point at $\tau_{1/2}$ given by Eq.~\eqref{tauHalf}, when the diagonal state becomes passive. However, since $\tau_{\gamma}\approx\tau_{1/2}$ for the chosen parameters, the decaying part of the dynamics and the non-monotonicity are not observed, i.e., the reversal weak measurement is performed around the maximum value achieved by the coherent ergotropy during the thermalization. Also, under these conditions and in contrast to the incoherent part, the action of $\hat{M}_{m}$ and $\hat{W}_{\tilde{w}}$ respectively increase and decrease the coherent ergotropy. Despite that, at the end of the protocol $\mathcal{R}_{iv}^{coh}$ still remains above the red curve, resulting in a small finite gain of $\overline{\mathcal{R}}^{coh}_{m\tilde{w}}(\tau_{\gamma})\approx 0.5\omega\times 10^{-2}$, equivalent to a $100\times\overline{\mathcal{R}}^{coh}_{m\tilde{w}}(\tau_{\gamma})/\mathcal{R}_{i}^{coh}\approx 5.66\%$ of extra charge.
\begin{figure*}
    \centering
    \includegraphics[width=0.9\textwidth]{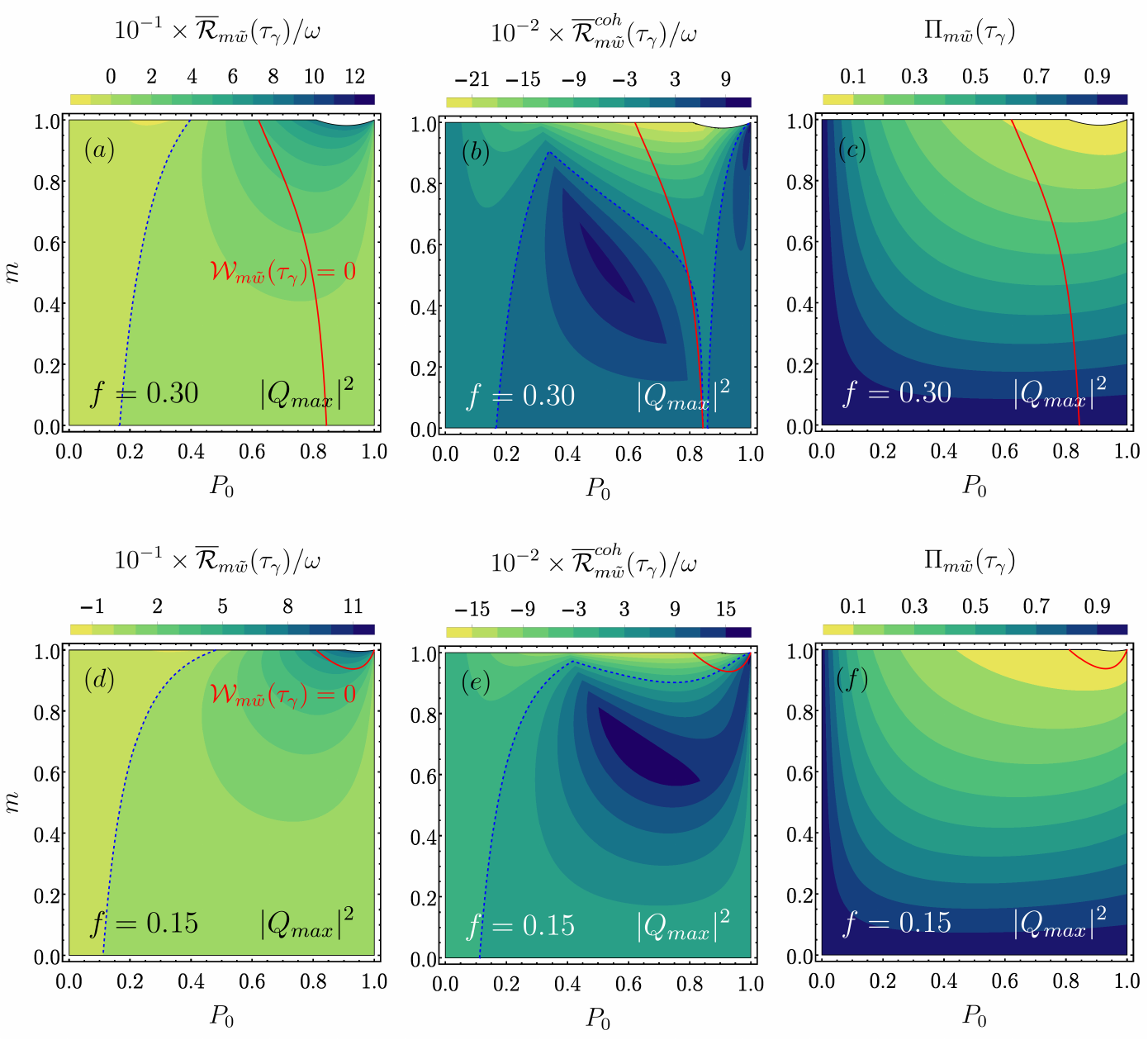}
    \caption{ \textbf{Protocol performance} in terms of the weak measurement strength $m$ and initial population $P_{0}$ for (a)-(c) $f=0.30$ and (d)-(f) $f=0.15$, higher and lower temperatures, respectively. These quantities were computed along the null energetic shift curve $\tilde{w}\in [0,1]$, considering $|Q_{0}|^{2}=|Q_{max}|^{2}$ for each $P_{0}$, and $\tau=\tau_{\gamma}$. The dashed blue curves indicate the points with null ergotropy gain, while the red lines highlight the points satisfying $\mathcal{W}_{m\tilde{w}}(\tau_{\gamma})=0$. (a) and (d) Total ergotropy gain; (b) and (e) Coherence contribution to the protocol, quantified by the coherent ergotropy gain; (c) and (f) Protocol's success probability.
    \justifying}
    \label{MainPlot1}
\end{figure*}

It is important to highlight that, although the incoherent and coherent components of the ergotropy gain behave differently according to the protocol parameters, the post-selection probabilities—and therefore the success probability—for a coherent battery and an equivalent incoherent one (with the same diagonal states) remain the same. In the following analysis, we examine the total ergotropy gain alongside the success probability to assess how likely it is to achieve the desired post-selections, while also weighing this against the benefits of mitigating the discharging dynamics.
\subsection{Total ergotropy gain and success probability}
\begin{figure}[htb]
    \center
    \includegraphics[width=1\columnwidth]{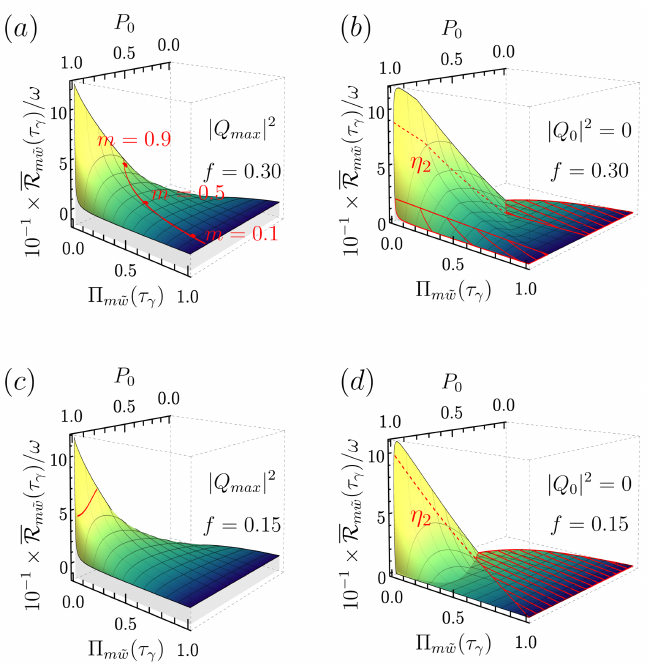}
    \caption{\textbf{Ergotropy gain $\times$ success probability} in terms of $P_{0}$ and $m$, computed at $\tau_{\gamma}$ and $\tilde{w}\in [0,1]$ for (a)-(b) $f=0.30$ and (c)-(d) $f=0.15$. In (a) and (c) it was assumed $|Q_{0}|^{2}=|Q_{max}|^{2}$, while in (b) and (d) $|Q_{0}|^{2}=0$. The red dashed and continuous lines, and the hatched areas indicate the points where $\mathcal{W}_{m\tilde{w}}(\tau_{\gamma})=0$. The null energetic shift curve $\eta_{2}$ is highlighted.
    \justifying}
    \label{fig:MainPlot2}
\end{figure}

To better assess the overall performance of the protocol, Fig.~\ref{MainPlot1} presents the relevant figures of merit for two different temperatures: higher temperature ($ f=0.30 $) in panels a–c and lower temperature ($ f=0.15 $) in panels d–f. Each row compares three key quantities as a function of the initial population $ P_{0} $ and the weak measurement strength $ m $. These quantities were computed along $\tilde{w}\in [0,1]$, dissipation time $\tau=\tau_{\gamma}$ and for the initial coherence it was assumed $|Q_{0}|^{2}=|Q_{max}|^{2}$ for each $P_{0}$. The blue dashed lines represent the points where the ergotropy gains are null, while the red continuous line highlights the protocol's working points, where $\mathcal{W}_{m\tilde{w}}(\tau_{\gamma})=0$ is also satisfied.
Figures~\ref{MainPlot1}(a) and (d) display the total ergotropy gain $\overline{\mathcal{R}}_{m\tilde{w}}(\tau_{\gamma})$. 
Notice that, despite being positive for most of the surface $P_{0}\times m$, for the chosen set of parameters the operational points of the protocol are approximately constrained to a-c $P_{0}\in[0.6,0.85]$ for all $m$ and d-f $P_{0}\in[0.8,1]$ for high values of $m\in[0.95,1]$. Also, it is possible to observe that one obtains higher gains for stronger measurements, even in regions where the coherence contribution is negative.
In this sense, Figures~\ref{MainPlot1}(b) and (e) show the coherent ergotropy gain $\overline{\mathcal{R}}^{\text{coh}}_{m\tilde{w}}(\tau_\gamma)$, revealing both positive and negative input in operational points, with stronger quantum coherence effects at lower temperatures (darker regions in panel (e)).
%
%
%
\begin{figure}[ht]
    \center
    \includegraphics[width=0.70\columnwidth]{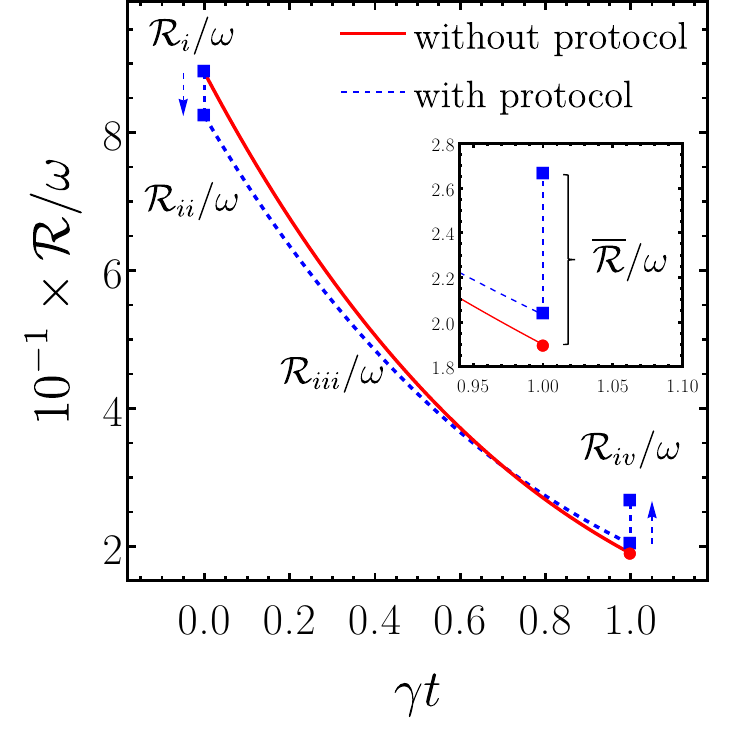}
    \caption{ \textbf{Ergotropy comparison.} Ergotropy changes with (dashed blue) and without (continuous red) the execution of the protocol for $P_{0}=0.9$, $|Q_{0}|^{2} \approx 7.67\times10^{-2}$, dissipation time $\tau=\tau_{\gamma}$, and strengths $m=0.4$ and $w=\tilde{w}\approx 0.2$. The blue arrows highlight the ergotropy jumps due to the measurements. The probability of success is $\Pi_{m\tilde{w}}(\tau_{\gamma})\approx0.57$.
    \justifying}
    \label{fig:ErgoTotalSteps}
\end{figure}
Finally, Figures~\ref{MainPlot1}(c) and (f) present the success probability $\Pi_{m\tilde{w}}(\tau_\gamma)$, showing a very similar behavior for both the temperatures considered. As a probabilistic protocol, its performance should be evaluated alongside its probability of success. In this sense, note that $\overline{\mathcal{R}}_{m\tilde{w}}(\tau_{\gamma})$ and $\Pi_{m\tilde{w}}(\tau_{\gamma})$ are inversely related in general, i.e., higher gains are obtained with lower probabilities of success and vice versa. Indeed, as described by Eq.~\eqref{TotalProbability}, stronger measurements are less probable. On the other hand, as shown in panel (e), the coherent ergotropy gain may present a positive contribution in the higher probabilities region.
It should be stressed that the observed probability landscape is a generic feature of the protocol, due to its independence of the initial coherence $Q_{0}$, i.e., the fact that the probability is maximized in the direction of lower $m$ and $P_{0}$ is observed regardless of the amount of coherence initially assumed.
%
%
%
%
%
%
Overall, this figure provides a more detailed understanding of the protocol’s behavior, emphasizing that temperature influences both the total and coherent ergotropy gains, as well as the likelihood of success. Indeed, it is possible to observe that -for the chosen parameters- the operational points are constrained in the high-gain-low-probability region for lower temperatures (lower $ f $), while for higher temperatures the number of operational points grows, suggesting that thermal fluctuations help balance the net ergotropy balance (see Appendix~\ref{EnergeticCostAppendix}).
%
%
%
%
%
%
%
%
%
%
%
%
%
%
%
%
%
%
%
%
%
%
%

Additionally, to better understand the relationship between gain and probability, Fig.~\ref{fig:MainPlot2} illustrates the ergotropy gain and the probability of success behavior assuming all values of $P_{0}$ and $m$ for $f=0.30$ in panels a-b and $f=0.15$ in panels c-d. Each row considers different values of initial coherence, with $|Q_{0}|^{2}=|Q_{max}|^{2}$ and $|Q_{0}|^{2}=0$ for the first and second row, respectively. The red lines and the hatched areas indicate the points where $\mathcal{W}_{m\tilde{w}}(\tau_{\gamma})=0$ is also satisfied. The red dots in (a) denote the working points with different values of $m$, while the red dashed line in (b) and (d) indicates the $\eta_{2}$ curve given by Eq.~\eqref{NullCostCurves}. 
The inverse relationship between $\overline{\mathcal{R}}_{m\tilde{w}}(\tau_{\gamma})$ and $\Pi_{m\tilde{w}}(\tau_{\gamma})$ is evident and is due to the unlikely nature of strong measurements and the decay of the ergotropy during dissipation.
When stronger measurements are applied (higher $m$ values), they create a more pronounced modification of the battery state, decreasing its ergotropy and leading to slower dissipation dynamics compared to the unmeasured case. Indeed, as discussed in Appendix~\ref{DissAndBatteryDischar}, the discharging rate for certain partially charged states is lower than that for charged ones. This enhanced state manipulation results in greater ergotropy gains by effectively modifying how the battery exchanges energy with its environment. However, these same stronger measurements substantially reduce the probability of obtaining the desired measurement outcomes, as they more definitively project the quantum state along specific measurement directions. This creates an inherent tradeoff, clearly illustrated in Fig. \ref{fig:MainPlot2}, where measurements with high strength values, e.g.  $m=0.9$ produce substantial ergotropy gains but with low success probabilities. In contrast, weaker measurements ($m = 0.1$) yield more modest gains with much higher reliability. This relationship suggests that practical implementations must balance effectiveness with reliability based on specific application requirements.

The total ergotropy changes induced during the protocol, composed of both the incoherent and coherent components, are depicted in Fig.~\ref{fig:ErgoTotalSteps} for the same parameters introduced previously. 
Note that, in this case, the resulting battery discharging process after the weak measurement (dashed blue line) is indeed slower than the one obtained without performing it (continuous red line). Such a difference in ergotropy loss leads to the crossing of both curves and, ultimately, more ergotropy is saved after some dissipation time.
Moreover, despite satisfying the constrain $\varepsilon_{m\tilde{w}}(\tau_{\gamma})=\mathcal{W}_{m\tilde{w}}(\tau_{\gamma})=0$, the reversal measurement gives even more ergotropy to the battery, providing a positive gain at the end of the protocol of $\overline{\mathcal{R}}_{mw}(\tau_{\gamma})\approx0.76\omega\times10^{-1}$, or $100\times\overline{\mathcal{R}}_{mw}(\tau_{\gamma})/\mathcal{R}_{i}\approx 8.57 \% $ of saved charge with the protocol. The probability of success for this particular example is $\Pi_{m\tilde{w}}(\tau_{\gamma})\approx0.57$.
%
%
%
%
%
%
%
%
%
%
%
%
%
%
%
%

\section{Extension to multi-qubit open quantum batteries}\label{MultiQubit}
Practical quantum batteries will likely require systems composed of multiple energy-storing units, or battery cells. Here, we extend our two-time weak measurement protocol to quantum batteries with $N$-qubit cells. To demonstrate this approach, we analyze a two-qubit interacting system and examine its behavior when quantum correlations are present.
For a quantum battery comprising $N$ identical qubits, the Hamiltonian takes the form $\hat{H}=\omega\sum_{k}^{N}|e\rangle\langle e|_{k}+\sum_{k\neq j}^{N}J_{kj}\hat{\sigma}_{+}^{k}\hat{\sigma}_{-}^{j}$, where $J_{kj}$ represents the coupling strength between the $k$-th and $j$-th cells, determined by the specific architecture.
The open battery dynamics are strongly influenced by parameter ranges and underlying assumptions. We adopt a local master equation derived under standard Born-Markov approximations with a common environment~\cite{breuer2002theory,cattaneo2019local}. This framework requires that internal couplings satisfy $J_{kj}\ll\omega$ for all $k, j$, allowing jump operators to be computed in their local bare Hamiltonian basis.
Under these conditions, the time evolution follows the structure of Eq.~\eqref{MasterEquation}, with dissipators given by:
\begin{equation}
    \begin{split}
        \mathcal{D}_{\downarrow}[\boldsymbol{\cdot}]&=\gamma(1-f)\left(\hat{S}_{+}\boldsymbol{\cdot}\hat{S}_{-}-\frac{1}{2}\left\{ \hat{S}_{-}\hat{S}_{+},\boldsymbol{\cdot}\right\} \right),\\\mathcal{D}_{\uparrow}[\boldsymbol{\cdot}]&=\gamma f\left(\hat{S}_{-}\boldsymbol{\cdot}\hat{S}_{+}-\frac{1}{2}\left\{ \hat{S}_{+}\hat{S}_{-},\boldsymbol{\cdot}\right\} \right),
    \end{split}\label{MasterEq2}
\end{equation}
where $\hat{S}_{\pm}=\sum_{k}^{N}\hat{\sigma}_{\pm}^{k}$ represent the raising and lowering ladder operators. This dynamics can be realized in atomic chains~\cite{PhysRevLett.125.073601, PhysRevLett.125.263601}, for instance. Notably, as a consequence of the local approach, the expression above is independent of internal couplings $J_{kj}$. Nevertheless, their effects are incorporated through the Hamiltonian component of the dynamics.
It is important to note that Eq. \eqref{MasterEq2} includes not only local dissipators but also cross-terms that couple different battery cells. Considering two qubits, for instance, these contributions may arise when the full secular approximation fails for resonant qubits, resulting in environment-induced coupling~\cite{cattaneo2019local}. Common environments have been shown to produce significant effects, including collective phenomena~\cite{PhysRevA.94.052118, Galve2017, Cattaneo2021} and steady-state entanglement~\cite{PhysRevA.85.062323, Hu2018, Ghasemian2020}, with various potential applications~\cite{PhysRevA.98.042102, Manzano_2019, e24010032, PhysRevB.107.075440,10.1063/5.0237842}. Also, it is important to highlight the dynamics induced by this master equation and, therefore, the stationary state strongly depends on the choice of the initial state.

For a $N$-cell battery, the TWM protocol follows a direct generalization of the procedure introduced in Sec.~\ref{IntroErgotropy}. The protocol involves applying multiple simultaneous local weak measurements during both the measurement and reversal steps. Specifically, for each $k$-th cell, measurements $\hat{M}_{k}$ and $\hat{W}_{k}$ are applied with respective forces $m_{k}$ and $w_{k}$, such that:
$\hat{M}=\bigotimes_{k}^{N}\hat{M}_{k}$ and $\hat{W}=\bigotimes_{k}^{N}\hat{W}_{k}$, where
\begin{equation}
    \begin{split}
        \hat{M}_{k}&=|g\rangle\langle g|_{k}+\sqrt{1-m_{k}}|e\rangle\langle e|_{k},\\\hat{W}_{k}&=\sqrt{1-w_{k}}|g\rangle\langle g|_{k}+|e\rangle\langle e|_{k}.
    \end{split}
\end{equation}
The operational points of the protocol are defined by regions in the parameter space that ensure null energetic and ergotropic shifts, i.e., $\varepsilon(\tau)=0$ and $\mathcal{W}(\tau)=0$, respectively. While the number of measurement strengths increases with the number of cells, s.t. $\textbf{m}=(m_{1},...,m_{N})$ and $\textbf{w}=(w_{1},...,w_{N})$, operational points remain achievable by relying on numerical methods. Furthermore, it is important to note that not all local measurements are required to obtain some gain, as long as the net effects of all local weak and reversal measurements compensate for each other. This implies that not all cells are required to be measured for a successful protocol execution.

To demonstrate this procedure, we will examine a two-qubit quantum battery, focusing on the feasibility of our protocol when quantum correlations, such as entanglement and quantum discord, are present. Without loss of generality, the following analysis will assume $\gamma=10^{-2}$ and $J=2\omega\gamma$, s.t. $J\ll\omega$.
\begin{figure}[ht]
    \center
    \includegraphics[width=1\columnwidth]{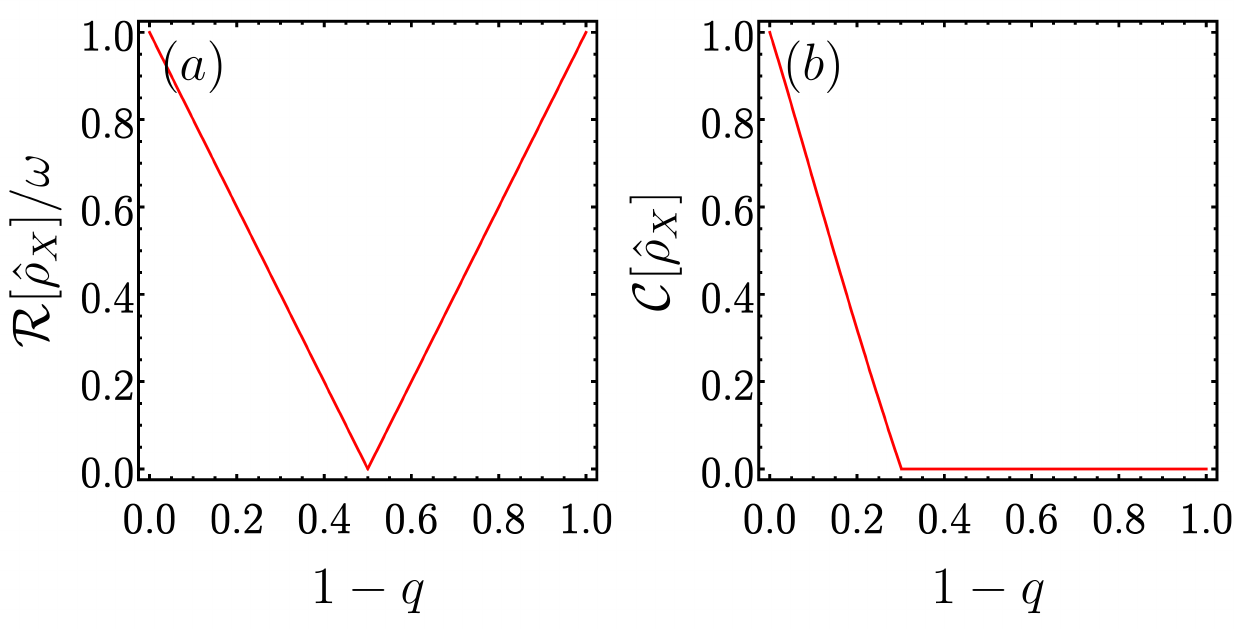}
    \caption{\textbf{$X$-state properties.} (a) Ergotropy and (b) concurrence in terms of $1-q$, assuming $J<\omega$. The symmetry of the ergotropy reveals parameter choices where systems share the same ergotropy but differ in the presence or absence of correlations.
    \justifying}
    \label{fig:ErgoConcXstate}
\end{figure}
\begin{figure}[ht]
    \center
    \includegraphics[width=1\columnwidth]{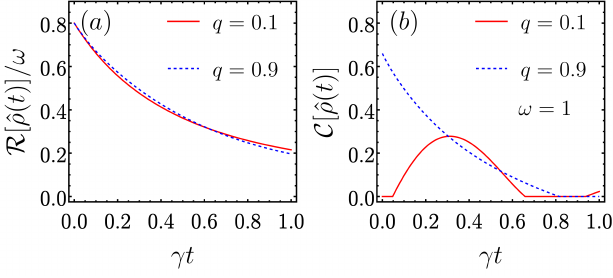}
    \caption{\textbf{Ergotropy and concurrence dynamics for the $X$-state.}  (a) Ergotropy and (b) concurrence for $\omega = 1$. The solid red and dashed blue curves represent $ q = 0.1 $ and $ q = 0.9 $, respectively. While the ergotropy decays rapidly in both cases, the concurrence exhibits different behaviors: for $ q = 0.1 $, it initially increases due to the dissipation dynamics given by Eq.~\eqref{MasterEq2} before eventually decreasing, whereas for $ q = 0.9 $, it decreases monotonically throughout the process. It was assumed
    $f=0.3$, $\gamma=10^{-2}$ and $J=2\omega\gamma$.
    \justifying}
    \label{fig:ErgoConcInTime}
\end{figure}

\subsection{Example: Two-qubit open quantum battery}\label{2qbits}
Given two uncoupled qubit batteries individually prepared in the same incoherent state $\hat{\rho}_{0}=q|g\rangle\langle g|+(1-q)|e\rangle\langle e|$, one can unitarily produce an $X$-state in the form of
\begin{equation}\label{StateXgen}
    \hat{\rho}_{X}(q)=\left(\begin{array}{cccc}
\rho_{11} & 0 & 0 & \rho_{14}\\
0 & \rho_{22} & \rho_{23} & 0\\
0 & \rho_{23}^* & \rho_{33} & 0\\
\rho_{14}^* & 0 & 0 & \rho_{44}
\end{array}\right),
\end{equation}
 that optimally enhances the quantum correlations between the cells. The resulting bipartite state $\hat{\rho}_{X}(q)$ is given by $\rho_{11}=\rho_{44}=q/2$, $\rho_{14}=q^2-q/2$, $\rho_{23}=0$, $\rho_{22}=q(1-q)$, and $\rho_{33}=(1-q)^2$~\cite{behzadi2018thermodynamic,gomes2022realism} (see Appendix~\ref{2QBdynamics} for more details).
%
%
%
%
Bipartite local weak measurements and their reversal have been shown to effectively restore quantum entanglement lost due to decoherence~\cite{kim2012protecting} and dissipation~\cite{e27040350}. Importantly, the symmetries of this model allow a meaningful comparison between two distinct scenarios with the same initial ergotropy—one where entanglement is present and another where it is absent due to entanglement sudden death.
%
%
%
%
%
%
\begin{figure}[ht]
    \center
    \includegraphics[width=0.9\columnwidth]{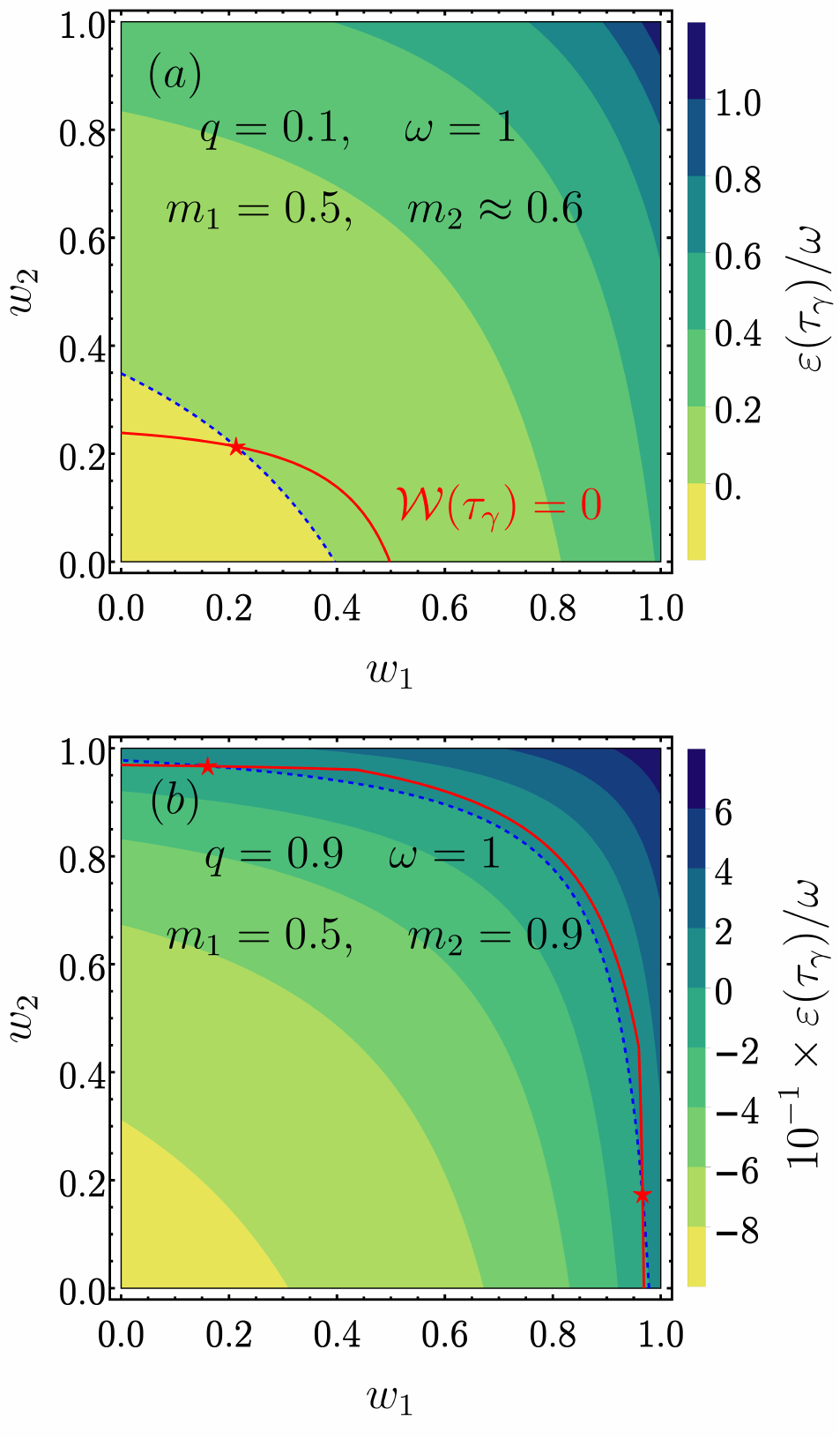}
    \caption{\textbf{Energy shift profile} at $\tau_{\gamma}$ in terms of the reversal measurement strengths $w_{1}$ and $w_{2}$ for (a) $q=0.1$ with weak measurement strengths $\textbf{m}\approx(0.5,0.6)$ and (b) $q=0.9$ with weak measurement strengths $\textbf{m}=(0.5,0.9)$. The red stars highlight the operational points where $\varepsilon(\tau_{\gamma})=0$ (dashed blue curve) and $\mathcal{W}(\tau_{\gamma})=0$ (continuous red curve) are simultaneously satisfied. It was assumed $f=0.3$, $\gamma=10^{-2}$, $J=2\omega\gamma$ and $\omega=1$.
    \justifying}
    \label{fig:OperationalPoints}
\end{figure}

Since $J<\omega$ by hypothesis, the ergotropy of $\hat{\rho}_{X}(q)$ is given by $\mathcal{R}\left[\hat{\rho}_{X}(q)\right]=\omega\left|1-2q\right|$, while the entanglement, quantified by the concurrence, is 
$\mathcal{C}[\hat{\rho}_{X}(q)]=\max\left[\left\{ 0,2q^{2}-q-2(1-q)\sqrt{q(1-q)}\right\}\right]$.
Figure~\ref{fig:ErgoConcXstate} illustrates the behavior of these quantities as functions of the excited population $1-q$. In Fig.~\ref{fig:ErgoConcXstate}(a), the ergotropy exhibits a symmetric profile with respect to $q$, such that pairs of different $X$-states share the same ergotropy: $\mathcal{R}[\hat{\rho}_{X}(1/2 - x)] = \mathcal{R}[\hat{\rho}_{X}(1/2 + x)] = 2\omega |x|$, for $x \in [-1/2, 1/2]$. Meanwhile, Fig.~\ref{fig:ErgoConcXstate}(b) reveals entanglement sudden death, with concurrence vanishing for $q \leq 0.7$.  
\begin{figure*}
    \centering
    \includegraphics[width=0.9\textwidth]{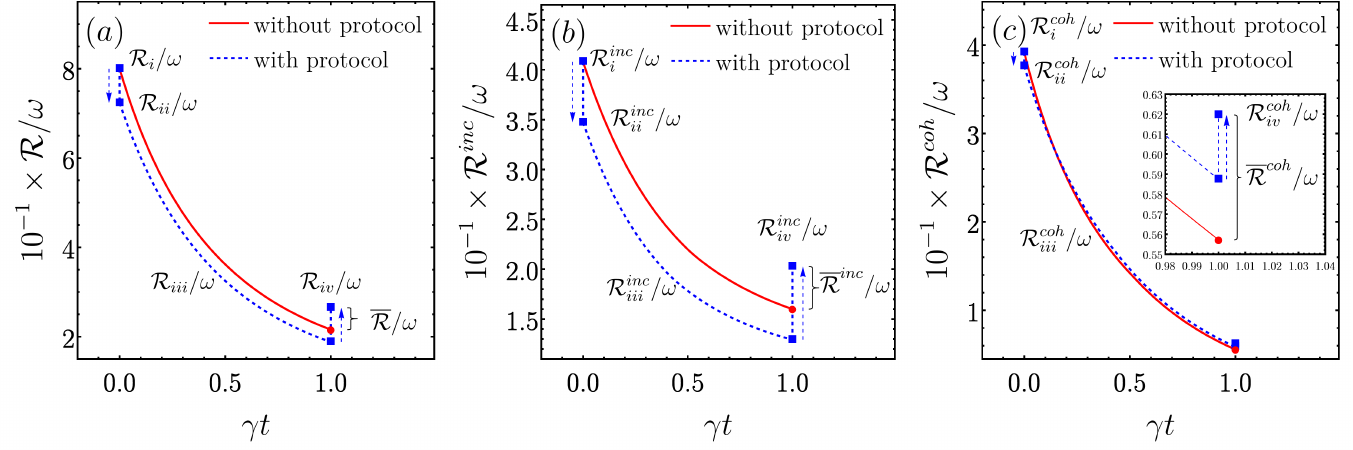}
    \caption{\textbf{Ergotropy comparison without initial entanglement.} Ergotropy changes with (dashed blue) and without (continuous red) the protocol for (a) total ergotropy, (b) incoherent ergotropy, and (c) coherent ergotropy assuming $q=0.1$, $m_{1}=0.5$, $m_{2}\approx 0.6$ and $w_{1}=w_{2}\approx 0.21$.  For this particular state, which initially has no entanglement, the protocol is more effective in recovering the incoherent part of the ergotropy.
    \justifying}
    \label{Ergotropy_q01}
\end{figure*}
\begin{figure*}
    \center
    \includegraphics[width=0.9\textwidth]{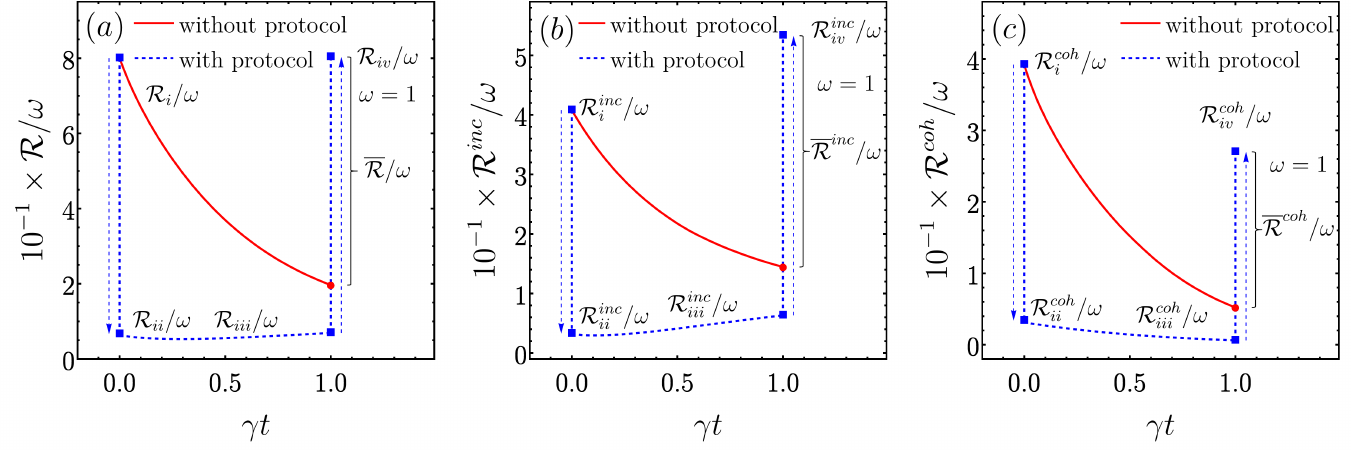}
    \caption{\textbf{Ergotropy comparison with initial entanglement.} Ergotropy changes with (dashed blue) and without (continuous red) the protocol for (a) total ergotropy, (b) incoherent ergotropy, and (c) coherent ergotropy assuming $q=0.9$, $m_{1}=0.5$, $m_{2}=0.9$, $w_{1}\approx 0.97$ and $w_{2}=0.17$. For the last step, it was assumed $\omega=1$. For the initially entangled state, the protocol performs effectively in recovering both the incoherent and coherent parts of the ergotropy, demonstrating its feasibility when quantum correlations are present.
    \justifying}
    \label{Ergotropy_q09}
\end{figure*}

In addition, assuming the initial charged state $\hat{\rho}_{X}(q)$, the battery dynamics under the master equation given by Eq.~\eqref{MasterEq2} can be shown to maintain the $X$ structure of Eq.~\eqref{StateXgen} invariant, such that
\begin{equation}
    \hat{\rho}(t)=\left(\begin{array}{cccc}
\rho_{11}(t) & 0 & 0 & \rho_{14}(t)\\
0 & \rho_{22}(t) & \rho_{23}(t) & 0\\
0 & \rho_{23}^{*}(t) & \rho_{33}(t) & 0\\
\rho_{14}^{*}(t) & 0 & 0 & \rho_{44}(t)
\end{array}\right).
\end{equation}
Due to ergotropy symmetry, distinct initial states can share the same ergotropy yet evolve differently, leading to separate steady states. Interestingly, certain initial conditions enable the generation of entanglement and/or an increase in ergotropy during both transient dynamics and in the steady state.
Figure~\ref{fig:ErgoConcInTime} depicts the time evolution of (a) ergotropy and (b) concurrence for $q = 0.1$ and $q = 0.9$ over the relevant protocol window $\gamma t \in [0,1]$. Initially, $\mathcal{R}[\hat{\rho}_{X}(0.1)] = \mathcal{R}[\hat{\rho}_{X}(0.9)] = 0.8\omega$, while the concurrence values differ:  
$\mathcal{C}[\hat{\rho}_{X}(0.1)] = 0$ and $\mathcal{C}[\hat{\rho}_{X}(0.9)] = 0.66$, for $\omega=1$.  
The ergotropy decays monotonically in both cases, but the concurrence evolves differently. For $q = 0.1$, the system generates correlations, leading to an initial increase in concurrence, followed by a decrease and slight oscillations. In contrast, for $q = 0.9$, the concurrence monotonically decreases due to environmental interactions.
\begin{figure}[ht]
    \center
    \includegraphics[width=0.8\columnwidth]{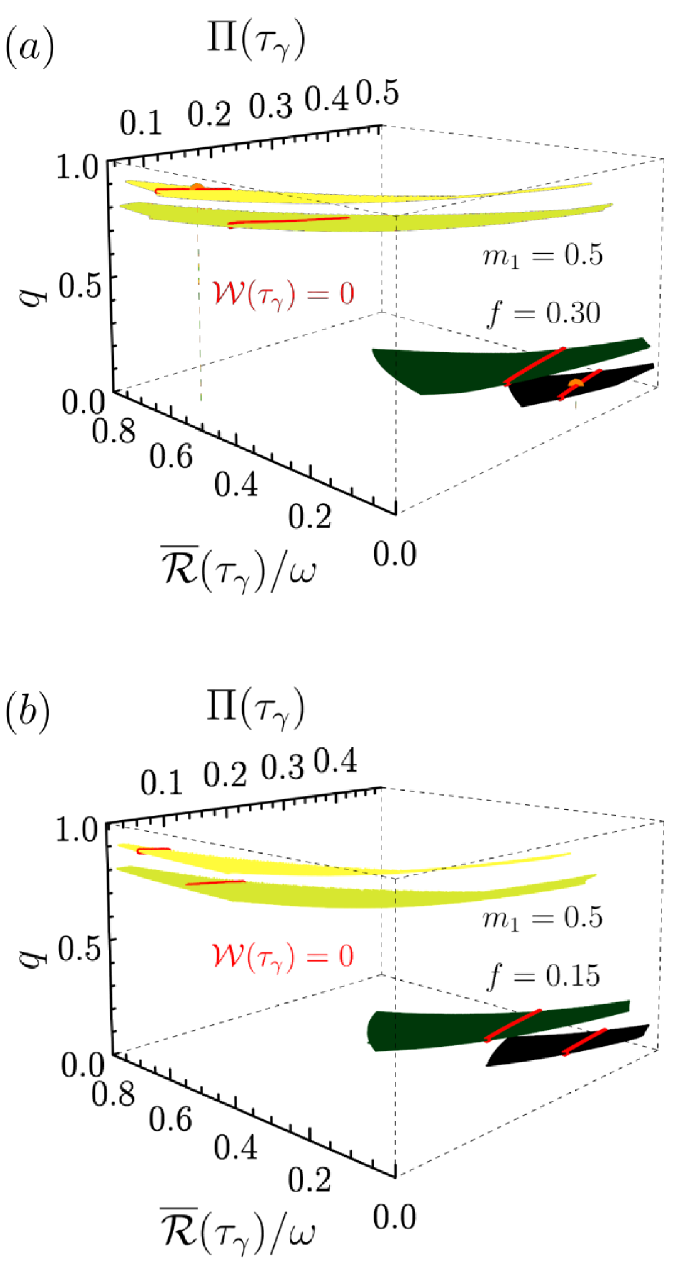}
    \caption{\textbf{Ergotropy gain $\times$ success probability} in terms of $q$ and $m_2$, $w_1$ and $w_2$, computed at $\tau_{\gamma}$, $\omega=1$ and $m_{1}=0.5$ for (a) $f=0.30$ (b) $f=0.15$. The red spots highlight the operational points, with zero ergotropy shift. The orange dots (vertical lines) at $q=0.1$ and $q=0.9$ in Fig. (a) highlight the operational points used for Figs.~\ref{Ergotropy_q01} and~\ref{Ergotropy_q09}, respectively.
    \justifying}
    \label{fig:2QubitRxP}
\end{figure}

Given the number of parameters involved, particularly the sets $\textbf{m}=(m_{1},m_{2})$ and $\textbf{w}=(w_{1},w_{2})$, determining the exact analog of Eq.~\eqref{NullCostW} for the null energetic shift surface is challenging. Nevertheless, as previously discussed, operational points can still be identified.
To illustrate this, Fig.~\ref{fig:OperationalPoints} presents the energetic shift $\varepsilon(\tau_{\gamma})/\omega$ profiles for $\omega=1$ as a function of the reversal measurement strengths $\textbf{w}$. Specifically, panel (a) corresponds to $q=0.1$ with weak measurement strengths $\textbf{m}\approx(0.5,0.6)$, while panel (b) shows the case for $q=0.9$ with $\textbf{m}=(0.5,0.9)$. In each plot, the dashed blue and continuous red curves indicate the points where $\varepsilon(\tau_{\gamma})=0$ and $\mathcal{W}(\tau_{\gamma})=0$, respectively. The intersections of these curves, marked by red stars, correspond to the operational points for the selected parameters and used for the following analysis.

Figures~\ref{Ergotropy_q01} and~\ref{Ergotropy_q09} confirm the effectiveness of the TWM protocol for two-qubit batteries in both initially non-entangled and entangled scenarios, demonstrating its potential for scaling to more complex coupled systems. In Fig.~\ref{Ergotropy_q01}, the (a) total, (b) incoherent, and (c) coherent ergotropy evolutions during the protocol (dashed blue curves) are compared to purely dissipative dynamics (continuous red curves) for $q = 0.1$, with measurement strengths $\mathbf{m} \approx (0.5, 0.6)$ and $\mathbf{w} \approx (0.21, 0.21)$. Across all ergotropy components, the weak measurement initially reduces the energy, followed by a monotonic decay, and a subsequent increase due to the reversal measurement. At the end of the protocol, a net ergotropy gain of  
$\overline{\mathcal{R}}(\tau_{\gamma}) \approx 0.49\omega \times 10^{-1}$ is achieved, with a success probability of $\Pi(\tau_{\gamma}) \approx 0.37$. This represents a $100\times\overline{\mathcal{R}}(\tau_{\gamma})/\mathcal{R}_{i}\approx 6.12\%$ of saved charge due to the protocol execution.
Similarly, Fig.~\ref{Ergotropy_q09} examines the same quantities for initially quantum correlated batteries with $q = 0.9$, using measurement strengths $\mathbf{m} = (0.5, 0.9)$ and $\mathbf{w} \approx (0.97, 0.17)$ and $\omega=1$. Despite having the same initial charge as in the $q=0.1$ case, the ergotropy evolution differs significantly. The strong weak measurement removes almost all ergotropy. Furthermore, due to collective dissipation in the shared reservoir, some initial states experience an ergotropy increase. Here, while the coherent ergotropy decreases, the incoherent and total ergotropy exhibit some growth.  
The overall ergotropy gain is $\overline{\mathcal{R}}(\tau_{\gamma}) \approx 6.1\omega \times 10^{-1}$, with a success probability of $\Pi(\tau_{\gamma}) \approx 0.09$. Interestingly, due to environment-assisted charging, the final ergotropy after the reversal measurement slightly exceeds the initial value, yielding $\mathcal{R}_{iv} - \mathcal{R}_{i} \approx 0.3\omega \times 10^{-2}$ and a recover of $100\times\overline{\mathcal{R}}(\tau_{\gamma})/\mathcal{R}_{i}\approx 75.81\%$ of the initial charge.
The protocol also affects the quantum correlations within the cells (see Appendix~\ref{2QBdynamics}). In particular, for $q=0.1$, the TWM enhances the final concurrence compared to the purely dissipative process, with gain 
$\overline{\mathcal{C}} = \mathcal{C}[\hat{\rho}_{mw}(\tau_{\gamma})] - \mathcal{C}[\hat{\rho}(\tau_{\gamma})] > 0$.

Finally, Fig.~\ref{fig:2QubitRxP} illustrates the relation between ergotropy gain $ \overline{\mathcal{R}}(\tau_\gamma) $, success probability $ \Pi(\tau_\gamma) $, and $ q $, evaluated at the operational time $ \tau_\gamma $, for fixed measurement strength $ m_1 = 0.5 $, and temperature parameter (a) $ f = 0.30 $, (b) $ f = 0.15 $. All displayed configurations satisfy the zero-energy shift condition. The red lines highlight representative operational pathways with zero ergotropy shift  $ \mathcal{W}(\tau_\gamma) = 0 $, while the vertical dashed orange lines in the  $ q = 0.1 $ and $ q = 0.9 $ surfaces in (a) mark the configurations analyzed in Figs.~\ref{Ergotropy_q01} and~\ref{Ergotropy_q09}, respectively.
From both panels of Fig.~\ref{fig:2QubitRxP}, we observe that temperature variations have a more pronounced effect on higher values of $q$.
Additionally, for lower $q$, the operational points are predominantly confined to the low-gain, high-success-probability regime, whereas higher $q$ values exhibit the opposite behavior, favoring high-gain but lower success probabilities.
In particular, for $ q \lesssim 0.7 $—where the initial $X$-states are non-entangled [cf. Fig.~\ref{fig:ErgoConcXstate}(b)]—a clear trade-off emerges: achieving higher success probabilities $ \Pi(\tau_\gamma) $ typically comes at the cost of lower ergotropy gains $ \overline{\mathcal{R}}(\tau_\gamma) $. This is visible in the downward-sloping red trajectories embedded within the green and black surfaces in that regime. In contrast, for $ q \gtrsim 0.7 $, where the initial states exhibit entanglement, this trade-off becomes less stringent—several operational points simultaneously offer both appreciable success probabilities and enhanced ergotropy gain. This indicates that initial entanglement may play a constructive role in alleviating the gain–probability trade-off.

Therefore, the TWM protocol can be effectively applied to multi-qubit quantum batteries. Specifically, we compared its behavior for two-qubit batteries in a common reservoir, where the initial states had the same charge but differed in their quantum correlations. Notably, due to collective dissipation, the initially correlated cells not only mitigated energy loss but even gained additional charge, with the cost of it being unlikely.  
Finally, we note that similar behavior and conclusions are expected to hold for other values of the internal coupling $J$, provided that $J \ll \omega$. In the local approach, the coupling primarily affects the unitary dynamics. Stronger couplings, however, fall outside the validity of Eq.~\eqref{MasterEq2}, requiring a global master equation for an accurate description.
%
%
%
%
%
%
%
%
%
%
%
%
%
%
%

\section{Discussion}\label{DiscussionSection}
Given recent advancements in quantum thermodynamics~\cite{Vinjanampathy2016} and the growing ability to manipulate quantum systems~\cite{myers2022quantum}, the development of quantum technologies for energy control and management has become increasingly crucial~\cite{werlang2014optimal,lobejko2020thermodynamics,bhattacharjee2021quantum,PhysRevE.109.064146}.
In this context, the design and analysis of efficient quantum devices for energy storage have gained significant interest, with proposals across different settings and physical platforms~\cite{RevModPhys.93.025005,PhysRevA.109.012204,batteries8050043,Hu_2022,PhysRevLett.131.260401,PhysRevA.110.032205,RevModPhys.92.015004,quach2022superabsorption,PhysRevA.106.042601,Cruz_2022,Zheng_2022}. Nevertheless, due to the fragile nature of quantum systems, realistic implementations must consider environmental interactions, which often lead to unwanted discharging. Developing methods to counteract such discharge effects has thus become a key focus for practical quantum battery applications~\cite{PhysRevResearch.2.013095}, and this work contributes to this ongoing effort.

Here, we introduced a protocol for mitigating the inevitable discharging process in open quantum batteries by applying two selective weak measurements. More specifically, a first weak measurement is performed right after the battery charging—before being subjected to environmental interaction—, followed by a reversal measurement after some dissipation time. The protocol’s operating points were set by thermodynamic constraints, namely by simultaneously meeting the conditions $\varepsilon_{mw}(\tau)=0$ and $\mathcal{W}_{mw}(\tau)=0$, which respectively define the proper parameter space for minimum external energetic and ergotropic influences. Such conditions can be achieved due to the symmetry of the weak measurement and its reversal. In this way, we showed that by properly executing the protocol, one can obtain positive gain, i.e., the final ergotropy can be higher compared to a purely discharged quantum battery, effectively mitigating the effects of environmental interactions. Nevertheless, it is worth mentioning the previous constraints can be relaxed, provided the ergotropy shift $\mathcal{W}_{mw}(\tau)$ remains below the protocol-induced ergotropy gain; in other words, even allowing for small extra recharging from the measurements can be advantageous, depending on the success probability. In the more restrictive case—where both measurements result in zero energy and ergotropy shifts—we analyzed the differences in behavior for the incoherent and coherent components of ergotropy.

Our results demonstrate that quantum coherence can be analyzed separately to optimize the performance of the protocol, allowing for better conditions for both the ergotropy gain and the probability of success. The optimal setup will naturally depend on the specific characteristics of the battery system and the environment. Our work characterizes a general scenario to facilitate adaptation to specific applications.

To validate the TWM protocol, we selected specific parameter examples. Since any quantum system decays over the time scale determined by the dissipation rate into the environment, we concentrated on the energy (ergotropy) mitigation in the period of $\tau_{\gamma}=1/\gamma$. It is also worth mentioning that the role of temperature, encoded in $f$, directly influences the dissipation process, affecting both the stationary state and the rate at which it is reached.  
Nevertheless, the choice of temperature does not fundamentally constrain the presented results. The findings are expected to hold for different temperature values, provided that the relevant parameters—particularly the weak and reversal measurement strengths, and dissipation time—are appropriately adapted.
In general, we observed that achieving higher ergotropy gains comes at the cost of lower probabilities of success, and vice versa, indicating an inverse relationship. Notably, the success probability landscape of the protocol for a single qubit battery remains invariant with respect to the initial coherence. Through our examples, we demonstrated how the coherent ergotropy gain can positively contribute at higher probabilities, showcasing how quantum features like coherence can be effectively harnessed to enhance the protocol's overall performance.

Moreover, we discussed how to extend the analysis to multi-cell quantum batteries, demonstrating how the two-time weak measurement protocol generalizes to systems composed of $N$ qubits. To validate this approach, we examined a specific scenario consisting of two interacting qubits sharing a common environment. This kind of dynamics captures non-trivial phenomena due to collective dissipation, such as entanglement and ergotropy increase. 
To contemplate the potential role of correlations, we considered $X$-states as a case study. In particular, we employed the symmetries within these states to compare two distinct scenarios with the same initial ergotropy, one not containing entanglement and the other containing it. The protocol is shown to work in both cases, mitigating the ergotropy loss compared to the purely dissipative case.

The results presented here can be straightforwardly expanded to consider multi-qubit batteries in different relevant regimes. In particular, future work should consider strongly coupled cells and characterize the role of internal coupling on the overall performance of the TWM protocol. We highlight that this regime is beyond the validity of Eq.~\eqref{MasterEq2}, requiring the use of a global master equation for an accurate dynamic description and thorough analysis of the coupling.
Additionally, it would be valuable to deeply understand the general features of quantum correlations such as entanglement and quantum discord~\cite{ollivier2001quantum,dieguez2018weak} within the overall performance of the TWM protocol, especially how it could scale with the number of cells and depend on the configuration of internal couplings.
Besides, it would be relevant to consider how the recently proposed notions of parallel ergotropy~\cite{castellano2024parallel}, local ergotropy~\cite{PhysRevA.107.012405,di2024local}, and extended local ergotropy~\cite{PhysRevLett.133.150402} change during such a procedure. Investigating its viability for strong system-environment couplings~\cite{di2024local} and non-Markovian environments~\cite{kamin2020non,selfdis} is also of fundamental interest. Examining the energetic cost of measurements, following Landauer’s principle~\cite{Georgescu2021}, would also be important, as this would provide insights into the thermodynamic cost associated with implementing the weak measurements, allowing for a more comprehensive assessment of the protocol's practicality.

Finally, optimal parameter choices for successfully applying the protocol will likely depend on system-specific and environmental constraints. In realistic setups, it may not always be possible to control factors like dissipation time and temperature or to precisely modulate measurement strength, which could limit the achievable gains. Future work should examine optimization techniques and consider these constraints to investigate adaptive protocols that can respond to variable conditions, enabling the practical application of this approach across a wider range of quantum battery designs and operational environments.
In this context, it is worth highlighting that the proposed protocol is practically feasible using state-of-the-art platforms. The implementation of weak measurements and measurement reversal operations for mitigating decoherence has already been demonstrated~\cite{kim2012protecting, protect1, Lee:11,protect5}, while quantum batteries have been realized in comparable settings. Thus, by considering the thermodynamic constraints, one could in principle achieve the protocol's working regime.
Considering recent experimental advancements in both areas, NMR~\cite{Cruz_2022, lisboa2022experimental} and superconducting circuits~\cite{Hu_2022,Zheng_2022} emerge as promising candidates for the application of the two-time weak measurement (TWM) protocol introduced in this work.

%
\section{Acknowledgements}
The authors thank Michał Horodecki and Paweł Mazurek for the insightful discussions and suggestions, and the referees for the valuable inputs. A.H.A.M. and R.S. acknowledge support from National Science Centre, Poland Grant OPUS-21 (No. 2021/41/B/ST2/03207). B.A. acknowledges support from National Science Center, Poland within the QuantERA II Programme (No 2021/03/Y/ST2/00178, acronym ExTRaQT) that has received funding from the European Union’s Horizon 2020. P.R.D. acknowledges support from the NCN Poland, ChistEra-2023/05/Y/ST2/00005 under the project Modern Device Independent Cryptography (MoDIC).
%
%
%
%
%
%
%
%
\appendix
\section{Dissipation and battery discharging}\label{DissAndBatteryDischar}
In the basis $\{\ket{g},\ket{e}\}$, the density matrix of the battery is written as
\begin{eqnarray}
\hat{\rho}(t)=\left(\begin{array}{cc}1-P(t) & Q(t) \\ Q^{*}(t) & P(t)\end{array}\right).
\end{eqnarray}
\begin{figure}[htb]
    \center
    \includegraphics[width=0.7\columnwidth]{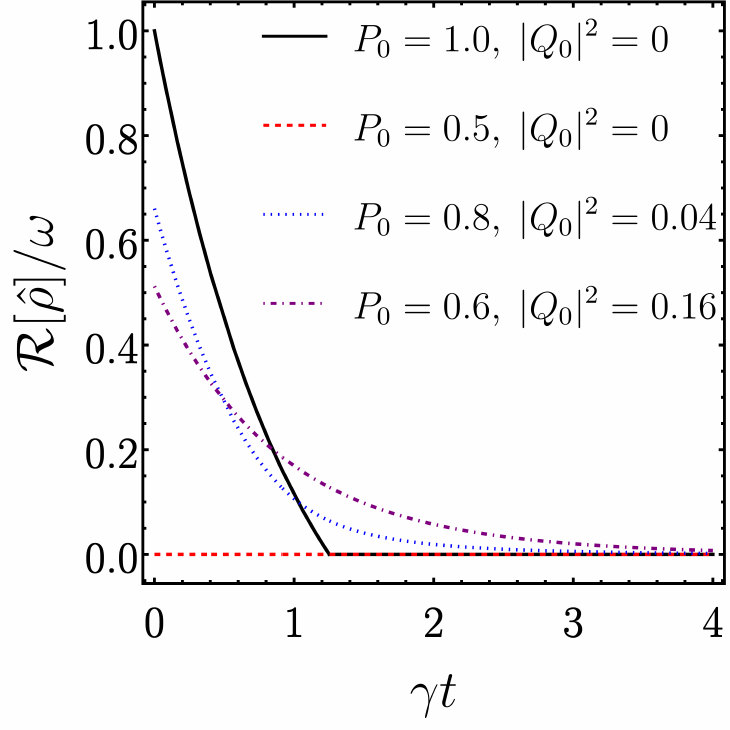}
    \caption{\textbf{Discharging process:} Ergotropy behavior as a function of the dissipation time for distinct initial states, characterized by $P_0$ and $|Q_0|^2$. It was assumed $f=0.3$ and $\gamma=10^{-2}$.\justifying}
    \label{fig:NewAppendixA0}
\end{figure}
Eq.~\eqref{MasterEquation} provides its time-evolution when the battery is exposed to an environment with the inverse of temperature $\beta$. In this case, it is straightforward to obtain the following set of uncoupled differential equations for the matrix elements $P(t)$ and $Q(t)$:
\begin{equation}
    \begin{split}
        \frac{d}{dt}P(t)&=\gamma(f-P(t)),\\
        \frac{d}{dt}Q(t)&=-\frac{1}{2}Q(t)(\gamma-2i\omega),
    \end{split}
\end{equation}
such that
\begin{equation}
    \begin{split}
        P(t)&=\left(P_{0}-f\right)e^{-\gamma t}+f,\\
        Q(t)&=Q_{0}e^{-\frac{\gamma t}{2}+it\omega},
    \end{split}
\end{equation}
where $P_0$ and $Q_0$ represent the initial population of the excited state and the initial coherence, respectively. It is important to note that these quantities evolve independently over time, without any coupling between them. Furthermore, in the long time limit ($t\rightarrow\infty$) we have $P_{\infty}=f$ and $Q_{\infty}=0$, i.e.,
\begin{equation}
    \lim_{t\rightarrow\infty}\hat{\rho}(t)=\hat{\tau}_{\textrm{th}},
\end{equation}
where $\hat{\tau}_{\textrm{th}}\coloneqq\left(1-f\right)|g\rangle\langle g|+f|e\rangle\langle e|$ is the thermal (Gibbs) state relative to $\beta$. Thus, it is clear that energy is irreversibly leaked into the environment during the battery thermalization, leading to its inevitable discharge.

This phenomenon is illustrated in Fig.~\ref{fig:NewAppendixA0} for various coherent and incoherent initial states, characterized by $P_0$ and $|Q_0|^2$. Notably, when the initial state is active, the ergotropy decreases monotonically and irreversibly over time. In contrast, for passive states, the ergotropy remains constant and zero, as indicated by the red dashed line. Additionally, note that the discharging rates vary significantly. For instance, in the case of a fully charged battery, represented by the solid black line, the decay is faster than for the initially coherent states. From this perspective, certain partially charged initial states may retain more ergotropy after a period of dissipation, presenting a potential advantage. This counterintuitive behavior arises from the complex interplay between the incoherent and coherent components of ergotropy during thermalization, resembling the ergotropic counterpart of the Mpemba effect~\cite{mpemba1969cool, medina2024anomalous}.
\begin{figure}[htb]
    \center
    \includegraphics[width=0.9\columnwidth]{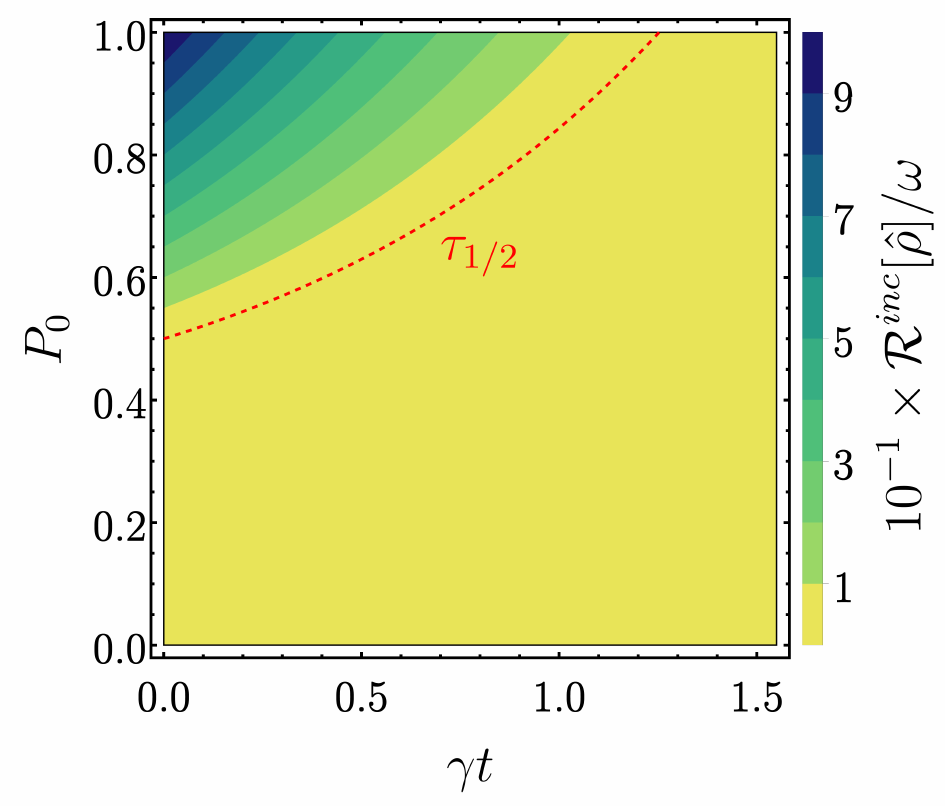}
    \caption{ \textbf{Incoherent ergotropy} in terms of the dissipation time and initial population $P_{0}$. The dashed red line highlights the curve $\tau_{1/2}$, representing the instant where the diagonal state becomes passive ($P(\tau_{1/2})=1/2$). It was assumed $f=0.3$, $\gamma=10^{-2}$.\justifying}
    \label{fig:NewAppendixA1}
\end{figure}

In such a scenario, the evolution of the incoherent ergotropy reads 
\begin{equation}\label{IncoErgoAppendixA}
\begin{split}
    \mathcal{R}^{inc}[\hat{\rho}(t)]&=\omega\left(2P(t)-1\right)\left[1-\Theta(1/2-P_{0})\right]\\&\times\left[1-\Theta(t-\tau_{1/2})\right],
\end{split}
\end{equation}
where $\Theta(x)$ is the Heaviside function, such that $\Theta(x)=1$ if $x\geq0$ and null otherwise, and
\begin{equation}
    \tau_{1/2}\coloneqq -\frac{1}{\gamma}ln\left[\left(\frac{1}{2}-f\right)\left(P_{0}-f\right)^{-1}\right]
\end{equation}
is the time instant when the diagonal state of the battery becomes passive ($P(\tau_{1/2})=1/2$) and the ergotropy reaches zero. In this sense, Fig.~\ref{fig:NewAppendixA1} shows the incoherent ergotropy landscape behavior relative to the dissipation time and initial population. For initially active diagonal states ($P_{0}>1/2$), the monotonicity of the ergotropy is evident. The dashed red line highlights the points where the battery is fully discharged, defined by $\tau_{1/2}$.

\begin{figure}[htb]
    \center
    \includegraphics[width=0.9\columnwidth]{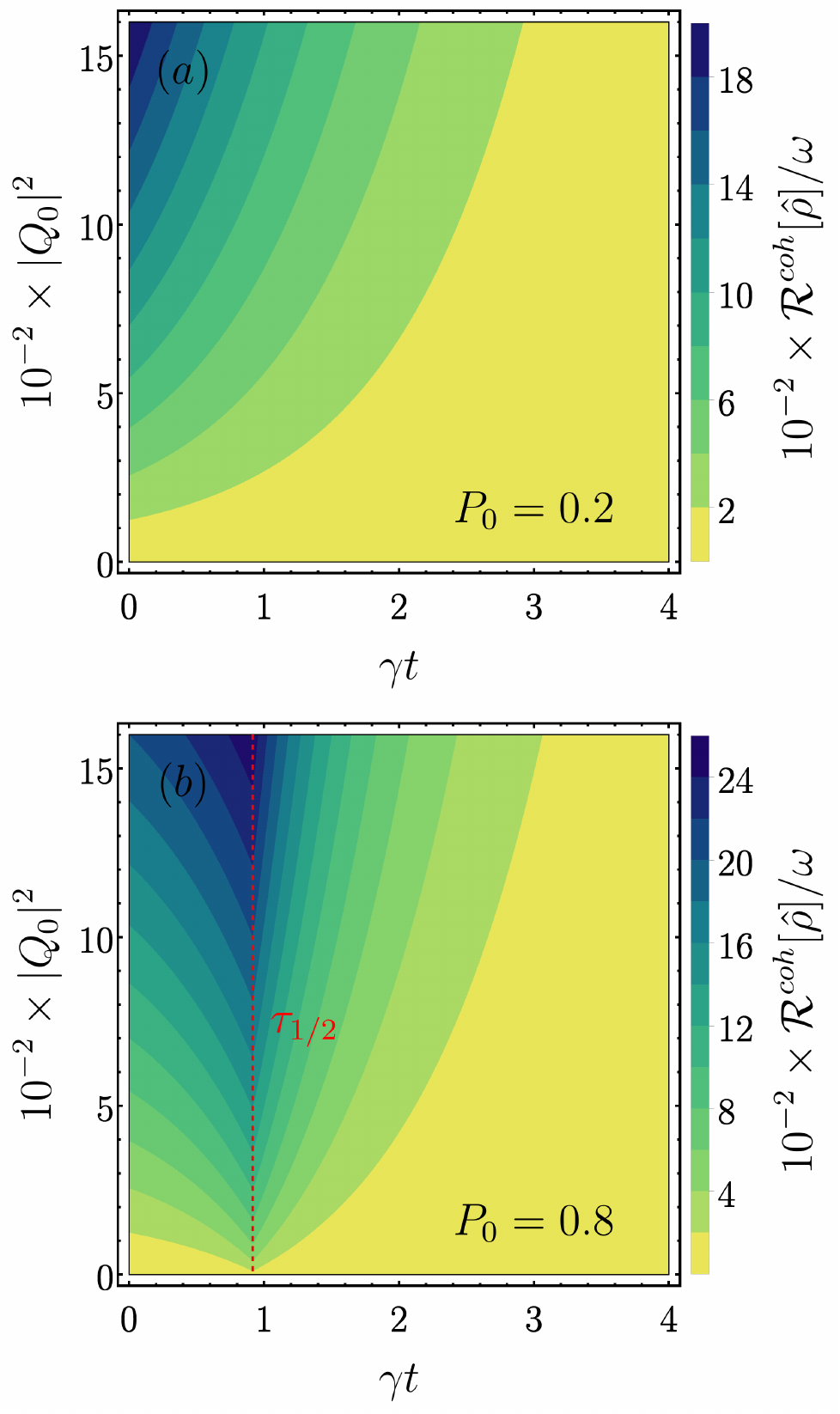}
    \caption{ \textbf{Coherent ergotropy} in terms of the dissipation time and initial coherence $|Q_{0}|^{2}\in[0,P_{0}(1-P_{0})]$ for (a) $P_{0}=0.2$ and (b) $P_{0}=0.8$. The dashed red line highlights the curve $\tau_{1/2}$. It was assumed $f=0.3$, $\gamma=10^{-2}$.\justifying}
    \label{fig:NewAppendixA2}
\end{figure}

On the other hand, the coherent ergotropy at time $t$ is written according to Eq.~\eqref{CoherentErgoFormula}.
As elucidated in the main text, it possesses a less trivial structure and dynamics, which is monotonic for $P_{0}\leq1/2$ and non-monotonic otherwise. Thus, despite environmental interaction, $\mathcal{R}^{coh}[\hat{\rho}(t)]$ may initially increase before eventually decreasing for $t>\tau_{1/2}$, which significantly influence the battery's discharging rate.
To better access the dynamics of the coherent ergotropy, Figure~\ref{fig:NewAppendixA2} illustrates its landscape as a function of dissipation time and initial coherence $|Q_{0}|^{2}\in[0,P_{0}(1-P_{0})]$ for (a) $P_0 = 0.2$ and (b) $P_0 = 0.8$. The dashed red lines indicate the curves $\tau_{1/2}$.
As expected, in Fig.~\ref{fig:NewAppendixA2}a, the coherent ergotropy decays monotonically over time while $P(t)$ approaches $f$. In contrast, in Fig.~\ref{fig:NewAppendixA2}b, the previously described behavior is observed: $\mathcal{R}^{coh}[\hat{\rho}(t)]$ increases over $t \in [0, \tau_{1/2}]$, followed by a subsequent decay for $t > \tau_{1/2}$. It is also noteworthy that the contribution of the coherent component to the total ergotropy is smaller than that of the incoherent component.

In short, although the discharging process due to environmental interaction has a relatively straightforward description, the decay of total ergotropy arises from a complex interplay between the incoherent and coherent components. While the incoherent component exhibits monotonic behavior, the coherent component may show a non-trivial time dependency within certain parameter ranges.
%
%
%
%
%
%
%
%
%
%
%
%
%
%
%
%
%
%
%
\section{Energy and ergotropy shifts}\label{EnergeticCostAppendix}
\begin{figure}[h]
    \center
    \includegraphics[width=0.9\columnwidth]{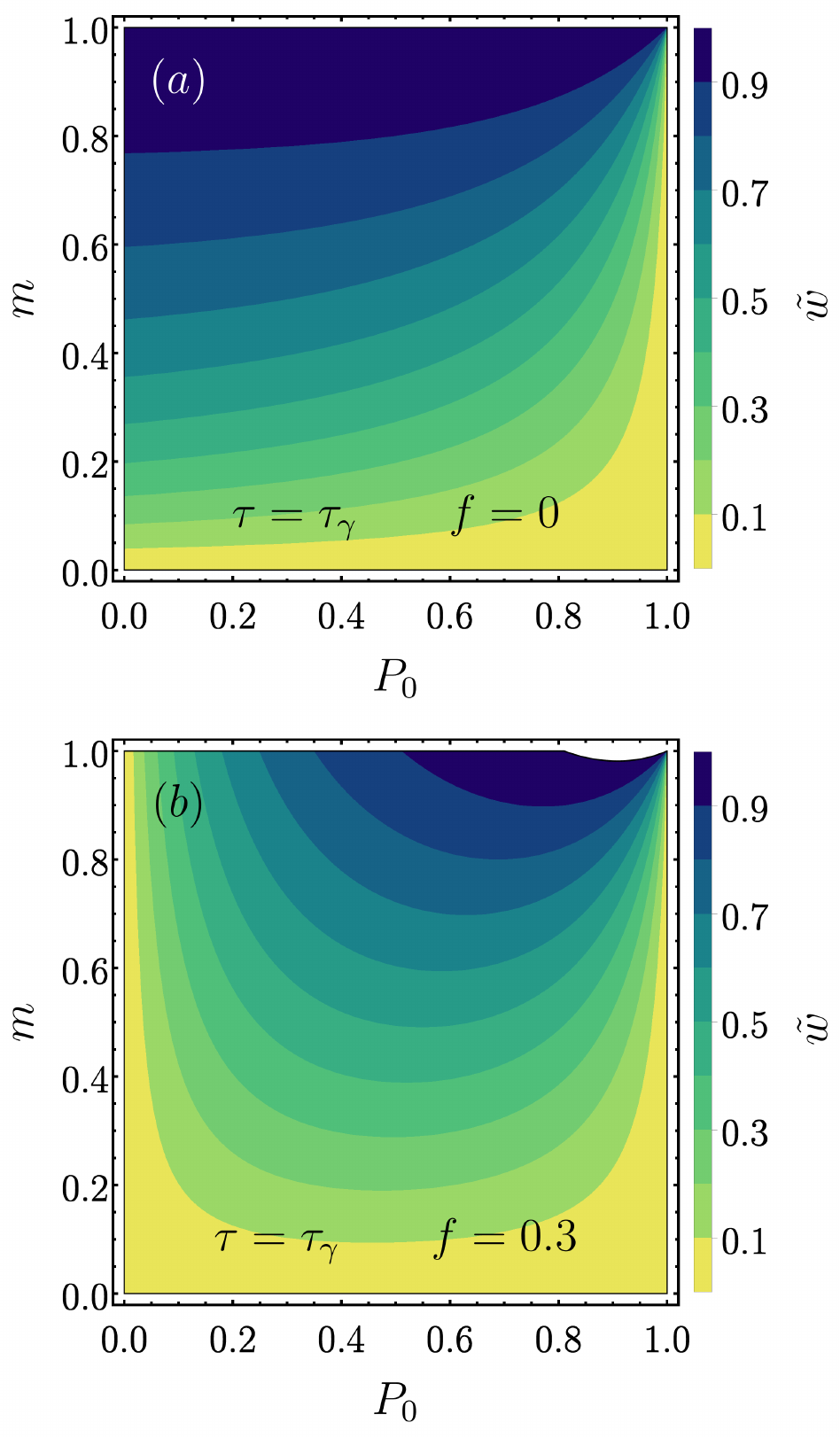}
    \caption{ \textbf{Null energetic shift $\tilde{w}$-surface} in terms of $P_{0}$ and $m$, computed at $\tau_{\gamma}$ for (a) $f=0$ and (b) $f=0.3$. The non-physical region, $\tilde{w}\notin[0,1]$, is suppressed. It was assumed $\gamma=10^{-2}$.\justifying}
    \label{NullCostPlot}
\end{figure}

By design, the described protocol requires the application of $\hat{M}_{m}$ and $\hat{W}_{w}$ to the battery state before and after some dissipation time. Both actions inherently cause energetic changes, resulting in additional charging or discharging of the battery. To ensure minimal external disturbance, it is essential that no net energy or ergotropy is imparted. In other words, the valid working regime of the protocol should be constrained to parameters that simultaneously satisfy $\varepsilon_{mw}(\tau) = 0$ and $\mathcal{W}_{mw}(\tau)=-\varepsilon_{mw}^{p}(\tau) = 0$. For fixed values of $\gamma$, $f$, $P_0$, $m$, and $\tau$, the former condition can be achieved by adjusting the reversal measurement strength to $\tilde{w}$, as defined in Eq.~\eqref{NullCostW}.
Figure~\ref{NullCostPlot} illustrates the behavior of the $\tilde{w}$-surface for a null energetic shift in terms of the initial excited population $P_0$ and strength $m$, evaluated at $\tau = \tau_{\gamma}$ for different temperatures (encoded in $f$) given by (a) $f=0$ and (b) $f=0.30$. The suppressed region in (b) represents non-physical values of measurement strength, where $\tilde{w} \notin [0,1]$. 
In general, higher values of $\tilde{w}$ are required as $m$ increases. However, its behavior in terms of $P_{0}$ clearly depends on the temperature, especially for $P_{0}<0.5$. In this sense, for $f=0.30$ the stronger values of $\tilde{w}$ are constrained to the intermediate values of $P_{0}$, while this region increases for lower temperatures toward small $P_{0}$.
Also, for both cases, one should have $\tilde{w} = 0$ to ensure that $\varepsilon_{m\tilde{w}}(\tau) = 0$ for $P_{0}=1$ and $P_{0}=0$ (not visible in (a)), i.e., since the post-weak measurement state remains invariant under the action of $\hat{M}_{m}$ no energy is transferred during the initial measurement, which means the reversal measurement strength should be null.
These general characteristics hold for different dissipation times $\tau$. Notably, for long times and $f\neq0$, the $\tilde{w}$-surface converges
\begin{equation}\label{WTildeLimit}
    \lim_{\tau\rightarrow\infty}\tilde{w}=\frac{(1-P_{0})(1-\textrm{N}_{m})}{(1-f)\left[(1-P_{0})(1-\textrm{N}_{m})+f\textrm{N}_{m}\right]}.
\end{equation}
For $f=0$ and $\tau\rightarrow\infty$, the battery thermalizes to the ground state, such that the reversal measurement has no effect and the only energetic contribution is due to the first weak measurement. In this case, the only way to guarantee a null energetic shift is by $m=0$ or $P_{0}=0,1$.

\begin{figure*}
    \centering
    \includegraphics[width=0.9\textwidth]{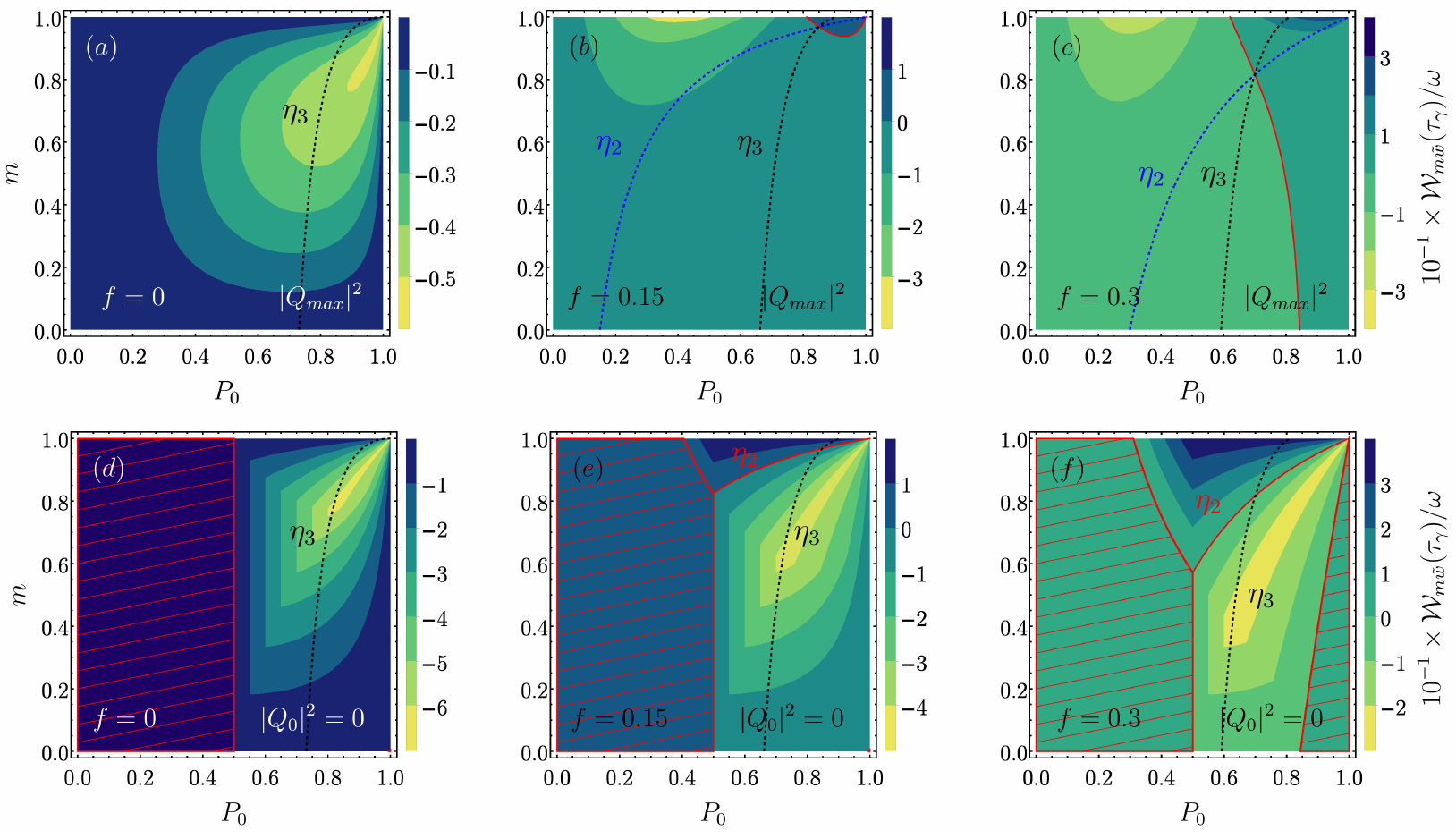}
    \caption{\textbf{Ergotropy shift profile} at time $\tau_{\gamma}$ along $\tilde{w}$. (a)-(c) Ergotropy shift in terms of the initial population $P_{0}$ and measurement strength $m$, assuming $|Q_{0}|^{2}=|Q_{max}|^{2}= P_{0}(1-P_{0})$ for each $P_{0}$, for (a) $f=0$, (b) $f=0.15$ and (c) $f=0.3$. (d)-(f) Ergotropy shift in terms of the initial population $P_{0}$ and measurement strength $m$, assuming $|Q_{0}|^{2}=0$ for (d) $f=0$, (e) $f=0.15$ and (f) $f=0.3$. The hatched red region and continuous red curves highlight where $\mathcal{W}_{m\tilde{w}}(\tau)=0$. The black and blue dashed curves represent contours of the energy shift $ \eta $, revealing the interplay between ergotropy and energy transfer. It was assumed $\gamma=10^{-2}$.
    \justifying}
    \label{fig:ErgotropyShift}
\end{figure*}

Nevertheless, since $\varepsilon_{m\tilde{w}}(\tau) = 0$, it is crucial to identify the parameters that also satisfy $\mathcal{W}_{m\tilde{w}}(\tau) =-\varepsilon_{m\tilde{w}}^{p}(\tau)= 0$.
To illustrate this and the influence of temperature, Fig.~\eqref{fig:ErgotropyShift} presents the ergotropy shift profile at $ \tau_\gamma $ for different parameter settings.  
Figures~\eqref{fig:ErgotropyShift}a–c show how the ergotropy shift depends on the initial population $ P_0 $ and measurement strength $ m $, assuming maximal initial coherence $ |Q_0|^2 = |Q_{\max}|^2 = P_0(1 - P_0) $. These plots correspond to different temperatures: (a) $ f = 0 $, (b) $ f = 0.15 $, and (c) $ f = 0.30 $. The desired condition $ \mathcal{W}_{m\tilde{w}}(\tau) = 0 $ (marked by red curves) is not satisfied for any $ P_0 $ and $ m $ when $ f = 0 $, indicating that at zero temperature, the protocol inevitably causes nonzero ergotropy shift. However, as $ f $ increases, the number of operational points grows, suggesting that thermal effects help to balance the ergotropy changes. Additionally, the black and blue dashed curves indicate the null energy shift curves $ \eta_{2,3} $, revealing the interplay between ergotropy and energy changes. 
Similarly, Figs.~\ref{fig:ErgotropyShift}d–f illustrate the ergotropy shift for the same temperatures but with $|Q_0|^2 = 0$, revealing a significantly different behavior.  As the temperature increases, the regions satisfying $\mathcal{W}_{m\tilde{w}}(\tau) = 0$ (red hatched regions) expand, particularly for larger $P_0$.
This suggests that higher temperatures introduce a stabilizing effect, enabling the protocol to operate within the required conditions over a broader parameter range, reinforcing the idea that thermal fluctuations can counteract undesired energy and ergotropy shifts.

\begin{figure*}
    \centering
    \includegraphics[width=0.9\textwidth]{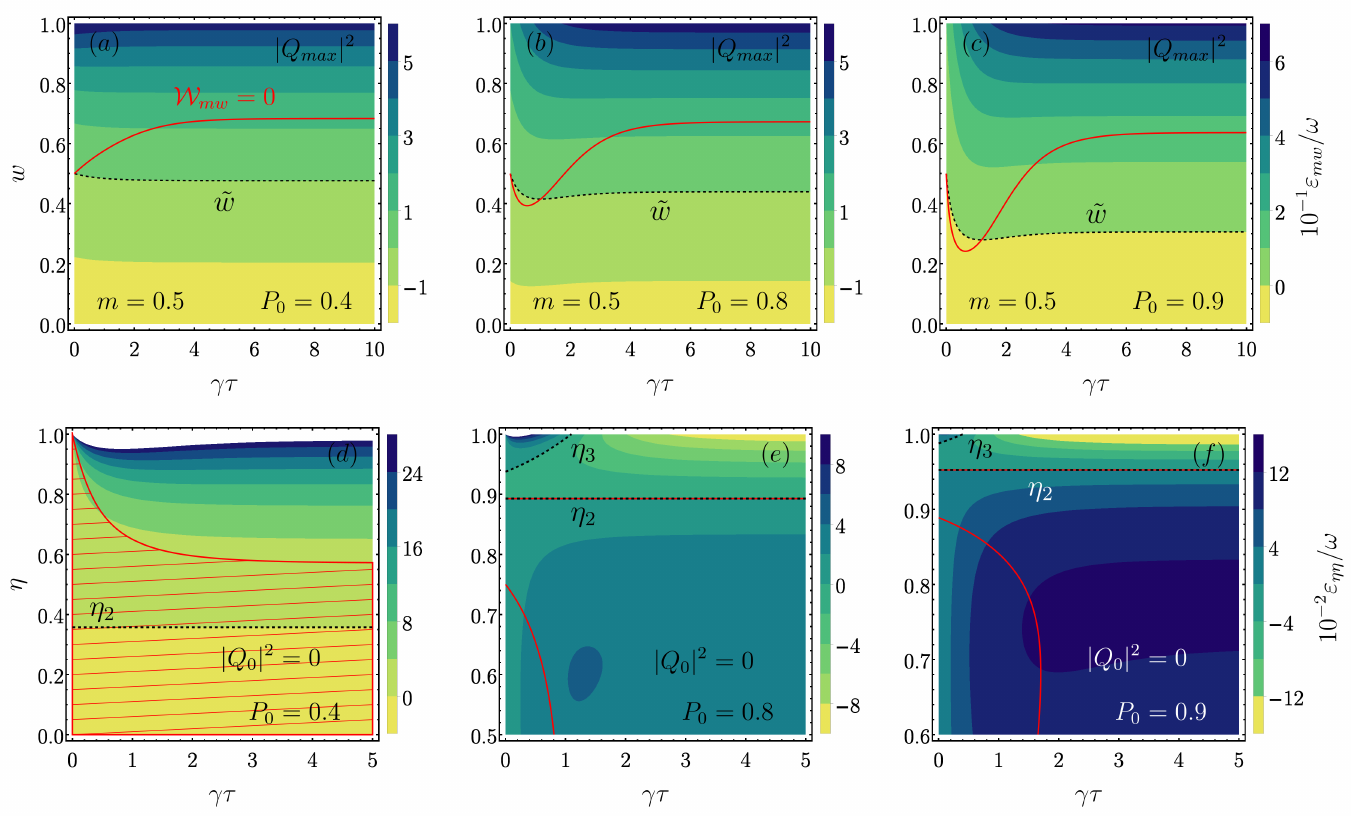}
    \caption{\textbf{Energy shift profile.} (a)-(c) Energy shift in terms of the reversal strength $w$ and dissipation time $\tau$, assuming fixed strength $m=0.5$ and $|Q_{0}|^{2}=|Q_{max}|^{2}= P_{0}(1-P_{0})$, for (a) $P_{0}=0.4$, (b) $P_{0}=0.8$ and (c) $P_{0}=0.9$. The dashed black lines represent the points along $\tilde{w}$, given by Eq.~\eqref{NullCostW}, where $\varepsilon_{mw}(\tau)=0$, while the continuous red curves highlight where $\mathcal{W}_{mw}(\tau)=0$. (d)-(f) Energy shift for equal measurement strengths $m = w = \eta$ relative to $\tau$, assuming $|Q_{0}|^{2}=0$ for (d) $P_{0}=0.4$, (e) $P_{0}=0.8$ and (f) $P_{0}=0.9$. The black dashed lines represent the curves $\eta_{2}$ and $\eta_{3}$ for a null energetic shift, given by Eq.~\eqref{NullCostCurves}. The hatched red region and continuous red curves represent the points where $\mathcal{W}_{mw}(\tau)=0$ is satisfied. It was assumed $f=0.3$, $\gamma=10^{-2}$.\justifying}
    \label{fig:EnergeticCost}
\end{figure*}

As mentioned above, the operational points of the TWM protocol are defined by the simultaneous fulfillment of both null energetic and ergotropic shifts.
In this sense, Fig.~\ref{fig:EnergeticCost} presents the energetic shift profile for different parameters.
Figures~\ref{fig:EnergeticCost}a–c display the behavior of the profile with respect to the reversal measurement strength $w$ and the dissipation time $\tau$, assuming $f=0.30$, an initial coherence $|Q_{0}|^{2}=|Q_{max}|^{2}= P_{0}(1-P_{0})$, and weak measurement strength $m = 0.5$, for different initial population values: (a) $P_0 = 0.4$, (b) $P_0 = 0.8$, and (c) $P_0 = 0.9$. As expected, most of the parameter space shows a non-zero net energetic exchange. However, the intersection of the dashed black and continuous red curves marks the operational points of the protocol. The former indicates a null energetic shift, characterized by $\tilde{w}$, while the latter identifies points where $\mathcal{W}_{mw}(\tau) = 0$ holds. For the chosen parameters, these intersections may only occur at specific dissipation times.
Analogously, Figs.~\ref{fig:EnergeticCost}d–f present the energy shift landscape for the same initial populations, assuming $|Q_{0}|^{2} = 0$ and equal measurement strengths, $m = w = \eta$. The dashed black lines mark points defined by the null energetic shift curves $\eta_{2,3}$, while the continuous red lines and hatched red areas highlight where a null ergotropy shift is obtained. Again, the intersections represent the desired working regime. As discussed in the main text, the $\eta_2$ curve for incoherent states leads to a cyclic transformation, $\hat{\rho}_{0} \rightarrow \hat{\tau}_{th} \rightarrow \hat{\rho}_{0}$, ensuring that the battery state remains unchanged during dissipation, thus satisfying $\varepsilon_{\eta_{2}\eta_{2}} = \mathcal{W}_{\eta_{2}\eta_{2}} = 0$ for all $\tau$. However, for the selected parameters, the $\eta_3$ curve does not exhibit any operational points.

It is important to emphasize that Fig.~\ref{fig:EnergeticCost} serves as a proof-of-concept illustration of the proposed protocol. The operational points depend heavily on the selected values. Nevertheless, some general insights can be observed: (i) a finite energetic change exists over most of the parameter space; (ii) the conditions required to minimize external influence significantly constrain the working regime. Nonetheless, it is still possible to identify intersection points where $\varepsilon_{mw}(\tau) = 0$ and $\mathcal{W}_{mw}(\tau) = 0$ are simultaneously satisfied.
Therefore, in principle, even with additional external constraints in experimental scenarios (such as temperature or measurement strength), it is feasible to adapt the protocol accordingly.
%
%
%
%
%
%
%
%
%
%
%
%
%
%
%
%
%
%
%
%
%
%
%
%
%
\section{Two-qubit open quantum battery dynamics: The $X$-state}\label{2QBdynamics}
\begin{figure}[ht]
    \center
    \includegraphics[width=1\columnwidth]{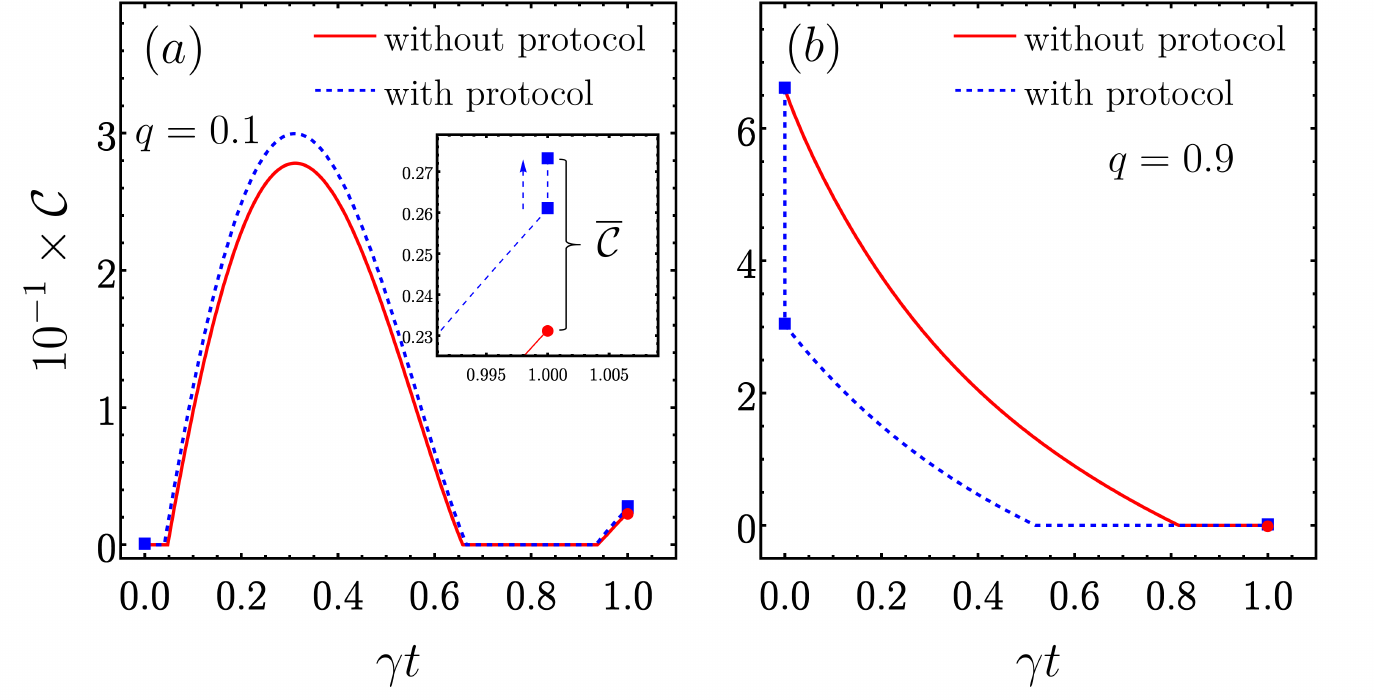}
    \caption{\textbf{Concurrence dynamics} for (a) $q=0.1$, $m_{1}=0.5$, $m_{2} \approx 0.6$, $w_{1}=w_{2}\approx 0.21$. Changes with (dashed blue) and without (continuous red) the protocol. The result demonstrates that weak measurements can enhance the entanglement generated by dissipation. (b) $q=0.9$, $m_{1}=0.5$, $m_{2}=0.9$, $w_{1}\approx 0.97$, $w_{2}=0.17$. Changes with (dashed blue) and without (continuous red) the protocol. The result shows that weak measurements while preserving ergotropy in this specific state, diminish entanglement when it is already present before dissipation takes place.
    \justifying}
    \label{fig:ConcurrenceDynamics_q01}
\end{figure}

Given two uncoupled qubit batteries individually prepared in the same incoherent state $\hat{\rho}_{1,2}=q|g\rangle\langle g|_{1,2}+(1-q)|e\rangle\langle e|_{1,2}$, one can ask what is the optimal unitary operation that can be realized in order to maximize quantum correlations between them. Note that, whenever $q\geq 1/2$, this is equivalent to a thermal equilibrium system with a reservoir at the inverse of temperature $\beta$, such that the bipartite state reads $\hat{\rho}_{0} = \hat{\tau}_\beta \otimes \hat{\tau}_\beta$, where $  \hat{\tau}_\beta = Z_{\beta}^{-1} e^{-\beta \hat{H}} $, with local Hamiltonians $ \hat{H} = \sum_{k=g,e} E_k \ket{k}\bra{k}$ and partition function $Z_{\beta}=\textrm{Tr}[e^{-\beta \hat{H}}]$.
Assuming the ground and excited state energies are $E_g = 0$ and $E_e = \omega$, respectively, the probability of occupying the ground state is
$ q = (1 + e^{-\beta \omega})^{-1} $. On the other hand, whenever
$q<1/2$, the state is active and, therefore, possesses some finite ergotropy. Interestingly, by applying the mentioned optimal unitary transformation, as we show in the following (see Ref.~\cite{behzadi2018thermodynamic} for details), one can construct a thermally correlated state that achieves the same ergotropy while allowing for different parameter choices that either produce or avoid quantum entanglement. 
To introduce thermal correlations, we apply a unitary operation $\hat{U} = \hat{V}_2 \hat{V}_1$ where
$ \hat{V}_1 = \ket{gg}\bra{gg} + \ket{ge}\bra{ge} + \ket{ee}\bra{eg} + \ket{eg}\bra{ee}, $
$ \hat{V}_2 = \ket{\phi^+}\bra{gg} + \ket{ge}\bra{ge} + \ket{eg}\bra{eg} + \ket{\phi^-}\bra{ee}, $
with $\ket{\phi^+} = (\ket{gg} + \ket{ee})/\sqrt{2}$ and $\ket{\phi^-} = (\ket{gg} - \ket{ee})/\sqrt{2}$. The resulting correlated state is
$ \hat{\rho}_{X}(q) = \hat{U} \hat{\rho}_{0} \hat{U}^{\dagger}, $
which, in the basis $\{\ket{gg}, \ket{ge}, \ket{eg}, \ket{ee}\}$, is represented as
\begin{equation}\label{StateX}
    \hat{\rho}_{X}(q)=\left(\begin{array}{cccc}
q/2 & 0 & 0 & q^{2}-q/2\\
0 & q(1-q) & 0 & 0\\
0 & 0 & (1-q)^{2} & 0\\
q^{2}-q/2 & 0 & 0 & q/2
\end{array}\right).
\end{equation}
This particular unitary operation was shown to maximize the quantum correlations presented in the final state. Moreover, entanglement (measured using concurrence, relative entropy, or entanglement negativity) vanishes at a finite critical temperature, other notions of quantum correlations—such as global quantum discord, local quantum uncertainty~\cite{behzadi2018thermodynamic}, and realism-based nonlocality~\cite{gomes2022realism}—gradually decay as the temperature increases.

As discussed in the main text, the dynamics of a two-qubit quantum battery in contact with a common environment is given by the following master equation
\begin{equation}\label{masterEqApp}
\frac{d}{dt} \hat{\rho}(t) = -i \left[\hat{H}, \hat{\rho}(t)\right] + \mathcal{D}_{\downarrow}[\hat{\rho}(t)] + \mathcal{D}_{\uparrow}[\hat{\rho}(t)],
\end{equation}
where the Hamiltonian reads $\hat{H}=\omega|e\rangle\langle e|_{1}+\omega|e\rangle\langle e|_{2}+J\left(\sigma_{+}^{1}\sigma_{-}^{2}+\sigma_{-}^{1}\sigma_{+}^{2}\right)$, with $J$ being the qubit's coupling strength, and the dissipators written as
\begin{equation}
    \begin{split}
        \mathcal{D}_{\downarrow}[\boldsymbol{\cdot}]&\coloneqq\gamma(1-f)\left(\hat{S}_{+}\boldsymbol{\cdot}\hat{S}_{-}-\frac{1}{2}\left\{ \hat{S}_{-}\hat{S}_{+},\boldsymbol{\cdot}\right\} \right),\\\mathcal{D}_{\uparrow}[\boldsymbol{\cdot}]&\coloneqq\gamma f\left(\hat{S}_{-}\boldsymbol{\cdot}\hat{S}_{+}-\frac{1}{2}\left\{ \hat{S}_{+}\hat{S}_{-},\boldsymbol{\cdot}\right\} \right),
    \end{split}
\end{equation}
where $\hat{S}_{\pm}=\hat{\sigma}_{\pm}^{1}+\hat{\sigma}_{\pm}^{2}$ are the raising and lowering ladder operators. Assuming as initial state $\hat{\rho}_{0}$, the $X$-state from Eq.~\eqref{StateX}
one can show that for all $t \geq 0$, the density matrix is written as
\begin{equation}
    \hat{\rho}(t)=\left(\begin{array}{cccc}
\rho_{11}(t) & 0 & 0 & \rho_{14}(t)\\
0 & \rho_{22}(t) & \rho_{23}(t) & 0\\
0 & \rho_{23}^{*}(t) & \rho_{33}(t) & 0\\
\rho_{14}^{*}(t) & 0 & 0 & \rho_{44}(t)
\end{array}\right),
\end{equation}
where, according to Eq.~\eqref{masterEqApp}, the dynamics is given by the following set of coupled differential equations:
\begin{widetext}
\begin{equation}
    \begin{split}
        \frac{d\rho_{11}(t)}{dt}&=-2\gamma f\rho_{11}(t)-\gamma(f-1)\left[\rho_{22}(t)+2\textrm{Re}\left[\rho_{23}(t)\right]+\rho_{33}(t)\right],\\\frac{d\rho_{14}(t)}{dt}&=-\left(\gamma-2i\omega\right)\rho_{14}(t),\\\frac{d\rho_{22}(t)}{dt}&=\gamma f\rho_{11}(t)-\gamma\left[\rho_{22}(t)+\left(f-1\right)\rho_{44}(t)\right]-\left(\gamma+2iJ\right)\textrm{Re}\left[\rho_{23}(t)\right]+2iJ\rho_{23}(t),\\\frac{d\rho_{23}(t)}{dt}&=\frac{1}{2}\gamma(2f\rho_{11}(t)-\rho_{22}(t)-\rho_{33}(t)-2(f-1)\rho_{44}(t)-2\rho_{23}(t))+iJ(\rho_{22}(t)-\rho_{33}(t)),\\\frac{d\rho_{33}(t)}{dt}&=\gamma f\rho_{11}(t)-\gamma\left[\rho_{33}(t)+\left(f-1\right)\rho_{44}(t)\right]-\left(\gamma-2iJ\right)\textrm{Re}\left[\rho_{23}(t)\right]-2iJ\rho_{23}(t),\\\frac{d\rho_{44}(t)}{dt}&=2\gamma\left(f-1\right)\rho_{44}(t)+\gamma f\left(\rho_{22}(t)+2\textrm{Re}\left[\rho_{23}(t)\right]+\rho_{33}(t)\right).
    \end{split}
\end{equation}
\end{widetext}
Note that the non-diagonal elements $\rho_{23}(t)$ and $\rho_{23}^{*}(t)$ are, in general, non-null during the dissipative dynamics. More importantly, the solution of previous equations allows one to compute the time-evolution of the relevant quantities in terms of $q$, in particular ergotropy, as presented in Section~\ref{2qbits}.
In this sense, the concurrence change along the protocol execution for the examples without ($q=0.1$) and with ($q=0.9$) initial entanglement is presented in Figs.~\ref{fig:ConcurrenceDynamics_q01} (a) and (b), respectively. For the former, the dynamic-induced entanglement observed for the purely dissipative time evolution is shown to be enhanced with the execution of the TWM protocol. For the latter, the entanglement is shown to be consumed faster compared to the case without the weak measurements. Nevertheless, in both scenarios, the use of the TWM protocol mitigates ergotropy loss.
%
%
%
%
%
%
%
%
%
%
%
%
%
%
%
%
%
%
%
%
%
%
%
%
%
%
%
%
%
%
%
%
%
%
%
%
%
%
%
%
%
%
%
%
%
%
%
%
%
%
%
%
%
%
%
%
%
%
%
%
%
%
%
%
%
%
%
%
%
%
%
%
%
%
%
%
%
%
%
%
%
%
%
%
%
%
%
%
%
%
%
%
%
%
%
%
%
%
%
%
%
%
%
%
%
%
%
%
%
%
%
%
%
%
%
%
%
%
%
%
%
%
%
%
%
%
%
%
%
%
%
%
%
%
%
%
%
%
%
%
%
%
%
%
%
%
%
%
%
%
%
%
%
%
%
%
%
\bibliography{References}

\providecommand{\noopsort}[1]{}\providecommand{\singleletter}[1]{#1}%
\begin{thebibliography}{172}%
\makeatletter
\providecommand \@ifxundefined [1]{%
 \@ifx{#1\undefined}
}%
\providecommand \@ifnum [1]{%
 \ifnum #1\expandafter \@firstoftwo
 \else \expandafter \@secondoftwo
 \fi
}%
\providecommand \@ifx [1]{%
 \ifx #1\expandafter \@firstoftwo
 \else \expandafter \@secondoftwo
 \fi
}%
\providecommand \natexlab [1]{#1}%
\providecommand \enquote  [1]{``#1''}%
\providecommand \bibnamefont  [1]{#1}%
\providecommand \bibfnamefont [1]{#1}%
\providecommand \citenamefont [1]{#1}%
\providecommand \href@noop [0]{\@secondoftwo}%
\providecommand \href [0]{\begingroup \@sanitize@url \@href}%
\providecommand \@href[1]{\@@startlink{#1}\@@href}%
\providecommand \@@href[1]{\endgroup#1\@@endlink}%
\providecommand \@sanitize@url [0]{\catcode `\\12\catcode `\$12\catcode `\&12\catcode `\#12\catcode `\^12\catcode `\_12\catcode `\%12\relax}%
\providecommand \@@startlink[1]{}%
\providecommand \@@endlink[0]{}%
\providecommand \url  [0]{\begingroup\@sanitize@url \@url }%
\providecommand \@url [1]{\endgroup\@href {#1}{\urlprefix }}%
\providecommand \urlprefix  [0]{URL }%
\providecommand \Eprint [0]{\href }%
\providecommand \doibase [0]{https://doi.org/}%
\providecommand \selectlanguage [0]{\@gobble}%
\providecommand \bibinfo  [0]{\@secondoftwo}%
\providecommand \bibfield  [0]{\@secondoftwo}%
\providecommand \translation [1]{[#1]}%
\providecommand \BibitemOpen [0]{}%
\providecommand \bibitemStop [0]{}%
\providecommand \bibitemNoStop [0]{.\EOS\space}%
\providecommand \EOS [0]{\spacefactor3000\relax}%
\providecommand \BibitemShut  [1]{\csname bibitem#1\endcsname}%
\let\auto@bib@innerbib\@empty
\bibitem [{\citenamefont {Alicki}\ and\ \citenamefont {Fannes}(2013)}]{PhysRevE.87.042123}%
  \BibitemOpen
  \bibfield  {author} {\bibinfo {author} {\bibfnamefont {R.}~\bibnamefont {Alicki}}\ and\ \bibinfo {author} {\bibfnamefont {M.}~\bibnamefont {Fannes}},\ }\bibfield  {title} {\bibinfo {title} {Entanglement boost for extractable work from ensembles of quantum batteries},\ }\href {https://doi.org/10.1103/PhysRevE.87.042123} {\bibfield  {journal} {\bibinfo  {journal} {Phys. Rev. E}\ }\textbf {\bibinfo {volume} {87}},\ \bibinfo {pages} {042123} (\bibinfo {year} {2013})}\BibitemShut {NoStop}%
\bibitem [{\citenamefont {Binder}\ \emph {et~al.}(2015)\citenamefont {Binder}, \citenamefont {Vinjanampathy}, \citenamefont {Modi},\ and\ \citenamefont {Goold}}]{binder2015quantacell}%
  \BibitemOpen
  \bibfield  {author} {\bibinfo {author} {\bibfnamefont {F.~C.}\ \bibnamefont {Binder}}, \bibinfo {author} {\bibfnamefont {S.}~\bibnamefont {Vinjanampathy}}, \bibinfo {author} {\bibfnamefont {K.}~\bibnamefont {Modi}},\ and\ \bibinfo {author} {\bibfnamefont {J.}~\bibnamefont {Goold}},\ }\bibfield  {title} {\bibinfo {title} {Quantacell: powerful charging of quantum batteries},\ }\href {https://doi.org/10.1088/1367-2630/17/7/075015} {\bibfield  {journal} {\bibinfo  {journal} {New Journal of Physics}\ }\textbf {\bibinfo {volume} {17}},\ \bibinfo {pages} {075015} (\bibinfo {year} {2015})}\BibitemShut {NoStop}%
\bibitem [{\citenamefont {Campaioli}\ \emph {et~al.}(2024)\citenamefont {Campaioli}, \citenamefont {Gherardini}, \citenamefont {Quach}, \citenamefont {Polini},\ and\ \citenamefont {Andolina}}]{RevModPhys.96.031001}%
  \BibitemOpen
  \bibfield  {author} {\bibinfo {author} {\bibfnamefont {F.}~\bibnamefont {Campaioli}}, \bibinfo {author} {\bibfnamefont {S.}~\bibnamefont {Gherardini}}, \bibinfo {author} {\bibfnamefont {J.~Q.}\ \bibnamefont {Quach}}, \bibinfo {author} {\bibfnamefont {M.}~\bibnamefont {Polini}},\ and\ \bibinfo {author} {\bibfnamefont {G.~M.}\ \bibnamefont {Andolina}},\ }\bibfield  {title} {\bibinfo {title} {Colloquium: Quantum batteries},\ }\href {https://doi.org/10.1103/RevModPhys.96.031001} {\bibfield  {journal} {\bibinfo  {journal} {Rev. Mod. Phys.}\ }\textbf {\bibinfo {volume} {96}},\ \bibinfo {pages} {031001} (\bibinfo {year} {2024})}\BibitemShut {NoStop}%
\bibitem [{\citenamefont {Quach}\ \emph {et~al.}(2022)\citenamefont {Quach}, \citenamefont {McGhee}, \citenamefont {Ganzer}, \citenamefont {Rouse}, \citenamefont {Lovett}, \citenamefont {Gauger}, \citenamefont {Keeling}, \citenamefont {Cerullo}, \citenamefont {Lidzey},\ and\ \citenamefont {Virgili}}]{quach2022superabsorption}%
  \BibitemOpen
  \bibfield  {author} {\bibinfo {author} {\bibfnamefont {J.~Q.}\ \bibnamefont {Quach}}, \bibinfo {author} {\bibfnamefont {K.~E.}\ \bibnamefont {McGhee}}, \bibinfo {author} {\bibfnamefont {L.}~\bibnamefont {Ganzer}}, \bibinfo {author} {\bibfnamefont {D.~M.}\ \bibnamefont {Rouse}}, \bibinfo {author} {\bibfnamefont {B.~W.}\ \bibnamefont {Lovett}}, \bibinfo {author} {\bibfnamefont {E.~M.}\ \bibnamefont {Gauger}}, \bibinfo {author} {\bibfnamefont {J.}~\bibnamefont {Keeling}}, \bibinfo {author} {\bibfnamefont {G.}~\bibnamefont {Cerullo}}, \bibinfo {author} {\bibfnamefont {D.~G.}\ \bibnamefont {Lidzey}},\ and\ \bibinfo {author} {\bibfnamefont {T.}~\bibnamefont {Virgili}},\ }\bibfield  {title} {\bibinfo {title} {Superabsorption in an organic microcavity: Toward a quantum battery},\ }\href {https://doi.org/10.1126/sciadv.abk3160} {\bibfield  {journal} {\bibinfo  {journal} {Science advances}\ }\textbf {\bibinfo {volume} {8}},\ \bibinfo {pages} {3160} (\bibinfo {year} {2022})}\BibitemShut {NoStop}%
\bibitem [{\citenamefont {Andolina}\ \emph {et~al.}(2018)\citenamefont {Andolina}, \citenamefont {Farina}, \citenamefont {Mari}, \citenamefont {Pellegrini}, \citenamefont {Giovannetti},\ and\ \citenamefont {Polini}}]{Andolina2018}%
  \BibitemOpen
  \bibfield  {author} {\bibinfo {author} {\bibfnamefont {G.~M.}\ \bibnamefont {Andolina}}, \bibinfo {author} {\bibfnamefont {D.}~\bibnamefont {Farina}}, \bibinfo {author} {\bibfnamefont {A.}~\bibnamefont {Mari}}, \bibinfo {author} {\bibfnamefont {V.}~\bibnamefont {Pellegrini}}, \bibinfo {author} {\bibfnamefont {V.}~\bibnamefont {Giovannetti}},\ and\ \bibinfo {author} {\bibfnamefont {M.}~\bibnamefont {Polini}},\ }\bibfield  {title} {\bibinfo {title} {Charger-mediated energy transfer in exactly solvable models for quantum batteries},\ }\href {https://doi.org/10.1103/PhysRevB.98.205423} {\bibfield  {journal} {\bibinfo  {journal} {Phys. Rev. B}\ }\textbf {\bibinfo {volume} {98}},\ \bibinfo {pages} {205423} (\bibinfo {year} {2018})}\BibitemShut {NoStop}%
\bibitem [{\citenamefont {Andolina}\ \emph {et~al.}(2019)\citenamefont {Andolina}, \citenamefont {Keck}, \citenamefont {Mari}, \citenamefont {Giovannetti},\ and\ \citenamefont {Polini}}]{PhysRevB.99.205437}%
  \BibitemOpen
  \bibfield  {author} {\bibinfo {author} {\bibfnamefont {G.~M.}\ \bibnamefont {Andolina}}, \bibinfo {author} {\bibfnamefont {M.}~\bibnamefont {Keck}}, \bibinfo {author} {\bibfnamefont {A.}~\bibnamefont {Mari}}, \bibinfo {author} {\bibfnamefont {V.}~\bibnamefont {Giovannetti}},\ and\ \bibinfo {author} {\bibfnamefont {M.}~\bibnamefont {Polini}},\ }\bibfield  {title} {\bibinfo {title} {Quantum versus classical many-body batteries},\ }\href {https://doi.org/10.1103/PhysRevB.99.205437} {\bibfield  {journal} {\bibinfo  {journal} {Phys. Rev. B}\ }\textbf {\bibinfo {volume} {99}},\ \bibinfo {pages} {205437} (\bibinfo {year} {2019})}\BibitemShut {NoStop}%
\bibitem [{\citenamefont {Gyhm}\ and\ \citenamefont {Fischer}(2024)}]{gyhm2024beneficial}%
  \BibitemOpen
  \bibfield  {author} {\bibinfo {author} {\bibfnamefont {J.-Y.}\ \bibnamefont {Gyhm}}\ and\ \bibinfo {author} {\bibfnamefont {U.~R.}\ \bibnamefont {Fischer}},\ }\bibfield  {title} {\bibinfo {title} {Beneficial and detrimental entanglement for quantum battery charging},\ }\href {https://doi.org/10.1116/5.0184903} {\bibfield  {journal} {\bibinfo  {journal} {AVS Quantum Science}\ }\textbf {\bibinfo {volume} {6}},\ \bibinfo {pages} {012001} (\bibinfo {year} {2024})}\BibitemShut {NoStop}%
\bibitem [{\citenamefont {Monsel}\ \emph {et~al.}(2020)\citenamefont {Monsel}, \citenamefont {Fellous-Asiani}, \citenamefont {Huard},\ and\ \citenamefont {Auff\`eves}}]{PhysRevLett.124.130601}%
  \BibitemOpen
  \bibfield  {author} {\bibinfo {author} {\bibfnamefont {J.}~\bibnamefont {Monsel}}, \bibinfo {author} {\bibfnamefont {M.}~\bibnamefont {Fellous-Asiani}}, \bibinfo {author} {\bibfnamefont {B.}~\bibnamefont {Huard}},\ and\ \bibinfo {author} {\bibfnamefont {A.}~\bibnamefont {Auff\`eves}},\ }\bibfield  {title} {\bibinfo {title} {The energetic cost of work extraction},\ }\href {https://doi.org/10.1103/PhysRevLett.124.130601} {\bibfield  {journal} {\bibinfo  {journal} {Phys. Rev. Lett.}\ }\textbf {\bibinfo {volume} {124}},\ \bibinfo {pages} {130601} (\bibinfo {year} {2020})}\BibitemShut {NoStop}%
\bibitem [{\citenamefont {Zhang}\ \emph {et~al.}(2019)\citenamefont {Zhang}, \citenamefont {Yang}, \citenamefont {Fu},\ and\ \citenamefont {Wang}}]{PhysRevE.99.052106}%
  \BibitemOpen
  \bibfield  {author} {\bibinfo {author} {\bibfnamefont {Y.-Y.}\ \bibnamefont {Zhang}}, \bibinfo {author} {\bibfnamefont {T.-R.}\ \bibnamefont {Yang}}, \bibinfo {author} {\bibfnamefont {L.}~\bibnamefont {Fu}},\ and\ \bibinfo {author} {\bibfnamefont {X.}~\bibnamefont {Wang}},\ }\bibfield  {title} {\bibinfo {title} {Powerful harmonic charging in a quantum battery},\ }\href {https://doi.org/10.1103/PhysRevE.99.052106} {\bibfield  {journal} {\bibinfo  {journal} {Phys. Rev. E}\ }\textbf {\bibinfo {volume} {99}},\ \bibinfo {pages} {052106} (\bibinfo {year} {2019})}\BibitemShut {NoStop}%
\bibitem [{\citenamefont {{\c{C}}akmak}(2020)}]{PhysRevE.102.042111}%
  \BibitemOpen
  \bibfield  {author} {\bibinfo {author} {\bibfnamefont {B.}~\bibnamefont {{\c{C}}akmak}},\ }\bibfield  {title} {\bibinfo {title} {Ergotropy from coherences in an open quantum system},\ }\href {https://doi.org/10.1103/PhysRevE.102.042111} {\bibfield  {journal} {\bibinfo  {journal} {Phys. Rev. E}\ }\textbf {\bibinfo {volume} {102}},\ \bibinfo {pages} {042111} (\bibinfo {year} {2020})}\BibitemShut {NoStop}%
\bibitem [{\citenamefont {Ferraro}\ \emph {et~al.}(2018)\citenamefont {Ferraro}, \citenamefont {Campisi}, \citenamefont {Andolina}, \citenamefont {Pellegrini},\ and\ \citenamefont {Polini}}]{PhysRevLett.120.117702}%
  \BibitemOpen
  \bibfield  {author} {\bibinfo {author} {\bibfnamefont {D.}~\bibnamefont {Ferraro}}, \bibinfo {author} {\bibfnamefont {M.}~\bibnamefont {Campisi}}, \bibinfo {author} {\bibfnamefont {G.~M.}\ \bibnamefont {Andolina}}, \bibinfo {author} {\bibfnamefont {V.}~\bibnamefont {Pellegrini}},\ and\ \bibinfo {author} {\bibfnamefont {M.}~\bibnamefont {Polini}},\ }\bibfield  {title} {\bibinfo {title} {High-power collective charging of a solid-state quantum battery},\ }\href {https://doi.org/10.1103/PhysRevLett.120.117702} {\bibfield  {journal} {\bibinfo  {journal} {Phys. Rev. Lett.}\ }\textbf {\bibinfo {volume} {120}},\ \bibinfo {pages} {117702} (\bibinfo {year} {2018})}\BibitemShut {NoStop}%
\bibitem [{\citenamefont {Campaioli}\ \emph {et~al.}(2017)\citenamefont {Campaioli}, \citenamefont {Pollock}, \citenamefont {Binder}, \citenamefont {C\'eleri}, \citenamefont {Goold}, \citenamefont {Vinjanampathy},\ and\ \citenamefont {Modi}}]{PhysRevLett.118.150601}%
  \BibitemOpen
  \bibfield  {author} {\bibinfo {author} {\bibfnamefont {F.}~\bibnamefont {Campaioli}}, \bibinfo {author} {\bibfnamefont {F.~A.}\ \bibnamefont {Pollock}}, \bibinfo {author} {\bibfnamefont {F.~C.}\ \bibnamefont {Binder}}, \bibinfo {author} {\bibfnamefont {L.}~\bibnamefont {C\'eleri}}, \bibinfo {author} {\bibfnamefont {J.}~\bibnamefont {Goold}}, \bibinfo {author} {\bibfnamefont {S.}~\bibnamefont {Vinjanampathy}},\ and\ \bibinfo {author} {\bibfnamefont {K.}~\bibnamefont {Modi}},\ }\bibfield  {title} {\bibinfo {title} {Enhancing the charging power of quantum batteries},\ }\href {https://doi.org/10.1103/PhysRevLett.118.150601} {\bibfield  {journal} {\bibinfo  {journal} {Phys. Rev. Lett.}\ }\textbf {\bibinfo {volume} {118}},\ \bibinfo {pages} {150601} (\bibinfo {year} {2017})}\BibitemShut {NoStop}%
\bibitem [{\citenamefont {Dou}\ \emph {et~al.}(2021)\citenamefont {Dou}, \citenamefont {Wang},\ and\ \citenamefont {Sun}}]{Dou2021}%
  \BibitemOpen
  \bibfield  {author} {\bibinfo {author} {\bibfnamefont {F.-Q.}\ \bibnamefont {Dou}}, \bibinfo {author} {\bibfnamefont {Y.-J.}\ \bibnamefont {Wang}},\ and\ \bibinfo {author} {\bibfnamefont {J.-A.}\ \bibnamefont {Sun}},\ }\bibfield  {title} {\bibinfo {title} {Highly efficient charging and discharging of three-level quantum batteries through shortcuts to adiabaticity},\ }\href {https://doi.org/10.1007/s11467-021-1130-5} {\bibfield  {journal} {\bibinfo  {journal} {Frontiers of Physics}\ }\textbf {\bibinfo {volume} {17}},\ \bibinfo {pages} {31503} (\bibinfo {year} {2021})}\BibitemShut {NoStop}%
\bibitem [{\citenamefont {Dou}\ \emph {et~al.}(2022{\natexlab{a}})\citenamefont {Dou}, \citenamefont {Zhou},\ and\ \citenamefont {Sun}}]{PhysRevA.106.032212}%
  \BibitemOpen
  \bibfield  {author} {\bibinfo {author} {\bibfnamefont {F.-Q.}\ \bibnamefont {Dou}}, \bibinfo {author} {\bibfnamefont {H.}~\bibnamefont {Zhou}},\ and\ \bibinfo {author} {\bibfnamefont {J.-A.}\ \bibnamefont {Sun}},\ }\bibfield  {title} {\bibinfo {title} {Cavity heisenberg-spin-chain quantum battery},\ }\href {https://doi.org/10.1103/PhysRevA.106.032212} {\bibfield  {journal} {\bibinfo  {journal} {Phys. Rev. A}\ }\textbf {\bibinfo {volume} {106}},\ \bibinfo {pages} {032212} (\bibinfo {year} {2022}{\natexlab{a}})}\BibitemShut {NoStop}%
\bibitem [{\citenamefont {Dou}\ \emph {et~al.}(2022{\natexlab{b}})\citenamefont {Dou}, \citenamefont {Lu}, \citenamefont {Wang},\ and\ \citenamefont {Sun}}]{PhysRevB.105.115405}%
  \BibitemOpen
  \bibfield  {author} {\bibinfo {author} {\bibfnamefont {F.-Q.}\ \bibnamefont {Dou}}, \bibinfo {author} {\bibfnamefont {Y.-Q.}\ \bibnamefont {Lu}}, \bibinfo {author} {\bibfnamefont {Y.-J.}\ \bibnamefont {Wang}},\ and\ \bibinfo {author} {\bibfnamefont {J.-A.}\ \bibnamefont {Sun}},\ }\bibfield  {title} {\bibinfo {title} {Extended dicke quantum battery with interatomic interactions and driving field},\ }\href {https://doi.org/10.1103/PhysRevB.105.115405} {\bibfield  {journal} {\bibinfo  {journal} {Phys. Rev. B}\ }\textbf {\bibinfo {volume} {105}},\ \bibinfo {pages} {115405} (\bibinfo {year} {2022}{\natexlab{b}})}\BibitemShut {NoStop}%
\bibitem [{\citenamefont {Gyhm}\ \emph {et~al.}(2022)\citenamefont {Gyhm}, \citenamefont {\ifmmode~\check{S}\else \v{S}\fi{}afr\'anek},\ and\ \citenamefont {Rosa}}]{PhysRevLett.128.140501}%
  \BibitemOpen
  \bibfield  {author} {\bibinfo {author} {\bibfnamefont {J.-Y.}\ \bibnamefont {Gyhm}}, \bibinfo {author} {\bibfnamefont {D.}~\bibnamefont {\ifmmode~\check{S}\else \v{S}\fi{}afr\'anek}},\ and\ \bibinfo {author} {\bibfnamefont {D.}~\bibnamefont {Rosa}},\ }\bibfield  {title} {\bibinfo {title} {Quantum charging advantage cannot be extensive without global operations},\ }\href {https://doi.org/10.1103/PhysRevLett.128.140501} {\bibfield  {journal} {\bibinfo  {journal} {Phys. Rev. Lett.}\ }\textbf {\bibinfo {volume} {128}},\ \bibinfo {pages} {140501} (\bibinfo {year} {2022})}\BibitemShut {NoStop}%
\bibitem [{\citenamefont {de~Moraes}\ \emph {et~al.}(2024)\citenamefont {de~Moraes}, \citenamefont {Duriez}, \citenamefont {Saguia}, \citenamefont {Santos},\ and\ \citenamefont {Sarandy}}]{deMoraes_2024}%
  \BibitemOpen
  \bibfield  {author} {\bibinfo {author} {\bibfnamefont {L.~F.~C.}\ \bibnamefont {de~Moraes}}, \bibinfo {author} {\bibfnamefont {A.~C.}\ \bibnamefont {Duriez}}, \bibinfo {author} {\bibfnamefont {A.}~\bibnamefont {Saguia}}, \bibinfo {author} {\bibfnamefont {A.~C.}\ \bibnamefont {Santos}},\ and\ \bibinfo {author} {\bibfnamefont {M.~S.}\ \bibnamefont {Sarandy}},\ }\bibfield  {title} {\bibinfo {title} {Quantum battery supercharging via counter-diabatic dynamics},\ }\href {https://doi.org/10.1088/2058-9565/ad71ed} {\bibfield  {journal} {\bibinfo  {journal} {Quantum Science and Technology}\ }\textbf {\bibinfo {volume} {9}},\ \bibinfo {pages} {045033} (\bibinfo {year} {2024})}\BibitemShut {NoStop}%
\bibitem [{\citenamefont {Zhang}\ and\ \citenamefont {Blaauboer}(2023)}]{10.3389/fphy.2022.1097564}%
  \BibitemOpen
  \bibfield  {author} {\bibinfo {author} {\bibfnamefont {X.}~\bibnamefont {Zhang}}\ and\ \bibinfo {author} {\bibfnamefont {M.}~\bibnamefont {Blaauboer}},\ }\bibfield  {title} {\bibinfo {title} {Enhanced energy transfer in a dicke quantum battery},\ }\href {https://doi.org/10.3389/fphy.2022.1097564} {\bibfield  {journal} {\bibinfo  {journal} {Frontiers in Physics}\ }\textbf {\bibinfo {volume} {10}},\ \bibinfo {pages} {1097564} (\bibinfo {year} {2023})}\BibitemShut {NoStop}%
\bibitem [{\citenamefont {Guo}\ \emph {et~al.}(2024)\citenamefont {Guo}, \citenamefont {Yang},\ and\ \citenamefont {Dou}}]{PhysRevA.109.032201}%
  \BibitemOpen
  \bibfield  {author} {\bibinfo {author} {\bibfnamefont {W.-X.}\ \bibnamefont {Guo}}, \bibinfo {author} {\bibfnamefont {F.-M.}\ \bibnamefont {Yang}},\ and\ \bibinfo {author} {\bibfnamefont {F.-Q.}\ \bibnamefont {Dou}},\ }\bibfield  {title} {\bibinfo {title} {Analytically solvable many-body rosen-zener quantum battery},\ }\href {https://doi.org/10.1103/PhysRevA.109.032201} {\bibfield  {journal} {\bibinfo  {journal} {Phys. Rev. A}\ }\textbf {\bibinfo {volume} {109}},\ \bibinfo {pages} {032201} (\bibinfo {year} {2024})}\BibitemShut {NoStop}%
\bibitem [{\citenamefont {Le}\ \emph {et~al.}(2018)\citenamefont {Le}, \citenamefont {Levinsen}, \citenamefont {Modi}, \citenamefont {Parish},\ and\ \citenamefont {Pollock}}]{PhysRevA.97.022106}%
  \BibitemOpen
  \bibfield  {author} {\bibinfo {author} {\bibfnamefont {T.~P.}\ \bibnamefont {Le}}, \bibinfo {author} {\bibfnamefont {J.}~\bibnamefont {Levinsen}}, \bibinfo {author} {\bibfnamefont {K.}~\bibnamefont {Modi}}, \bibinfo {author} {\bibfnamefont {M.~M.}\ \bibnamefont {Parish}},\ and\ \bibinfo {author} {\bibfnamefont {F.~A.}\ \bibnamefont {Pollock}},\ }\bibfield  {title} {\bibinfo {title} {Spin-chain model of a many-body quantum battery},\ }\href {https://doi.org/10.1103/PhysRevA.97.022106} {\bibfield  {journal} {\bibinfo  {journal} {Phys. Rev. A}\ }\textbf {\bibinfo {volume} {97}},\ \bibinfo {pages} {022106} (\bibinfo {year} {2018})}\BibitemShut {NoStop}%
\bibitem [{\citenamefont {Zhao}\ \emph {et~al.}(2022)\citenamefont {Zhao}, \citenamefont {Dou},\ and\ \citenamefont {Zhao}}]{PhysRevResearch.4.013172}%
  \BibitemOpen
  \bibfield  {author} {\bibinfo {author} {\bibfnamefont {F.}~\bibnamefont {Zhao}}, \bibinfo {author} {\bibfnamefont {F.-Q.}\ \bibnamefont {Dou}},\ and\ \bibinfo {author} {\bibfnamefont {Q.}~\bibnamefont {Zhao}},\ }\bibfield  {title} {\bibinfo {title} {Charging performance of the su-schrieffer-heeger quantum battery},\ }\href {https://doi.org/10.1103/PhysRevResearch.4.013172} {\bibfield  {journal} {\bibinfo  {journal} {Phys. Rev. Res.}\ }\textbf {\bibinfo {volume} {4}},\ \bibinfo {pages} {013172} (\bibinfo {year} {2022})}\BibitemShut {NoStop}%
\bibitem [{\citenamefont {Dou}\ \emph {et~al.}(2020)\citenamefont {Dou}, \citenamefont {Wang},\ and\ \citenamefont {Sun}}]{Dou_2020}%
  \BibitemOpen
  \bibfield  {author} {\bibinfo {author} {\bibfnamefont {F.-Q.}\ \bibnamefont {Dou}}, \bibinfo {author} {\bibfnamefont {Y.-J.}\ \bibnamefont {Wang}},\ and\ \bibinfo {author} {\bibfnamefont {J.-A.}\ \bibnamefont {Sun}},\ }\bibfield  {title} {\bibinfo {title} {Closed-loop three-level charged quantum battery},\ }\href {https://doi.org/10.1209/0295-5075/131/43001} {\bibfield  {journal} {\bibinfo  {journal} {Europhysics Letters}\ }\textbf {\bibinfo {volume} {131}},\ \bibinfo {pages} {43001} (\bibinfo {year} {2020})}\BibitemShut {NoStop}%
\bibitem [{\citenamefont {Yang}\ \emph {et~al.}(2024{\natexlab{a}})\citenamefont {Yang}, \citenamefont {Yang},\ and\ \citenamefont {Dou}}]{PhysRevB.109.235432}%
  \BibitemOpen
  \bibfield  {author} {\bibinfo {author} {\bibfnamefont {D.-L.}\ \bibnamefont {Yang}}, \bibinfo {author} {\bibfnamefont {F.-M.}\ \bibnamefont {Yang}},\ and\ \bibinfo {author} {\bibfnamefont {F.-Q.}\ \bibnamefont {Dou}},\ }\bibfield  {title} {\bibinfo {title} {Three-level dicke quantum battery},\ }\href {https://doi.org/10.1103/PhysRevB.109.235432} {\bibfield  {journal} {\bibinfo  {journal} {Phys. Rev. B}\ }\textbf {\bibinfo {volume} {109}},\ \bibinfo {pages} {235432} (\bibinfo {year} {2024}{\natexlab{a}})}\BibitemShut {NoStop}%
\bibitem [{\citenamefont {Ali}\ \emph {et~al.}(2024)\citenamefont {Ali}, \citenamefont {Al-Kuwari}, \citenamefont {Hussain}, \citenamefont {Byrnes}, \citenamefont {Rahim}, \citenamefont {Quach}, \citenamefont {Ghominejad},\ and\ \citenamefont {Haddadi}}]{PhysRevA.110.052404}%
  \BibitemOpen
  \bibfield  {author} {\bibinfo {author} {\bibfnamefont {A.}~\bibnamefont {Ali}}, \bibinfo {author} {\bibfnamefont {S.}~\bibnamefont {Al-Kuwari}}, \bibinfo {author} {\bibfnamefont {M.~I.}\ \bibnamefont {Hussain}}, \bibinfo {author} {\bibfnamefont {T.}~\bibnamefont {Byrnes}}, \bibinfo {author} {\bibfnamefont {M.~T.}\ \bibnamefont {Rahim}}, \bibinfo {author} {\bibfnamefont {J.~Q.}\ \bibnamefont {Quach}}, \bibinfo {author} {\bibfnamefont {M.}~\bibnamefont {Ghominejad}},\ and\ \bibinfo {author} {\bibfnamefont {S.}~\bibnamefont {Haddadi}},\ }\bibfield  {title} {\bibinfo {title} {Ergotropy and capacity optimization in heisenberg spin-chain quantum batteries},\ }\href {https://doi.org/10.1103/PhysRevA.110.052404} {\bibfield  {journal} {\bibinfo  {journal} {Phys. Rev. A}\ }\textbf {\bibinfo {volume} {110}},\ \bibinfo {pages} {052404} (\bibinfo {year} {2024})}\BibitemShut {NoStop}%
\bibitem [{\citenamefont {Hadipour}\ and\ \citenamefont {Haseli}(2025)}]{hadipour2025nonequilibrium}%
  \BibitemOpen
  \bibfield  {author} {\bibinfo {author} {\bibfnamefont {M.}~\bibnamefont {Hadipour}}\ and\ \bibinfo {author} {\bibfnamefont {S.}~\bibnamefont {Haseli}},\ }\bibfield  {title} {\bibinfo {title} {Nonequilibrium quantum batteries: Amplified work extraction through thermal bath modulation},\ }\href {https://doi.org/10.48550/arXiv.2502.05508} {\bibfield  {journal} {\bibinfo  {journal} {arXiv:2502.05508}\ } (\bibinfo {year} {2025})}\BibitemShut {NoStop}%
\bibitem [{\citenamefont {Seah}\ \emph {et~al.}(2021)\citenamefont {Seah}, \citenamefont {Perarnau-Llobet}, \citenamefont {Haack}, \citenamefont {Brunner},\ and\ \citenamefont {Nimmrichter}}]{PhysRevLett.127.100601}%
  \BibitemOpen
  \bibfield  {author} {\bibinfo {author} {\bibfnamefont {S.}~\bibnamefont {Seah}}, \bibinfo {author} {\bibfnamefont {M.}~\bibnamefont {Perarnau-Llobet}}, \bibinfo {author} {\bibfnamefont {G.}~\bibnamefont {Haack}}, \bibinfo {author} {\bibfnamefont {N.}~\bibnamefont {Brunner}},\ and\ \bibinfo {author} {\bibfnamefont {S.}~\bibnamefont {Nimmrichter}},\ }\bibfield  {title} {\bibinfo {title} {Quantum speed-up in collisional battery charging},\ }\href {https://doi.org/10.1103/PhysRevLett.127.100601} {\bibfield  {journal} {\bibinfo  {journal} {Phys. Rev. Lett.}\ }\textbf {\bibinfo {volume} {127}},\ \bibinfo {pages} {100601} (\bibinfo {year} {2021})}\BibitemShut {NoStop}%
\bibitem [{\citenamefont {Salvia}\ \emph {et~al.}(2023{\natexlab{a}})\citenamefont {Salvia}, \citenamefont {Perarnau-Llobet}, \citenamefont {Haack}, \citenamefont {Brunner},\ and\ \citenamefont {Nimmrichter}}]{PhysRevResearch.5.013155}%
  \BibitemOpen
  \bibfield  {author} {\bibinfo {author} {\bibfnamefont {R.}~\bibnamefont {Salvia}}, \bibinfo {author} {\bibfnamefont {M.}~\bibnamefont {Perarnau-Llobet}}, \bibinfo {author} {\bibfnamefont {G.}~\bibnamefont {Haack}}, \bibinfo {author} {\bibfnamefont {N.}~\bibnamefont {Brunner}},\ and\ \bibinfo {author} {\bibfnamefont {S.}~\bibnamefont {Nimmrichter}},\ }\bibfield  {title} {\bibinfo {title} {Quantum advantage in charging cavity and spin batteries by repeated interactions},\ }\href {https://doi.org/10.1103/PhysRevResearch.5.013155} {\bibfield  {journal} {\bibinfo  {journal} {Phys. Rev. Res.}\ }\textbf {\bibinfo {volume} {5}},\ \bibinfo {pages} {013155} (\bibinfo {year} {2023}{\natexlab{a}})}\BibitemShut {NoStop}%
\bibitem [{\citenamefont {Glicenstein}\ \emph {et~al.}(2022)\citenamefont {Glicenstein}, \citenamefont {Ferioli}, \citenamefont {Browaeys},\ and\ \citenamefont {Ferrier-Barbut}}]{Glicenstein2022}%
  \BibitemOpen
  \bibfield  {author} {\bibinfo {author} {\bibfnamefont {A.}~\bibnamefont {Glicenstein}}, \bibinfo {author} {\bibfnamefont {G.}~\bibnamefont {Ferioli}}, \bibinfo {author} {\bibfnamefont {A.}~\bibnamefont {Browaeys}},\ and\ \bibinfo {author} {\bibfnamefont {I.}~\bibnamefont {Ferrier-Barbut}},\ }\bibfield  {title} {\bibinfo {title} {From superradiance to subradiance: exploring the many-body dicke ladder},\ }\href {https://doi.org/10.1364/OL.451903} {\bibfield  {journal} {\bibinfo  {journal} {Opt. Lett.}\ }\textbf {\bibinfo {volume} {47}},\ \bibinfo {pages} {1541} (\bibinfo {year} {2022})}\BibitemShut {NoStop}%
\bibitem [{\citenamefont {Ahmadi}\ \emph {et~al.}(2024)\citenamefont {Ahmadi}, \citenamefont {Mazurek}, \citenamefont {Horodecki},\ and\ \citenamefont {Barzanjeh}}]{PhysRevLett.132.210402}%
  \BibitemOpen
  \bibfield  {author} {\bibinfo {author} {\bibfnamefont {B.}~\bibnamefont {Ahmadi}}, \bibinfo {author} {\bibfnamefont {P.}~\bibnamefont {Mazurek}}, \bibinfo {author} {\bibfnamefont {P.}~\bibnamefont {Horodecki}},\ and\ \bibinfo {author} {\bibfnamefont {S.}~\bibnamefont {Barzanjeh}},\ }\bibfield  {title} {\bibinfo {title} {Nonreciprocal quantum batteries},\ }\href {https://doi.org/10.1103/PhysRevLett.132.210402} {\bibfield  {journal} {\bibinfo  {journal} {Phys. Rev. Lett.}\ }\textbf {\bibinfo {volume} {132}},\ \bibinfo {pages} {210402} (\bibinfo {year} {2024})}\BibitemShut {NoStop}%
\bibitem [{\citenamefont {Ahmadi}\ \emph {et~al.}(2025)\citenamefont {Ahmadi}, \citenamefont {Mazurek}, \citenamefont {Barzanjeh},\ and\ \citenamefont {Horodecki}}]{PhysRevApplied.23.024010}%
  \BibitemOpen
  \bibfield  {author} {\bibinfo {author} {\bibfnamefont {B.}~\bibnamefont {Ahmadi}}, \bibinfo {author} {\bibfnamefont {P.}~\bibnamefont {Mazurek}}, \bibinfo {author} {\bibfnamefont {S.}~\bibnamefont {Barzanjeh}},\ and\ \bibinfo {author} {\bibfnamefont {P.}~\bibnamefont {Horodecki}},\ }\bibfield  {title} {\bibinfo {title} {Superoptimal charging of quantum batteries via reservoir engineering: Arbitrary energy transfer unlocked},\ }\href {https://doi.org/10.1103/PhysRevApplied.23.024010} {\bibfield  {journal} {\bibinfo  {journal} {Phys. Rev. Appl.}\ }\textbf {\bibinfo {volume} {23}},\ \bibinfo {pages} {024010} (\bibinfo {year} {2025})}\BibitemShut {NoStop}%
\bibitem [{\citenamefont {Barra}\ \emph {et~al.}(2022)\citenamefont {Barra}, \citenamefont {Hovhannisyan},\ and\ \citenamefont {Imparato}}]{barra2022quantum}%
  \BibitemOpen
  \bibfield  {author} {\bibinfo {author} {\bibfnamefont {F.}~\bibnamefont {Barra}}, \bibinfo {author} {\bibfnamefont {K.~V.}\ \bibnamefont {Hovhannisyan}},\ and\ \bibinfo {author} {\bibfnamefont {A.}~\bibnamefont {Imparato}},\ }\bibfield  {title} {\bibinfo {title} {Quantum batteries at the verge of a phase transition},\ }\href {https://iopscience.iop.org/article/10.1088/1367-2630/ac43ed} {\bibfield  {journal} {\bibinfo  {journal} {New Journal of Physics}\ }\textbf {\bibinfo {volume} {24}},\ \bibinfo {pages} {015003} (\bibinfo {year} {2022})}\BibitemShut {NoStop}%
\bibitem [{\citenamefont {Shaghaghi}\ \emph {et~al.}(2022)\citenamefont {Shaghaghi}, \citenamefont {Singh}, \citenamefont {Benenti},\ and\ \citenamefont {Rosa}}]{Shaghaghi2022}%
  \BibitemOpen
  \bibfield  {author} {\bibinfo {author} {\bibfnamefont {V.}~\bibnamefont {Shaghaghi}}, \bibinfo {author} {\bibfnamefont {V.}~\bibnamefont {Singh}}, \bibinfo {author} {\bibfnamefont {G.}~\bibnamefont {Benenti}},\ and\ \bibinfo {author} {\bibfnamefont {D.}~\bibnamefont {Rosa}},\ }\bibfield  {title} {\bibinfo {title} {Micromasers as quantum batteries},\ }\href {https://doi.org/10.1088/2058-9565/ac8829} {\bibfield  {journal} {\bibinfo  {journal} {Quantum Science and Technology}\ }\textbf {\bibinfo {volume} {7}},\ \bibinfo {pages} {04LT01} (\bibinfo {year} {2022})}\BibitemShut {NoStop}%
\bibitem [{\citenamefont {Shaghaghi}\ \emph {et~al.}(2023)\citenamefont {Shaghaghi}, \citenamefont {Singh}, \citenamefont {Carrega}, \citenamefont {Rosa},\ and\ \citenamefont {Benenti}}]{Shaghaghi2023}%
  \BibitemOpen
  \bibfield  {author} {\bibinfo {author} {\bibfnamefont {V.}~\bibnamefont {Shaghaghi}}, \bibinfo {author} {\bibfnamefont {V.}~\bibnamefont {Singh}}, \bibinfo {author} {\bibfnamefont {M.}~\bibnamefont {Carrega}}, \bibinfo {author} {\bibfnamefont {D.}~\bibnamefont {Rosa}},\ and\ \bibinfo {author} {\bibfnamefont {G.}~\bibnamefont {Benenti}},\ }\bibfield  {title} {\bibinfo {title} {Lossy micromaser battery: Almost pure states in the jaynes–cummings regime},\ }\href {https://www.mdpi.com/1099-4300/25/3/430} {\bibfield  {journal} {\bibinfo  {journal} {Entropy}\ }\textbf {\bibinfo {volume} {25}},\ \bibinfo {pages} {430} (\bibinfo {year} {2023})}\BibitemShut {NoStop}%
\bibitem [{\citenamefont {Lu}\ \emph {et~al.}(2024)\citenamefont {Lu}, \citenamefont {Tian}, \citenamefont {L{\"u}},\ and\ \citenamefont {Shang}}]{lu2024topological}%
  \BibitemOpen
  \bibfield  {author} {\bibinfo {author} {\bibfnamefont {Z.-G.}\ \bibnamefont {Lu}}, \bibinfo {author} {\bibfnamefont {G.}~\bibnamefont {Tian}}, \bibinfo {author} {\bibfnamefont {X.-Y.}\ \bibnamefont {L{\"u}}},\ and\ \bibinfo {author} {\bibfnamefont {C.}~\bibnamefont {Shang}},\ }\bibfield  {title} {\bibinfo {title} {Topological quantum batteries},\ }\href {https://arxiv.org/abs/2405.03675} {\bibfield  {journal} {\bibinfo  {journal} {arXiv:2405.03675}\ } (\bibinfo {year} {2024})}\BibitemShut {NoStop}%
\bibitem [{\citenamefont {Dennis}\ \emph {et~al.}(2002)\citenamefont {Dennis}, \citenamefont {Kitaev}, \citenamefont {Landahl},\ and\ \citenamefont {Preskill}}]{Dennis2002}%
  \BibitemOpen
  \bibfield  {author} {\bibinfo {author} {\bibfnamefont {E.}~\bibnamefont {Dennis}}, \bibinfo {author} {\bibfnamefont {A.}~\bibnamefont {Kitaev}}, \bibinfo {author} {\bibfnamefont {A.}~\bibnamefont {Landahl}},\ and\ \bibinfo {author} {\bibfnamefont {J.}~\bibnamefont {Preskill}},\ }\bibfield  {title} {\bibinfo {title} {Topological quantum memory},\ }\href {https://doi.org/10.1063/1.1499754} {\bibfield  {journal} {\bibinfo  {journal} {Journal of Mathematical Physics}\ }\textbf {\bibinfo {volume} {43}},\ \bibinfo {pages} {4452} (\bibinfo {year} {2002})}\BibitemShut {NoStop}%
\bibitem [{\citenamefont {Rodr\'{\i}guez}\ \emph {et~al.}(2023)\citenamefont {Rodr\'{\i}guez}, \citenamefont {Ahmadi}, \citenamefont {Mazurek}, \citenamefont {Barzanjeh}, \citenamefont {Alicki},\ and\ \citenamefont {Horodecki}}]{PhysRevA.107.042419}%
  \BibitemOpen
  \bibfield  {author} {\bibinfo {author} {\bibfnamefont {R.~R.}\ \bibnamefont {Rodr\'{\i}guez}}, \bibinfo {author} {\bibfnamefont {B.}~\bibnamefont {Ahmadi}}, \bibinfo {author} {\bibfnamefont {P.}~\bibnamefont {Mazurek}}, \bibinfo {author} {\bibfnamefont {S.}~\bibnamefont {Barzanjeh}}, \bibinfo {author} {\bibfnamefont {R.}~\bibnamefont {Alicki}},\ and\ \bibinfo {author} {\bibfnamefont {P.}~\bibnamefont {Horodecki}},\ }\bibfield  {title} {\bibinfo {title} {Catalysis in charging quantum batteries},\ }\href {https://doi.org/10.1103/PhysRevA.107.042419} {\bibfield  {journal} {\bibinfo  {journal} {Phys. Rev. A}\ }\textbf {\bibinfo {volume} {107}},\ \bibinfo {pages} {042419} (\bibinfo {year} {2023})}\BibitemShut {NoStop}%
\bibitem [{\citenamefont {Kamin}\ \emph {et~al.}(2020)\citenamefont {Kamin}, \citenamefont {Tabesh}, \citenamefont {Salimi}, \citenamefont {Kheirandish},\ and\ \citenamefont {Santos}}]{kamin2020non}%
  \BibitemOpen
  \bibfield  {author} {\bibinfo {author} {\bibfnamefont {F.}~\bibnamefont {Kamin}}, \bibinfo {author} {\bibfnamefont {F.}~\bibnamefont {Tabesh}}, \bibinfo {author} {\bibfnamefont {S.}~\bibnamefont {Salimi}}, \bibinfo {author} {\bibfnamefont {F.}~\bibnamefont {Kheirandish}},\ and\ \bibinfo {author} {\bibfnamefont {A.~C.}\ \bibnamefont {Santos}},\ }\bibfield  {title} {\bibinfo {title} {Non-markovian effects on charging and self-discharging process of quantum batteries},\ }\href {https://doi.org/10.1088/1367-2630/ab9ee2} {\bibfield  {journal} {\bibinfo  {journal} {New Journal of Physics}\ }\textbf {\bibinfo {volume} {22}},\ \bibinfo {pages} {083007} (\bibinfo {year} {2020})}\BibitemShut {NoStop}%
\bibitem [{\citenamefont {Song}\ \emph {et~al.}(2024)\citenamefont {Song}, \citenamefont {Liu}, \citenamefont {Zhou}, \citenamefont {Yang},\ and\ \citenamefont {An}}]{PhysRevLett.132.090401}%
  \BibitemOpen
  \bibfield  {author} {\bibinfo {author} {\bibfnamefont {W.-L.}\ \bibnamefont {Song}}, \bibinfo {author} {\bibfnamefont {H.-B.}\ \bibnamefont {Liu}}, \bibinfo {author} {\bibfnamefont {B.}~\bibnamefont {Zhou}}, \bibinfo {author} {\bibfnamefont {W.-L.}\ \bibnamefont {Yang}},\ and\ \bibinfo {author} {\bibfnamefont {J.-H.}\ \bibnamefont {An}},\ }\bibfield  {title} {\bibinfo {title} {Remote charging and degradation suppression for the quantum battery},\ }\href {https://doi.org/10.1103/PhysRevLett.132.090401} {\bibfield  {journal} {\bibinfo  {journal} {Phys. Rev. Lett.}\ }\textbf {\bibinfo {volume} {132}},\ \bibinfo {pages} {090401} (\bibinfo {year} {2024})}\BibitemShut {NoStop}%
\bibitem [{\citenamefont {Mojaveri}\ \emph {et~al.}(2024)\citenamefont {Mojaveri}, \citenamefont {Jafarzadeh~Bahrbeig},\ and\ \citenamefont {Fasihi}}]{PhysRevE.110.064107}%
  \BibitemOpen
  \bibfield  {author} {\bibinfo {author} {\bibfnamefont {B.}~\bibnamefont {Mojaveri}}, \bibinfo {author} {\bibfnamefont {R.}~\bibnamefont {Jafarzadeh~Bahrbeig}},\ and\ \bibinfo {author} {\bibfnamefont {M.~A.}\ \bibnamefont {Fasihi}},\ }\bibfield  {title} {\bibinfo {title} {Charging a quantum battery mediated by parity-deformed fields},\ }\href {https://doi.org/10.1103/PhysRevE.110.064107} {\bibfield  {journal} {\bibinfo  {journal} {Phys. Rev. E}\ }\textbf {\bibinfo {volume} {110}},\ \bibinfo {pages} {064107} (\bibinfo {year} {2024})}\BibitemShut {NoStop}%
\bibitem [{\citenamefont {Hu}\ \emph {et~al.}(2025)\citenamefont {Hu}, \citenamefont {Gao},\ and\ \citenamefont {Fan}}]{hu2025efficient}%
  \BibitemOpen
  \bibfield  {author} {\bibinfo {author} {\bibfnamefont {M.-L.}\ \bibnamefont {Hu}}, \bibinfo {author} {\bibfnamefont {T.}~\bibnamefont {Gao}},\ and\ \bibinfo {author} {\bibfnamefont {H.}~\bibnamefont {Fan}},\ }\bibfield  {title} {\bibinfo {title} {Efficient wireless charging of a quantum battery},\ }\href {https://doi.org/10.48550/arXiv.2501.08843} {\bibfield  {journal} {\bibinfo  {journal} {arXiv:2501.08843}\ } (\bibinfo {year} {2025})}\BibitemShut {NoStop}%
\bibitem [{\citenamefont {Sun}\ \emph {et~al.}(2024)\citenamefont {Sun}, \citenamefont {Zhou},\ and\ \citenamefont {Dou}}]{sun2024cavity}%
  \BibitemOpen
  \bibfield  {author} {\bibinfo {author} {\bibfnamefont {P.-Y.}\ \bibnamefont {Sun}}, \bibinfo {author} {\bibfnamefont {H.}~\bibnamefont {Zhou}},\ and\ \bibinfo {author} {\bibfnamefont {F.-Q.}\ \bibnamefont {Dou}},\ }\bibfield  {title} {\bibinfo {title} {Cavity-heisenberg spin-$ j $ chain quantum battery and reinforcement learning optimization},\ }\href {https://doi.org/10.48550/arXiv.2412.01442} {\bibfield  {journal} {\bibinfo  {journal} {arXiv:2412.01442}\ } (\bibinfo {year} {2024})}\BibitemShut {NoStop}%
\bibitem [{\citenamefont {Erdman}\ \emph {et~al.}(2024)\citenamefont {Erdman}, \citenamefont {Andolina}, \citenamefont {Giovannetti},\ and\ \citenamefont {No\'e}}]{PhysRevLett.133.243602}%
  \BibitemOpen
  \bibfield  {author} {\bibinfo {author} {\bibfnamefont {P.~A.}\ \bibnamefont {Erdman}}, \bibinfo {author} {\bibfnamefont {G.~M.}\ \bibnamefont {Andolina}}, \bibinfo {author} {\bibfnamefont {V.}~\bibnamefont {Giovannetti}},\ and\ \bibinfo {author} {\bibfnamefont {F.}~\bibnamefont {No\'e}},\ }\bibfield  {title} {\bibinfo {title} {Reinforcement learning optimization of the charging of a dicke quantum battery},\ }\href {https://doi.org/10.1103/PhysRevLett.133.243602} {\bibfield  {journal} {\bibinfo  {journal} {Phys. Rev. Lett.}\ }\textbf {\bibinfo {volume} {133}},\ \bibinfo {pages} {243602} (\bibinfo {year} {2024})}\BibitemShut {NoStop}%
\bibitem [{\citenamefont {Yan}\ and\ \citenamefont {Jing}(2023)}]{PhysRevApplied.19.064069}%
  \BibitemOpen
  \bibfield  {author} {\bibinfo {author} {\bibfnamefont {J.-s.}\ \bibnamefont {Yan}}\ and\ \bibinfo {author} {\bibfnamefont {J.}~\bibnamefont {Jing}},\ }\bibfield  {title} {\bibinfo {title} {Charging by quantum measurement},\ }\href {https://doi.org/10.1103/PhysRevApplied.19.064069} {\bibfield  {journal} {\bibinfo  {journal} {Phys. Rev. Appl.}\ }\textbf {\bibinfo {volume} {19}},\ \bibinfo {pages} {064069} (\bibinfo {year} {2023})}\BibitemShut {NoStop}%
\bibitem [{\citenamefont {Rodríguez}\ \emph {et~al.}(2024)\citenamefont {Rodríguez}, \citenamefont {Ahmadi}, \citenamefont {Suárez}, \citenamefont {Mazurek}, \citenamefont {Barzanjeh},\ and\ \citenamefont {Horodecki}}]{Rodríguez_2024}%
  \BibitemOpen
  \bibfield  {author} {\bibinfo {author} {\bibfnamefont {R.~R.}\ \bibnamefont {Rodríguez}}, \bibinfo {author} {\bibfnamefont {B.}~\bibnamefont {Ahmadi}}, \bibinfo {author} {\bibfnamefont {G.}~\bibnamefont {Suárez}}, \bibinfo {author} {\bibfnamefont {P.}~\bibnamefont {Mazurek}}, \bibinfo {author} {\bibfnamefont {S.}~\bibnamefont {Barzanjeh}},\ and\ \bibinfo {author} {\bibfnamefont {P.}~\bibnamefont {Horodecki}},\ }\bibfield  {title} {\bibinfo {title} {Optimal quantum control of charging quantum batteries},\ }\href {https://doi.org/10.1088/1367-2630/ad3843} {\bibfield  {journal} {\bibinfo  {journal} {New Journal of Physics}\ }\textbf {\bibinfo {volume} {26}},\ \bibinfo {pages} {043004} (\bibinfo {year} {2024})}\BibitemShut {NoStop}%
\bibitem [{\citenamefont {Mazzoncini}\ \emph {et~al.}(2023)\citenamefont {Mazzoncini}, \citenamefont {Cavina}, \citenamefont {Andolina}, \citenamefont {Erdman},\ and\ \citenamefont {Giovannetti}}]{PhysRevA.107.032218}%
  \BibitemOpen
  \bibfield  {author} {\bibinfo {author} {\bibfnamefont {F.}~\bibnamefont {Mazzoncini}}, \bibinfo {author} {\bibfnamefont {V.}~\bibnamefont {Cavina}}, \bibinfo {author} {\bibfnamefont {G.~M.}\ \bibnamefont {Andolina}}, \bibinfo {author} {\bibfnamefont {P.~A.}\ \bibnamefont {Erdman}},\ and\ \bibinfo {author} {\bibfnamefont {V.}~\bibnamefont {Giovannetti}},\ }\bibfield  {title} {\bibinfo {title} {Optimal control methods for quantum batteries},\ }\href {https://doi.org/10.1103/PhysRevA.107.032218} {\bibfield  {journal} {\bibinfo  {journal} {Phys. Rev. A}\ }\textbf {\bibinfo {volume} {107}},\ \bibinfo {pages} {032218} (\bibinfo {year} {2023})}\BibitemShut {NoStop}%
\bibitem [{\citenamefont {Hu}\ \emph {et~al.}(2022)\citenamefont {Hu}, \citenamefont {Qiu}, \citenamefont {Souza}, \citenamefont {Yuan}, \citenamefont {Zhou}, \citenamefont {Zhang}, \citenamefont {Chu}, \citenamefont {Pan}, \citenamefont {Hu}, \citenamefont {Li}, \citenamefont {Xu}, \citenamefont {Zhong}, \citenamefont {Liu}, \citenamefont {Yan}, \citenamefont {Tan}, \citenamefont {Bachelard}, \citenamefont {Villas-Boas}, \citenamefont {Santos},\ and\ \citenamefont {Yu}}]{Hu_2022}%
  \BibitemOpen
  \bibfield  {author} {\bibinfo {author} {\bibfnamefont {C.-K.}\ \bibnamefont {Hu}}, \bibinfo {author} {\bibfnamefont {J.}~\bibnamefont {Qiu}}, \bibinfo {author} {\bibfnamefont {P.~J.~P.}\ \bibnamefont {Souza}}, \bibinfo {author} {\bibfnamefont {J.}~\bibnamefont {Yuan}}, \bibinfo {author} {\bibfnamefont {Y.}~\bibnamefont {Zhou}}, \bibinfo {author} {\bibfnamefont {L.}~\bibnamefont {Zhang}}, \bibinfo {author} {\bibfnamefont {J.}~\bibnamefont {Chu}}, \bibinfo {author} {\bibfnamefont {X.}~\bibnamefont {Pan}}, \bibinfo {author} {\bibfnamefont {L.}~\bibnamefont {Hu}}, \bibinfo {author} {\bibfnamefont {J.}~\bibnamefont {Li}}, \bibinfo {author} {\bibfnamefont {Y.}~\bibnamefont {Xu}}, \bibinfo {author} {\bibfnamefont {Y.}~\bibnamefont {Zhong}}, \bibinfo {author} {\bibfnamefont {S.}~\bibnamefont {Liu}}, \bibinfo {author} {\bibfnamefont {F.}~\bibnamefont {Yan}}, \bibinfo {author} {\bibfnamefont {D.}~\bibnamefont {Tan}}, \bibinfo {author} {\bibfnamefont {R.}~\bibnamefont {Bachelard}}, \bibinfo {author} {\bibfnamefont
  {C.~J.}\ \bibnamefont {Villas-Boas}}, \bibinfo {author} {\bibfnamefont {A.~C.}\ \bibnamefont {Santos}},\ and\ \bibinfo {author} {\bibfnamefont {D.}~\bibnamefont {Yu}},\ }\bibfield  {title} {\bibinfo {title} {Optimal charging of a superconducting quantum battery},\ }\href {https://doi.org/10.1088/2058-9565/ac8444} {\bibfield  {journal} {\bibinfo  {journal} {Quantum Science and Technology}\ }\textbf {\bibinfo {volume} {7}},\ \bibinfo {pages} {045018} (\bibinfo {year} {2022})}\BibitemShut {NoStop}%
\bibitem [{\citenamefont {Blais}\ \emph {et~al.}(2021)\citenamefont {Blais}, \citenamefont {Grimsmo}, \citenamefont {Girvin},\ and\ \citenamefont {Wallraff}}]{RevModPhys.93.025005}%
  \BibitemOpen
  \bibfield  {author} {\bibinfo {author} {\bibfnamefont {A.}~\bibnamefont {Blais}}, \bibinfo {author} {\bibfnamefont {A.~L.}\ \bibnamefont {Grimsmo}}, \bibinfo {author} {\bibfnamefont {S.~M.}\ \bibnamefont {Girvin}},\ and\ \bibinfo {author} {\bibfnamefont {A.}~\bibnamefont {Wallraff}},\ }\bibfield  {title} {\bibinfo {title} {Circuit quantum electrodynamics},\ }\href {https://doi.org/10.1103/RevModPhys.93.025005} {\bibfield  {journal} {\bibinfo  {journal} {Rev. Mod. Phys.}\ }\textbf {\bibinfo {volume} {93}},\ \bibinfo {pages} {025005} (\bibinfo {year} {2021})}\BibitemShut {NoStop}%
\bibitem [{\citenamefont {Yang}\ \emph {et~al.}(2024{\natexlab{b}})\citenamefont {Yang}, \citenamefont {Shi}, \citenamefont {Wan}, \citenamefont {Zhang}, \citenamefont {Wang},\ and\ \citenamefont {Yang}}]{PhysRevA.109.012204}%
  \BibitemOpen
  \bibfield  {author} {\bibinfo {author} {\bibfnamefont {H.-Y.}\ \bibnamefont {Yang}}, \bibinfo {author} {\bibfnamefont {H.-L.}\ \bibnamefont {Shi}}, \bibinfo {author} {\bibfnamefont {Q.-K.}\ \bibnamefont {Wan}}, \bibinfo {author} {\bibfnamefont {K.}~\bibnamefont {Zhang}}, \bibinfo {author} {\bibfnamefont {X.-H.}\ \bibnamefont {Wang}},\ and\ \bibinfo {author} {\bibfnamefont {W.-L.}\ \bibnamefont {Yang}},\ }\bibfield  {title} {\bibinfo {title} {Optimal energy storage in the tavis-cummings quantum battery},\ }\href {https://doi.org/10.1103/PhysRevA.109.012204} {\bibfield  {journal} {\bibinfo  {journal} {Phys. Rev. A}\ }\textbf {\bibinfo {volume} {109}},\ \bibinfo {pages} {012204} (\bibinfo {year} {2024}{\natexlab{b}})}\BibitemShut {NoStop}%
\bibitem [{\citenamefont {Gemme}\ \emph {et~al.}(2022)\citenamefont {Gemme}, \citenamefont {Grossi}, \citenamefont {Ferraro}, \citenamefont {Vallecorsa},\ and\ \citenamefont {Sassetti}}]{batteries8050043}%
  \BibitemOpen
  \bibfield  {author} {\bibinfo {author} {\bibfnamefont {G.}~\bibnamefont {Gemme}}, \bibinfo {author} {\bibfnamefont {M.}~\bibnamefont {Grossi}}, \bibinfo {author} {\bibfnamefont {D.}~\bibnamefont {Ferraro}}, \bibinfo {author} {\bibfnamefont {S.}~\bibnamefont {Vallecorsa}},\ and\ \bibinfo {author} {\bibfnamefont {M.}~\bibnamefont {Sassetti}},\ }\bibfield  {title} {\bibinfo {title} {Ibm quantum platforms: A quantum battery perspective},\ }\href {https://www.mdpi.com/2313-0105/8/5/43} {\bibfield  {journal} {\bibinfo  {journal} {Batteries}\ }\textbf {\bibinfo {volume} {8}},\ \bibinfo {pages} {43} (\bibinfo {year} {2022})}\BibitemShut {NoStop}%
\bibitem [{\citenamefont {Dou}\ and\ \citenamefont {Yang}(2023)}]{PhysRevA.107.023725}%
  \BibitemOpen
  \bibfield  {author} {\bibinfo {author} {\bibfnamefont {F.-Q.}\ \bibnamefont {Dou}}\ and\ \bibinfo {author} {\bibfnamefont {F.-M.}\ \bibnamefont {Yang}},\ }\bibfield  {title} {\bibinfo {title} {Superconducting transmon qubit-resonator quantum battery},\ }\href {https://doi.org/10.1103/PhysRevA.107.023725} {\bibfield  {journal} {\bibinfo  {journal} {Phys. Rev. A}\ }\textbf {\bibinfo {volume} {107}},\ \bibinfo {pages} {023725} (\bibinfo {year} {2023})}\BibitemShut {NoStop}%
\bibitem [{\citenamefont {Yang}\ and\ \citenamefont {Dou}(2024)}]{PhysRevA.109.062432}%
  \BibitemOpen
  \bibfield  {author} {\bibinfo {author} {\bibfnamefont {F.-M.}\ \bibnamefont {Yang}}\ and\ \bibinfo {author} {\bibfnamefont {F.-Q.}\ \bibnamefont {Dou}},\ }\bibfield  {title} {\bibinfo {title} {Resonator-qutrit quantum battery},\ }\href {https://doi.org/10.1103/PhysRevA.109.062432} {\bibfield  {journal} {\bibinfo  {journal} {Phys. Rev. A}\ }\textbf {\bibinfo {volume} {109}},\ \bibinfo {pages} {062432} (\bibinfo {year} {2024})}\BibitemShut {NoStop}%
\bibitem [{\citenamefont {Zheng}\ \emph {et~al.}(2022)\citenamefont {Zheng}, \citenamefont {Ning}, \citenamefont {Yang}, \citenamefont {Xia},\ and\ \citenamefont {Zheng}}]{Zheng_2022}%
  \BibitemOpen
  \bibfield  {author} {\bibinfo {author} {\bibfnamefont {R.-H.}\ \bibnamefont {Zheng}}, \bibinfo {author} {\bibfnamefont {W.}~\bibnamefont {Ning}}, \bibinfo {author} {\bibfnamefont {Z.-B.}\ \bibnamefont {Yang}}, \bibinfo {author} {\bibfnamefont {Y.}~\bibnamefont {Xia}},\ and\ \bibinfo {author} {\bibfnamefont {S.-B.}\ \bibnamefont {Zheng}},\ }\bibfield  {title} {\bibinfo {title} {Demonstration of dynamical control of three-level open systems with a superconducting qutrit},\ }\href {https://doi.org/10.1088/1367-2630/ac788f} {\bibfield  {journal} {\bibinfo  {journal} {New Journal of Physics}\ }\textbf {\bibinfo {volume} {24}},\ \bibinfo {pages} {063031} (\bibinfo {year} {2022})}\BibitemShut {NoStop}%
\bibitem [{\citenamefont {Maillette~de Buy~Wenniger}\ \emph {et~al.}(2023)\citenamefont {Maillette~de Buy~Wenniger}, \citenamefont {Thomas}, \citenamefont {Maffei}, \citenamefont {Wein}, \citenamefont {Pont}, \citenamefont {Belabas}, \citenamefont {Prasad}, \citenamefont {Harouri}, \citenamefont {Lema\^{\i}tre}, \citenamefont {Sagnes}, \citenamefont {Somaschi}, \citenamefont {Auff\`eves},\ and\ \citenamefont {Senellart}}]{PhysRevLett.131.260401}%
  \BibitemOpen
  \bibfield  {author} {\bibinfo {author} {\bibfnamefont {I.}~\bibnamefont {Maillette~de Buy~Wenniger}}, \bibinfo {author} {\bibfnamefont {S.~E.}\ \bibnamefont {Thomas}}, \bibinfo {author} {\bibfnamefont {M.}~\bibnamefont {Maffei}}, \bibinfo {author} {\bibfnamefont {S.~C.}\ \bibnamefont {Wein}}, \bibinfo {author} {\bibfnamefont {M.}~\bibnamefont {Pont}}, \bibinfo {author} {\bibfnamefont {N.}~\bibnamefont {Belabas}}, \bibinfo {author} {\bibfnamefont {S.}~\bibnamefont {Prasad}}, \bibinfo {author} {\bibfnamefont {A.}~\bibnamefont {Harouri}}, \bibinfo {author} {\bibfnamefont {A.}~\bibnamefont {Lema\^{\i}tre}}, \bibinfo {author} {\bibfnamefont {I.}~\bibnamefont {Sagnes}}, \bibinfo {author} {\bibfnamefont {N.}~\bibnamefont {Somaschi}}, \bibinfo {author} {\bibfnamefont {A.}~\bibnamefont {Auff\`eves}},\ and\ \bibinfo {author} {\bibfnamefont {P.}~\bibnamefont {Senellart}},\ }\bibfield  {title} {\bibinfo {title} {Experimental analysis of energy transfers between a quantum emitter and light fields},\ }\href
  {https://doi.org/10.1103/PhysRevLett.131.260401} {\bibfield  {journal} {\bibinfo  {journal} {Phys. Rev. Lett.}\ }\textbf {\bibinfo {volume} {131}},\ \bibinfo {pages} {260401} (\bibinfo {year} {2023})}\BibitemShut {NoStop}%
\bibitem [{\citenamefont {Rojo-Franc\`as}\ \emph {et~al.}(2024)\citenamefont {Rojo-Franc\`as}, \citenamefont {Isaule}, \citenamefont {Santos}, \citenamefont {Juli\'a-D\'{\i}az},\ and\ \citenamefont {Zinner}}]{PhysRevA.110.032205}%
  \BibitemOpen
  \bibfield  {author} {\bibinfo {author} {\bibfnamefont {A.}~\bibnamefont {Rojo-Franc\`as}}, \bibinfo {author} {\bibfnamefont {F.}~\bibnamefont {Isaule}}, \bibinfo {author} {\bibfnamefont {A.~C.}\ \bibnamefont {Santos}}, \bibinfo {author} {\bibfnamefont {B.}~\bibnamefont {Juli\'a-D\'{\i}az}},\ and\ \bibinfo {author} {\bibfnamefont {N.~T.}\ \bibnamefont {Zinner}},\ }\bibfield  {title} {\bibinfo {title} {Stable collective charging of ultracold-atom quantum batteries},\ }\href {https://doi.org/10.1103/PhysRevA.110.032205} {\bibfield  {journal} {\bibinfo  {journal} {Phys. Rev. A}\ }\textbf {\bibinfo {volume} {110}},\ \bibinfo {pages} {032205} (\bibinfo {year} {2024})}\BibitemShut {NoStop}%
\bibitem [{\citenamefont {Yang}\ \emph {et~al.}(2024{\natexlab{c}})\citenamefont {Yang}, \citenamefont {Yang}, \citenamefont {Liu}, \citenamefont {Jiang}, \citenamefont {Zheng}, \citenamefont {Fei},\ and\ \citenamefont {Luo}}]{YANG2024102300}%
  \BibitemOpen
  \bibfield  {author} {\bibinfo {author} {\bibfnamefont {X.}~\bibnamefont {Yang}}, \bibinfo {author} {\bibfnamefont {Y.-H.}\ \bibnamefont {Yang}}, \bibinfo {author} {\bibfnamefont {X.-Z.}\ \bibnamefont {Liu}}, \bibinfo {author} {\bibfnamefont {J.-L.}\ \bibnamefont {Jiang}}, \bibinfo {author} {\bibfnamefont {X.-Z.}\ \bibnamefont {Zheng}}, \bibinfo {author} {\bibfnamefont {S.-M.}\ \bibnamefont {Fei}},\ and\ \bibinfo {author} {\bibfnamefont {M.-X.}\ \bibnamefont {Luo}},\ }\bibfield  {title} {\bibinfo {title} {Experimental verification of quantum battery capacity with an optical platform},\ }\href {https://doi.org/https://doi.org/10.1016/j.xcrp.2024.102300} {\bibfield  {journal} {\bibinfo  {journal} {Cell Reports Physical Science}\ }\textbf {\bibinfo {volume} {5}},\ \bibinfo {pages} {102300} (\bibinfo {year} {2024}{\natexlab{c}})}\BibitemShut {NoStop}%
\bibitem [{\citenamefont {Barry}\ \emph {et~al.}(2020)\citenamefont {Barry}, \citenamefont {Schloss}, \citenamefont {Bauch}, \citenamefont {Turner}, \citenamefont {Hart}, \citenamefont {Pham},\ and\ \citenamefont {Walsworth}}]{RevModPhys.92.015004}%
  \BibitemOpen
  \bibfield  {author} {\bibinfo {author} {\bibfnamefont {J.~F.}\ \bibnamefont {Barry}}, \bibinfo {author} {\bibfnamefont {J.~M.}\ \bibnamefont {Schloss}}, \bibinfo {author} {\bibfnamefont {E.}~\bibnamefont {Bauch}}, \bibinfo {author} {\bibfnamefont {M.~J.}\ \bibnamefont {Turner}}, \bibinfo {author} {\bibfnamefont {C.~A.}\ \bibnamefont {Hart}}, \bibinfo {author} {\bibfnamefont {L.~M.}\ \bibnamefont {Pham}},\ and\ \bibinfo {author} {\bibfnamefont {R.~L.}\ \bibnamefont {Walsworth}},\ }\bibfield  {title} {\bibinfo {title} {Sensitivity optimization for nv-diamond magnetometry},\ }\href {https://doi.org/10.1103/RevModPhys.92.015004} {\bibfield  {journal} {\bibinfo  {journal} {Rev. Mod. Phys.}\ }\textbf {\bibinfo {volume} {92}},\ \bibinfo {pages} {015004} (\bibinfo {year} {2020})}\BibitemShut {NoStop}%
\bibitem [{\citenamefont {Joshi}\ and\ \citenamefont {Mahesh}(2022)}]{PhysRevA.106.042601}%
  \BibitemOpen
  \bibfield  {author} {\bibinfo {author} {\bibfnamefont {J.}~\bibnamefont {Joshi}}\ and\ \bibinfo {author} {\bibfnamefont {T.~S.}\ \bibnamefont {Mahesh}},\ }\bibfield  {title} {\bibinfo {title} {Experimental investigation of a quantum battery using star-topology nmr spin systems},\ }\href {https://doi.org/10.1103/PhysRevA.106.042601} {\bibfield  {journal} {\bibinfo  {journal} {Phys. Rev. A}\ }\textbf {\bibinfo {volume} {106}},\ \bibinfo {pages} {042601} (\bibinfo {year} {2022})}\BibitemShut {NoStop}%
\bibitem [{\citenamefont {Cruz}\ \emph {et~al.}(2022)\citenamefont {Cruz}, \citenamefont {Anka}, \citenamefont {Reis}, \citenamefont {Bachelard},\ and\ \citenamefont {Santos}}]{Cruz_2022}%
  \BibitemOpen
  \bibfield  {author} {\bibinfo {author} {\bibfnamefont {C.}~\bibnamefont {Cruz}}, \bibinfo {author} {\bibfnamefont {M.~F.}\ \bibnamefont {Anka}}, \bibinfo {author} {\bibfnamefont {M.~S.}\ \bibnamefont {Reis}}, \bibinfo {author} {\bibfnamefont {R.}~\bibnamefont {Bachelard}},\ and\ \bibinfo {author} {\bibfnamefont {A.~C.}\ \bibnamefont {Santos}},\ }\bibfield  {title} {\bibinfo {title} {Quantum battery based on quantum discord at room temperature},\ }\href {https://doi.org/10.1088/2058-9565/ac57f3} {\bibfield  {journal} {\bibinfo  {journal} {Quantum Science and Technology}\ }\textbf {\bibinfo {volume} {7}},\ \bibinfo {pages} {025020} (\bibinfo {year} {2022})}\BibitemShut {NoStop}%
\bibitem [{\citenamefont {Breuer}\ \emph {et~al.}(2007)\citenamefont {Breuer}, \citenamefont {Petruccione} \emph {et~al.}}]{breuer2002theory}%
  \BibitemOpen
  \bibfield  {author} {\bibinfo {author} {\bibfnamefont {H.-P.}\ \bibnamefont {Breuer}}, \bibinfo {author} {\bibfnamefont {F.}~\bibnamefont {Petruccione}}, \emph {et~al.},\ }\href {https://doi.org/10.1093/acprof:oso/9780199213900.001.0001} {\emph {\bibinfo {title} {The theory of open quantum systems}}}\ (\bibinfo  {publisher} {Oxford University, Oxford},\ \bibinfo {year} {2007})\BibitemShut {NoStop}%
\bibitem [{\citenamefont {Pirmoradian}\ and\ \citenamefont {Molmer}(2019)}]{PhysRevA.100.043833}%
  \BibitemOpen
  \bibfield  {author} {\bibinfo {author} {\bibfnamefont {F.}~\bibnamefont {Pirmoradian}}\ and\ \bibinfo {author} {\bibfnamefont {K.}~\bibnamefont {Molmer}},\ }\bibfield  {title} {\bibinfo {title} {Aging of a quantum battery},\ }\href {https://doi.org/10.1103/PhysRevA.100.043833} {\bibfield  {journal} {\bibinfo  {journal} {Phys. Rev. A}\ }\textbf {\bibinfo {volume} {100}},\ \bibinfo {pages} {043833} (\bibinfo {year} {2019})}\BibitemShut {NoStop}%
\bibitem [{\citenamefont {Santos}(2021)}]{PhysRevE.103.042118}%
  \BibitemOpen
  \bibfield  {author} {\bibinfo {author} {\bibfnamefont {A.~C.}\ \bibnamefont {Santos}},\ }\bibfield  {title} {\bibinfo {title} {Quantum advantage of two-level batteries in the self-discharging process},\ }\href {https://doi.org/10.1103/PhysRevE.103.042118} {\bibfield  {journal} {\bibinfo  {journal} {Phys. Rev. E}\ }\textbf {\bibinfo {volume} {103}},\ \bibinfo {pages} {042118} (\bibinfo {year} {2021})}\BibitemShut {NoStop}%
\bibitem [{\citenamefont {Downing}\ and\ \citenamefont {Ukhtary}(2024)}]{PhysRevA.109.052206}%
  \BibitemOpen
  \bibfield  {author} {\bibinfo {author} {\bibfnamefont {C.~A.}\ \bibnamefont {Downing}}\ and\ \bibinfo {author} {\bibfnamefont {M.~S.}\ \bibnamefont {Ukhtary}},\ }\bibfield  {title} {\bibinfo {title} {Hyperbolic enhancement of a quantum battery},\ }\href {https://doi.org/10.1103/PhysRevA.109.052206} {\bibfield  {journal} {\bibinfo  {journal} {Phys. Rev. A}\ }\textbf {\bibinfo {volume} {109}},\ \bibinfo {pages} {052206} (\bibinfo {year} {2024})}\BibitemShut {NoStop}%
\bibitem [{\citenamefont {Santos}\ \emph {et~al.}(2019)\citenamefont {Santos}, \citenamefont {Çakmak}, \citenamefont {Campbell},\ and\ \citenamefont {Zinner}}]{PhysRevE.100.032107}%
  \BibitemOpen
  \bibfield  {author} {\bibinfo {author} {\bibfnamefont {A.~C.}\ \bibnamefont {Santos}}, \bibinfo {author} {\bibfnamefont {B.}~\bibnamefont {Çakmak}}, \bibinfo {author} {\bibfnamefont {S.}~\bibnamefont {Campbell}},\ and\ \bibinfo {author} {\bibfnamefont {N.~T.}\ \bibnamefont {Zinner}},\ }\bibfield  {title} {\bibinfo {title} {Stable adiabatic quantum batteries},\ }\href {https://doi.org/10.1103/PhysRevE.100.032107} {\bibfield  {journal} {\bibinfo  {journal} {Phys. Rev. E}\ }\textbf {\bibinfo {volume} {100}},\ \bibinfo {pages} {032107} (\bibinfo {year} {2019})}\BibitemShut {NoStop}%
\bibitem [{\citenamefont {Mitchison}\ \emph {et~al.}(2021)\citenamefont {Mitchison}, \citenamefont {Goold},\ and\ \citenamefont {Prior}}]{Mitchison2021chargingquantum}%
  \BibitemOpen
  \bibfield  {author} {\bibinfo {author} {\bibfnamefont {M.~T.}\ \bibnamefont {Mitchison}}, \bibinfo {author} {\bibfnamefont {J.}~\bibnamefont {Goold}},\ and\ \bibinfo {author} {\bibfnamefont {J.}~\bibnamefont {Prior}},\ }\bibfield  {title} {\bibinfo {title} {Charging a quantum battery with linear feedback control},\ }\href {https://doi.org/10.22331/q-2021-07-13-500} {\bibfield  {journal} {\bibinfo  {journal} {{Quantum}}\ }\textbf {\bibinfo {volume} {5}},\ \bibinfo {pages} {500} (\bibinfo {year} {2021})}\BibitemShut {NoStop}%
\bibitem [{\citenamefont {Liu}\ \emph {et~al.}(2019)\citenamefont {Liu}, \citenamefont {Segal},\ and\ \citenamefont {Hanna}}]{Liu2019}%
  \BibitemOpen
  \bibfield  {author} {\bibinfo {author} {\bibfnamefont {J.}~\bibnamefont {Liu}}, \bibinfo {author} {\bibfnamefont {D.}~\bibnamefont {Segal}},\ and\ \bibinfo {author} {\bibfnamefont {G.}~\bibnamefont {Hanna}},\ }\bibfield  {title} {\bibinfo {title} {Loss-free excitonic quantum battery},\ }\href {https://doi.org/10.1021/acs.jpcc.9b06373} {\bibfield  {journal} {\bibinfo  {journal} {The Journal of Physical Chemistry C}\ }\textbf {\bibinfo {volume} {123}},\ \bibinfo {pages} {18303} (\bibinfo {year} {2019})}\BibitemShut {NoStop}%
\bibitem [{\citenamefont {Xu}\ \emph {et~al.}(2024)\citenamefont {Xu}, \citenamefont {Li}, \citenamefont {Zhu},\ and\ \citenamefont {Liu}}]{selfdis}%
  \BibitemOpen
  \bibfield  {author} {\bibinfo {author} {\bibfnamefont {K.}~\bibnamefont {Xu}}, \bibinfo {author} {\bibfnamefont {H.-G.}\ \bibnamefont {Li}}, \bibinfo {author} {\bibfnamefont {H.-J.}\ \bibnamefont {Zhu}},\ and\ \bibinfo {author} {\bibfnamefont {W.-M.}\ \bibnamefont {Liu}},\ }\bibfield  {title} {\bibinfo {title} {Inhibiting the self-discharging process of quantum batteries in non-markovian noises},\ }\href {https://doi.org/10.1103/PhysRevE.109.054132} {\bibfield  {journal} {\bibinfo  {journal} {Phys. Rev. E}\ }\textbf {\bibinfo {volume} {109}},\ \bibinfo {pages} {054132} (\bibinfo {year} {2024})}\BibitemShut {NoStop}%
\bibitem [{\citenamefont {Quach}\ and\ \citenamefont {Munro}(2020)}]{stable1}%
  \BibitemOpen
  \bibfield  {author} {\bibinfo {author} {\bibfnamefont {J.~Q.}\ \bibnamefont {Quach}}\ and\ \bibinfo {author} {\bibfnamefont {W.~J.}\ \bibnamefont {Munro}},\ }\bibfield  {title} {\bibinfo {title} {Using dark states to charge and stabilize open quantum batteries},\ }\href {https://doi.org/10.1103/PhysRevApplied.14.024092} {\bibfield  {journal} {\bibinfo  {journal} {Phys. Rev. Appl.}\ }\textbf {\bibinfo {volume} {14}},\ \bibinfo {pages} {024092} (\bibinfo {year} {2020})}\BibitemShut {NoStop}%
\bibitem [{\citenamefont {Carrega}\ \emph {et~al.}(2020)\citenamefont {Carrega}, \citenamefont {Crescente}, \citenamefont {Ferraro},\ and\ \citenamefont {Sassetti}}]{Carrega_2020}%
  \BibitemOpen
  \bibfield  {author} {\bibinfo {author} {\bibfnamefont {M.}~\bibnamefont {Carrega}}, \bibinfo {author} {\bibfnamefont {A.}~\bibnamefont {Crescente}}, \bibinfo {author} {\bibfnamefont {D.}~\bibnamefont {Ferraro}},\ and\ \bibinfo {author} {\bibfnamefont {M.}~\bibnamefont {Sassetti}},\ }\bibfield  {title} {\bibinfo {title} {Dissipative dynamics of an open quantum battery},\ }\href {https://doi.org/10.1088/1367-2630/abaa01} {\bibfield  {journal} {\bibinfo  {journal} {New Journal of Physics}\ }\textbf {\bibinfo {volume} {22}},\ \bibinfo {pages} {083085} (\bibinfo {year} {2020})}\BibitemShut {NoStop}%
\bibitem [{\citenamefont {Barra}(2019)}]{PhysRevLett.122.210601}%
  \BibitemOpen
  \bibfield  {author} {\bibinfo {author} {\bibfnamefont {F.}~\bibnamefont {Barra}},\ }\bibfield  {title} {\bibinfo {title} {Dissipative charging of a quantum battery},\ }\href {https://doi.org/10.1103/PhysRevLett.122.210601} {\bibfield  {journal} {\bibinfo  {journal} {Phys. Rev. Lett.}\ }\textbf {\bibinfo {volume} {122}},\ \bibinfo {pages} {210601} (\bibinfo {year} {2019})}\BibitemShut {NoStop}%
\bibitem [{\citenamefont {Farina}\ \emph {et~al.}(2019)\citenamefont {Farina}, \citenamefont {Andolina}, \citenamefont {Mari}, \citenamefont {Polini},\ and\ \citenamefont {Giovannetti}}]{PhysRevB.99.035421}%
  \BibitemOpen
  \bibfield  {author} {\bibinfo {author} {\bibfnamefont {D.}~\bibnamefont {Farina}}, \bibinfo {author} {\bibfnamefont {G.~M.}\ \bibnamefont {Andolina}}, \bibinfo {author} {\bibfnamefont {A.}~\bibnamefont {Mari}}, \bibinfo {author} {\bibfnamefont {M.}~\bibnamefont {Polini}},\ and\ \bibinfo {author} {\bibfnamefont {V.}~\bibnamefont {Giovannetti}},\ }\bibfield  {title} {\bibinfo {title} {Charger-mediated energy transfer for quantum batteries: An open-system approach},\ }\href {https://doi.org/10.1103/PhysRevB.99.035421} {\bibfield  {journal} {\bibinfo  {journal} {Phys. Rev. B}\ }\textbf {\bibinfo {volume} {99}},\ \bibinfo {pages} {035421} (\bibinfo {year} {2019})}\BibitemShut {NoStop}%
\bibitem [{\citenamefont {Bai}\ and\ \citenamefont {An}(2020)}]{PhysRevA.102.060201}%
  \BibitemOpen
  \bibfield  {author} {\bibinfo {author} {\bibfnamefont {S.-Y.}\ \bibnamefont {Bai}}\ and\ \bibinfo {author} {\bibfnamefont {J.-H.}\ \bibnamefont {An}},\ }\bibfield  {title} {\bibinfo {title} {Floquet engineering to reactivate a dissipative quantum battery},\ }\href {https://doi.org/10.1103/PhysRevA.102.060201} {\bibfield  {journal} {\bibinfo  {journal} {Phys. Rev. A}\ }\textbf {\bibinfo {volume} {102}},\ \bibinfo {pages} {060201} (\bibinfo {year} {2020})}\BibitemShut {NoStop}%
\bibitem [{\citenamefont {Zhao}\ \emph {et~al.}(2021)\citenamefont {Zhao}, \citenamefont {Dou},\ and\ \citenamefont {Zhao}}]{PhysRevA.103.033715}%
  \BibitemOpen
  \bibfield  {author} {\bibinfo {author} {\bibfnamefont {F.}~\bibnamefont {Zhao}}, \bibinfo {author} {\bibfnamefont {F.-Q.}\ \bibnamefont {Dou}},\ and\ \bibinfo {author} {\bibfnamefont {Q.}~\bibnamefont {Zhao}},\ }\bibfield  {title} {\bibinfo {title} {Quantum battery of interacting spins with environmental noise},\ }\href {https://doi.org/10.1103/PhysRevA.103.033715} {\bibfield  {journal} {\bibinfo  {journal} {Phys. Rev. A}\ }\textbf {\bibinfo {volume} {103}},\ \bibinfo {pages} {033715} (\bibinfo {year} {2021})}\BibitemShut {NoStop}%
\bibitem [{\citenamefont {Allahverdyan}\ \emph {et~al.}(2004)\citenamefont {Allahverdyan}, \citenamefont {Balian},\ and\ \citenamefont {Nieuwenhuizen}}]{A.E.Allahverdyan_2004}%
  \BibitemOpen
  \bibfield  {author} {\bibinfo {author} {\bibfnamefont {A.~E.}\ \bibnamefont {Allahverdyan}}, \bibinfo {author} {\bibfnamefont {R.}~\bibnamefont {Balian}},\ and\ \bibinfo {author} {\bibfnamefont {T.~M.}\ \bibnamefont {Nieuwenhuizen}},\ }\bibfield  {title} {\bibinfo {title} {Maximal work extraction from finite quantum systems},\ }\href {https://doi.org/10.1209/epl/i2004-10101-2} {\bibfield  {journal} {\bibinfo  {journal} {Europhysics Letters}\ }\textbf {\bibinfo {volume} {67}},\ \bibinfo {pages} {565} (\bibinfo {year} {2004})}\BibitemShut {NoStop}%
\bibitem [{\citenamefont {Pusz}\ and\ \citenamefont {Woronowicz}(1978)}]{Pusz1978}%
  \BibitemOpen
  \bibfield  {author} {\bibinfo {author} {\bibfnamefont {W.}~\bibnamefont {Pusz}}\ and\ \bibinfo {author} {\bibfnamefont {S.~L.}\ \bibnamefont {Woronowicz}},\ }\bibfield  {title} {\bibinfo {title} {Passive states and kms states for general quantum systems},\ }\href {https://doi.org/10.1007/BF01614224} {\bibfield  {journal} {\bibinfo  {journal} {Communications in Mathematical Physics}\ }\textbf {\bibinfo {volume} {58}},\ \bibinfo {pages} {273} (\bibinfo {year} {1978})}\BibitemShut {NoStop}%
\bibitem [{\citenamefont {Cai}\ \emph {et~al.}(2023)\citenamefont {Cai}, \citenamefont {Babbush}, \citenamefont {Benjamin}, \citenamefont {Endo}, \citenamefont {Huggins}, \citenamefont {Li}, \citenamefont {McClean},\ and\ \citenamefont {O'Brien}}]{RevModPhys.95.045005}%
  \BibitemOpen
  \bibfield  {author} {\bibinfo {author} {\bibfnamefont {Z.}~\bibnamefont {Cai}}, \bibinfo {author} {\bibfnamefont {R.}~\bibnamefont {Babbush}}, \bibinfo {author} {\bibfnamefont {S.~C.}\ \bibnamefont {Benjamin}}, \bibinfo {author} {\bibfnamefont {S.}~\bibnamefont {Endo}}, \bibinfo {author} {\bibfnamefont {W.~J.}\ \bibnamefont {Huggins}}, \bibinfo {author} {\bibfnamefont {Y.}~\bibnamefont {Li}}, \bibinfo {author} {\bibfnamefont {J.~R.}\ \bibnamefont {McClean}},\ and\ \bibinfo {author} {\bibfnamefont {T.~E.}\ \bibnamefont {O'Brien}},\ }\bibfield  {title} {\bibinfo {title} {Quantum error mitigation},\ }\href {https://doi.org/10.1103/RevModPhys.95.045005} {\bibfield  {journal} {\bibinfo  {journal} {Rev. Mod. Phys.}\ }\textbf {\bibinfo {volume} {95}},\ \bibinfo {pages} {045005} (\bibinfo {year} {2023})}\BibitemShut {NoStop}%
\bibitem [{\citenamefont {Galve}\ \emph {et~al.}(2017)\citenamefont {Galve}, \citenamefont {Mandarino}, \citenamefont {Paris}, \citenamefont {Benedetti},\ and\ \citenamefont {Zambrini}}]{Galve2017}%
  \BibitemOpen
  \bibfield  {author} {\bibinfo {author} {\bibfnamefont {F.}~\bibnamefont {Galve}}, \bibinfo {author} {\bibfnamefont {A.}~\bibnamefont {Mandarino}}, \bibinfo {author} {\bibfnamefont {M.~G.~A.}\ \bibnamefont {Paris}}, \bibinfo {author} {\bibfnamefont {C.}~\bibnamefont {Benedetti}},\ and\ \bibinfo {author} {\bibfnamefont {R.}~\bibnamefont {Zambrini}},\ }\bibfield  {title} {\bibinfo {title} {Microscopic description for the emergence of collective dissipation in extended quantum systems},\ }\href {https://doi.org/10.1038/srep42050} {\bibfield  {journal} {\bibinfo  {journal} {Scientific Reports}\ }\textbf {\bibinfo {volume} {7}},\ \bibinfo {pages} {42050} (\bibinfo {year} {2017})}\BibitemShut {NoStop}%
\bibitem [{\citenamefont {Etxezarreta~Martinez}\ \emph {et~al.}(2022)\citenamefont {Etxezarreta~Martinez}, \citenamefont {Fuentes}, \citenamefont {Crespo},\ and\ \citenamefont {Garcia-Frias}}]{PhysRevA.105.012432}%
  \BibitemOpen
  \bibfield  {author} {\bibinfo {author} {\bibfnamefont {J.}~\bibnamefont {Etxezarreta~Martinez}}, \bibinfo {author} {\bibfnamefont {P.}~\bibnamefont {Fuentes}}, \bibinfo {author} {\bibfnamefont {P.~M.}\ \bibnamefont {Crespo}},\ and\ \bibinfo {author} {\bibfnamefont {J.}~\bibnamefont {Garcia-Frias}},\ }\bibfield  {title} {\bibinfo {title} {Quantum outage probability for time-varying quantum channels},\ }\href {https://doi.org/10.1103/PhysRevA.105.012432} {\bibfield  {journal} {\bibinfo  {journal} {Phys. Rev. A}\ }\textbf {\bibinfo {volume} {105}},\ \bibinfo {pages} {012432} (\bibinfo {year} {2022})}\BibitemShut {NoStop}%
\bibitem [{\citenamefont {Molitor}\ \emph {et~al.}(2024)\citenamefont {Molitor}, \citenamefont {Malavazi}, \citenamefont {Baldij{\~a}o}, \citenamefont {Orthey}, \citenamefont {Paiva},\ and\ \citenamefont {Dieguez}}]{Molitor2024}%
  \BibitemOpen
  \bibfield  {author} {\bibinfo {author} {\bibfnamefont {O.~A.~D.}\ \bibnamefont {Molitor}}, \bibinfo {author} {\bibfnamefont {A.~H.~A.}\ \bibnamefont {Malavazi}}, \bibinfo {author} {\bibfnamefont {R.~D.}\ \bibnamefont {Baldij{\~a}o}}, \bibinfo {author} {\bibfnamefont {A.~C.}\ \bibnamefont {Orthey}}, \bibinfo {author} {\bibfnamefont {I.~L.}\ \bibnamefont {Paiva}},\ and\ \bibinfo {author} {\bibfnamefont {P.~R.}\ \bibnamefont {Dieguez}},\ }\bibfield  {title} {\bibinfo {title} {Quantum switch instabilities with an open control},\ }\href {https://doi.org/10.1038/s42005-024-01843-y} {\bibfield  {journal} {\bibinfo  {journal} {Communications Physics}\ }\textbf {\bibinfo {volume} {7}},\ \bibinfo {pages} {373} (\bibinfo {year} {2024})}\BibitemShut {NoStop}%
\bibitem [{\citenamefont {Aziz}\ \emph {et~al.}(2025)\citenamefont {Aziz}, \citenamefont {Malavazi},\ and\ \citenamefont {Dieguez}}]{aziz2025revival}%
  \BibitemOpen
  \bibfield  {author} {\bibinfo {author} {\bibfnamefont {S.}~\bibnamefont {Aziz}}, \bibinfo {author} {\bibfnamefont {A.~H.}\ \bibnamefont {Malavazi}},\ and\ \bibinfo {author} {\bibfnamefont {P.~R.}\ \bibnamefont {Dieguez}},\ }\bibfield  {title} {\bibinfo {title} {Revival and instabilities of entanglement in monitoring maps with indefinite causal order},\ }\href {https://doi.org/10.48550/arXiv.2503.13373} {\bibfield  {journal} {\bibinfo  {journal} {arXiv:2503.13373}\ } (\bibinfo {year} {2025})}\BibitemShut {NoStop}%
\bibitem [{\citenamefont {Gherardini}\ \emph {et~al.}(2020)\citenamefont {Gherardini}, \citenamefont {Campaioli}, \citenamefont {Caruso},\ and\ \citenamefont {Binder}}]{PhysRevResearch.2.013095}%
  \BibitemOpen
  \bibfield  {author} {\bibinfo {author} {\bibfnamefont {S.}~\bibnamefont {Gherardini}}, \bibinfo {author} {\bibfnamefont {F.}~\bibnamefont {Campaioli}}, \bibinfo {author} {\bibfnamefont {F.}~\bibnamefont {Caruso}},\ and\ \bibinfo {author} {\bibfnamefont {F.~C.}\ \bibnamefont {Binder}},\ }\bibfield  {title} {\bibinfo {title} {Stabilizing open quantum batteries by sequential measurements},\ }\href {https://doi.org/10.1103/PhysRevResearch.2.013095} {\bibfield  {journal} {\bibinfo  {journal} {Phys. Rev. Res.}\ }\textbf {\bibinfo {volume} {2}},\ \bibinfo {pages} {013095} (\bibinfo {year} {2020})}\BibitemShut {NoStop}%
\bibitem [{\citenamefont {Itano}\ \emph {et~al.}(1990)\citenamefont {Itano}, \citenamefont {Heinzen}, \citenamefont {Bollinger},\ and\ \citenamefont {Wineland}}]{PhysRevA.41.2295}%
  \BibitemOpen
  \bibfield  {author} {\bibinfo {author} {\bibfnamefont {W.~M.}\ \bibnamefont {Itano}}, \bibinfo {author} {\bibfnamefont {D.~J.}\ \bibnamefont {Heinzen}}, \bibinfo {author} {\bibfnamefont {J.~J.}\ \bibnamefont {Bollinger}},\ and\ \bibinfo {author} {\bibfnamefont {D.~J.}\ \bibnamefont {Wineland}},\ }\bibfield  {title} {\bibinfo {title} {Quantum zeno effect},\ }\href {https://doi.org/10.1103/PhysRevA.41.2295} {\bibfield  {journal} {\bibinfo  {journal} {Phys. Rev. A}\ }\textbf {\bibinfo {volume} {41}},\ \bibinfo {pages} {2295} (\bibinfo {year} {1990})}\BibitemShut {NoStop}%
\bibitem [{\citenamefont {Yi}\ \emph {et~al.}(2017)\citenamefont {Yi}, \citenamefont {Talkner},\ and\ \citenamefont {Kim}}]{yi2017single}%
  \BibitemOpen
  \bibfield  {author} {\bibinfo {author} {\bibfnamefont {J.}~\bibnamefont {Yi}}, \bibinfo {author} {\bibfnamefont {P.}~\bibnamefont {Talkner}},\ and\ \bibinfo {author} {\bibfnamefont {Y.~W.}\ \bibnamefont {Kim}},\ }\bibfield  {title} {\bibinfo {title} {Single-temperature quantum engine without feedback control},\ }\href {https://doi.org/https://10.1103/physreve.96.022108} {\bibfield  {journal} {\bibinfo  {journal} {Phys. Rev. E}\ }\textbf {\bibinfo {volume} {96}},\ \bibinfo {pages} {022108} (\bibinfo {year} {2017})}\BibitemShut {NoStop}%
\bibitem [{\citenamefont {Elouard}\ \emph {et~al.}(2017{\natexlab{a}})\citenamefont {Elouard}, \citenamefont {Herrera-Mart{\'\i}}, \citenamefont {Clusel},\ and\ \citenamefont {Auffeves}}]{elouard2017role}%
  \BibitemOpen
  \bibfield  {author} {\bibinfo {author} {\bibfnamefont {C.}~\bibnamefont {Elouard}}, \bibinfo {author} {\bibfnamefont {D.~A.}\ \bibnamefont {Herrera-Mart{\'\i}}}, \bibinfo {author} {\bibfnamefont {M.}~\bibnamefont {Clusel}},\ and\ \bibinfo {author} {\bibfnamefont {A.}~\bibnamefont {Auffeves}},\ }\bibfield  {title} {\bibinfo {title} {The role of quantum measurement in stochastic thermodynamics},\ }\href {https://doi.org/https://doi.org/10.1038/s41534-017-0008-4} {\bibfield  {journal} {\bibinfo  {journal} {npj Quan. Inf.}\ }\textbf {\bibinfo {volume} {3}},\ \bibinfo {pages} {1} (\bibinfo {year} {2017}{\natexlab{a}})}\BibitemShut {NoStop}%
\bibitem [{\citenamefont {Brandner}\ \emph {et~al.}(2015)\citenamefont {Brandner}, \citenamefont {Bauer}, \citenamefont {Schmid},\ and\ \citenamefont {Seifert}}]{brandner2015coherence}%
  \BibitemOpen
  \bibfield  {author} {\bibinfo {author} {\bibfnamefont {K.}~\bibnamefont {Brandner}}, \bibinfo {author} {\bibfnamefont {M.}~\bibnamefont {Bauer}}, \bibinfo {author} {\bibfnamefont {M.~T.}\ \bibnamefont {Schmid}},\ and\ \bibinfo {author} {\bibfnamefont {U.}~\bibnamefont {Seifert}},\ }\bibfield  {title} {\bibinfo {title} {Coherence-enhanced efficiency of feedback-driven quantum engines},\ }\href {https://doi.org/https://doi.org/10.1088/1367-2630/17/6/065006} {\bibfield  {journal} {\bibinfo  {journal} {New J. of Phys.}\ }\textbf {\bibinfo {volume} {17}},\ \bibinfo {pages} {065006} (\bibinfo {year} {2015})}\BibitemShut {NoStop}%
\bibitem [{\citenamefont {Campisi}\ \emph {et~al.}(2017)\citenamefont {Campisi}, \citenamefont {Pekola},\ and\ \citenamefont {Fazio}}]{campisi2017feedback}%
  \BibitemOpen
  \bibfield  {author} {\bibinfo {author} {\bibfnamefont {M.}~\bibnamefont {Campisi}}, \bibinfo {author} {\bibfnamefont {J.}~\bibnamefont {Pekola}},\ and\ \bibinfo {author} {\bibfnamefont {R.}~\bibnamefont {Fazio}},\ }\bibfield  {title} {\bibinfo {title} {Feedback-controlled heat transport in quantum devices: theory and solid-state experimental proposal},\ }\href {https://doi.org/https://doi.org/10.1088/1367-2630/aa6acb} {\bibfield  {journal} {\bibinfo  {journal} {New J. Phys.}\ }\textbf {\bibinfo {volume} {19}},\ \bibinfo {pages} {053027} (\bibinfo {year} {2017})}\BibitemShut {NoStop}%
\bibitem [{\citenamefont {Chand}\ and\ \citenamefont {Biswas}(2017{\natexlab{a}})}]{chand2017single}%
  \BibitemOpen
  \bibfield  {author} {\bibinfo {author} {\bibfnamefont {S.}~\bibnamefont {Chand}}\ and\ \bibinfo {author} {\bibfnamefont {A.}~\bibnamefont {Biswas}},\ }\bibfield  {title} {\bibinfo {title} {Single-ion quantum {Otto} engine with always-on bath interaction},\ }\href {https://doi.org/https://doi.org/10.1209/0295-5075/118/60003} {\bibfield  {journal} {\bibinfo  {journal} {Europhys. Lett.}\ }\textbf {\bibinfo {volume} {118}},\ \bibinfo {pages} {60003} (\bibinfo {year} {2017}{\natexlab{a}})}\BibitemShut {NoStop}%
\bibitem [{\citenamefont {Chand}\ and\ \citenamefont {Biswas}(2017{\natexlab{b}})}]{chand2017measurement}%
  \BibitemOpen
  \bibfield  {author} {\bibinfo {author} {\bibfnamefont {S.}~\bibnamefont {Chand}}\ and\ \bibinfo {author} {\bibfnamefont {A.}~\bibnamefont {Biswas}},\ }\bibfield  {title} {\bibinfo {title} {Measurement-induced operation of two-ion quantum heat machines},\ }\href {https://doi.org/https://doi.org/10.1103/PhysRevE.95.032111} {\bibfield  {journal} {\bibinfo  {journal} {Phys. Rev. E}\ }\textbf {\bibinfo {volume} {95}},\ \bibinfo {pages} {032111} (\bibinfo {year} {2017}{\natexlab{b}})}\BibitemShut {NoStop}%
\bibitem [{\citenamefont {Mohammady}\ and\ \citenamefont {Anders}(2017)}]{mohammady2017quantum}%
  \BibitemOpen
  \bibfield  {author} {\bibinfo {author} {\bibfnamefont {M.~H.}\ \bibnamefont {Mohammady}}\ and\ \bibinfo {author} {\bibfnamefont {J.}~\bibnamefont {Anders}},\ }\bibfield  {title} {\bibinfo {title} {A quantum {Szil\'{a}rd} engine without heat from a thermal reservoir},\ }\href {https://doi.org/https://doi.org/10.1088/1367-2630/aa8ba1} {\bibfield  {journal} {\bibinfo  {journal} {New J. Phys.}\ }\textbf {\bibinfo {volume} {19}},\ \bibinfo {pages} {113026} (\bibinfo {year} {2017})}\BibitemShut {NoStop}%
\bibitem [{\citenamefont {Elouard}\ \emph {et~al.}(2017{\natexlab{b}})\citenamefont {Elouard}, \citenamefont {Herrera-Mart{\'\i}}, \citenamefont {Huard},\ and\ \citenamefont {Auffeves}}]{elouard2017extracting}%
  \BibitemOpen
  \bibfield  {author} {\bibinfo {author} {\bibfnamefont {C.}~\bibnamefont {Elouard}}, \bibinfo {author} {\bibfnamefont {D.}~\bibnamefont {Herrera-Mart{\'\i}}}, \bibinfo {author} {\bibfnamefont {B.}~\bibnamefont {Huard}},\ and\ \bibinfo {author} {\bibfnamefont {A.}~\bibnamefont {Auffeves}},\ }\bibfield  {title} {\bibinfo {title} {Extracting work from quantum measurement in {Maxwell’s} demon engines},\ }\href {https://doi.org/https://doi.org/10.1103/PhysRevLett.118.260603} {\bibfield  {journal} {\bibinfo  {journal} {Phys. Rev. Lett.}\ }\textbf {\bibinfo {volume} {118}},\ \bibinfo {pages} {260603} (\bibinfo {year} {2017}{\natexlab{b}})}\BibitemShut {NoStop}%
\bibitem [{\citenamefont {Chand}\ and\ \citenamefont {Biswas}(2018)}]{chand2018critical}%
  \BibitemOpen
  \bibfield  {author} {\bibinfo {author} {\bibfnamefont {S.}~\bibnamefont {Chand}}\ and\ \bibinfo {author} {\bibfnamefont {A.}~\bibnamefont {Biswas}},\ }\bibfield  {title} {\bibinfo {title} {Critical-point behavior of a measurement-based quantum heat engine},\ }\href {https://doi.org/https://doi.org/10.1103/PhysRevE.98.052147} {\bibfield  {journal} {\bibinfo  {journal} {Phys. Rev. E}\ }\textbf {\bibinfo {volume} {98}},\ \bibinfo {pages} {052147} (\bibinfo {year} {2018})}\BibitemShut {NoStop}%
\bibitem [{\citenamefont {Ding}\ \emph {et~al.}(2018)\citenamefont {Ding}, \citenamefont {Yi}, \citenamefont {Kim},\ and\ \citenamefont {Talkner}}]{ding2018measurement}%
  \BibitemOpen
  \bibfield  {author} {\bibinfo {author} {\bibfnamefont {X.}~\bibnamefont {Ding}}, \bibinfo {author} {\bibfnamefont {J.}~\bibnamefont {Yi}}, \bibinfo {author} {\bibfnamefont {Y.~W.}\ \bibnamefont {Kim}},\ and\ \bibinfo {author} {\bibfnamefont {P.}~\bibnamefont {Talkner}},\ }\bibfield  {title} {\bibinfo {title} {Measurement-driven single temperature engine},\ }\href {https://doi.org/https://doi.org/10.1103/PhysRevE.98.042122} {\bibfield  {journal} {\bibinfo  {journal} {Phys. Rev. E}\ }\textbf {\bibinfo {volume} {98}},\ \bibinfo {pages} {042122} (\bibinfo {year} {2018})}\BibitemShut {NoStop}%
\bibitem [{\citenamefont {Elouard}\ and\ \citenamefont {Jordan}(2018)}]{elouard2018efficient}%
  \BibitemOpen
  \bibfield  {author} {\bibinfo {author} {\bibfnamefont {C.}~\bibnamefont {Elouard}}\ and\ \bibinfo {author} {\bibfnamefont {A.~N.}\ \bibnamefont {Jordan}},\ }\bibfield  {title} {\bibinfo {title} {Efficient quantum measurement engines},\ }\href {https://doi.org/https://doi.org/10.1103/PhysRevLett.120.260601} {\bibfield  {journal} {\bibinfo  {journal} {Phys. Rev. Lett.}\ }\textbf {\bibinfo {volume} {120}},\ \bibinfo {pages} {260601} (\bibinfo {year} {2018})}\BibitemShut {NoStop}%
\bibitem [{\citenamefont {Buffoni}\ \emph {et~al.}(2019)\citenamefont {Buffoni}, \citenamefont {Solfanelli}, \citenamefont {Verrucchi}, \citenamefont {Cuccoli},\ and\ \citenamefont {Campisi}}]{buffoni2019quantum}%
  \BibitemOpen
  \bibfield  {author} {\bibinfo {author} {\bibfnamefont {L.}~\bibnamefont {Buffoni}}, \bibinfo {author} {\bibfnamefont {A.}~\bibnamefont {Solfanelli}}, \bibinfo {author} {\bibfnamefont {P.}~\bibnamefont {Verrucchi}}, \bibinfo {author} {\bibfnamefont {A.}~\bibnamefont {Cuccoli}},\ and\ \bibinfo {author} {\bibfnamefont {M.}~\bibnamefont {Campisi}},\ }\bibfield  {title} {\bibinfo {title} {Quantum measurement cooling},\ }\href {https://doi.org/https://doi.org/10.1103/PhysRevLett.122.070603} {\bibfield  {journal} {\bibinfo  {journal} {Phys. Rev. Lett.}\ }\textbf {\bibinfo {volume} {122}},\ \bibinfo {pages} {070603} (\bibinfo {year} {2019})}\BibitemShut {NoStop}%
\bibitem [{\citenamefont {Solfanelli}\ \emph {et~al.}(2019)\citenamefont {Solfanelli}, \citenamefont {Buffoni}, \citenamefont {Cuccoli},\ and\ \citenamefont {Campisi}}]{solfanelli2019maximal}%
  \BibitemOpen
  \bibfield  {author} {\bibinfo {author} {\bibfnamefont {A.}~\bibnamefont {Solfanelli}}, \bibinfo {author} {\bibfnamefont {L.}~\bibnamefont {Buffoni}}, \bibinfo {author} {\bibfnamefont {A.}~\bibnamefont {Cuccoli}},\ and\ \bibinfo {author} {\bibfnamefont {M.}~\bibnamefont {Campisi}},\ }\bibfield  {title} {\bibinfo {title} {Maximal energy extraction via quantum measurement},\ }\href {https://doi.org/https://doi.org/10.1088/1742-5468/ab3721} {\bibfield  {journal} {\bibinfo  {journal} {J. Stat. Mech.}\ }\textbf {\bibinfo {volume} {2019}},\ \bibinfo {pages} {094003} (\bibinfo {year} {2019})}\BibitemShut {NoStop}%
\bibitem [{\citenamefont {Jordan}\ \emph {et~al.}(2020)\citenamefont {Jordan}, \citenamefont {Elouard},\ and\ \citenamefont {Auff{\`e}ves}}]{jordan2020quantum}%
  \BibitemOpen
  \bibfield  {author} {\bibinfo {author} {\bibfnamefont {A.~N.}\ \bibnamefont {Jordan}}, \bibinfo {author} {\bibfnamefont {C.}~\bibnamefont {Elouard}},\ and\ \bibinfo {author} {\bibfnamefont {A.}~\bibnamefont {Auff{\`e}ves}},\ }\bibfield  {title} {\bibinfo {title} {Quantum measurement engines and their relevance for quantum interpretations},\ }\href {https://doi.org/https://doi.org/10.1007/s40509-019-00217-2} {\bibfield  {journal} {\bibinfo  {journal} {Quantum Stud. Math. and Found.}\ }\textbf {\bibinfo {volume} {7}},\ \bibinfo {pages} {203} (\bibinfo {year} {2020})}\BibitemShut {NoStop}%
\bibitem [{\citenamefont {Behzadi}(2020)}]{behzadi2020quantum}%
  \BibitemOpen
  \bibfield  {author} {\bibinfo {author} {\bibfnamefont {N.}~\bibnamefont {Behzadi}},\ }\bibfield  {title} {\bibinfo {title} {Quantum engine based on general measurements},\ }\href {https://doi.org/https://doi.org/10.1088/1751-8121/abca74} {\bibfield  {journal} {\bibinfo  {journal} {J. of Phys. A: Math. and Theo.}\ }\textbf {\bibinfo {volume} {54}},\ \bibinfo {pages} {015304} (\bibinfo {year} {2020})}\BibitemShut {NoStop}%
\bibitem [{\citenamefont {Seah}\ \emph {et~al.}(2020)\citenamefont {Seah}, \citenamefont {Nimmrichter},\ and\ \citenamefont {Scarani}}]{seah2020maxwell}%
  \BibitemOpen
  \bibfield  {author} {\bibinfo {author} {\bibfnamefont {S.}~\bibnamefont {Seah}}, \bibinfo {author} {\bibfnamefont {S.}~\bibnamefont {Nimmrichter}},\ and\ \bibinfo {author} {\bibfnamefont {V.}~\bibnamefont {Scarani}},\ }\bibfield  {title} {\bibinfo {title} {{Maxwell’s} lesser demon: A quantum engine driven by pointer measurements},\ }\href {https://doi.org/https://doi.org/10.1103/PhysRevLett.124.100603} {\bibfield  {journal} {\bibinfo  {journal} {Phys. Rev. Lett.}\ }\textbf {\bibinfo {volume} {124}},\ \bibinfo {pages} {100603} (\bibinfo {year} {2020})}\BibitemShut {NoStop}%
\bibitem [{\citenamefont {Bresque}\ \emph {et~al.}(2021)\citenamefont {Bresque}, \citenamefont {Camati}, \citenamefont {Rogers}, \citenamefont {Murch}, \citenamefont {Jordan},\ and\ \citenamefont {Auff{\`e}ves}}]{bresque2021two}%
  \BibitemOpen
  \bibfield  {author} {\bibinfo {author} {\bibfnamefont {L.}~\bibnamefont {Bresque}}, \bibinfo {author} {\bibfnamefont {P.~A.}\ \bibnamefont {Camati}}, \bibinfo {author} {\bibfnamefont {S.}~\bibnamefont {Rogers}}, \bibinfo {author} {\bibfnamefont {K.}~\bibnamefont {Murch}}, \bibinfo {author} {\bibfnamefont {A.~N.}\ \bibnamefont {Jordan}},\ and\ \bibinfo {author} {\bibfnamefont {A.}~\bibnamefont {Auff{\`e}ves}},\ }\bibfield  {title} {\bibinfo {title} {Two-qubit engine fueled by entanglement and local measurements},\ }\href {https://doi.org/https://doi.org/10.1103/PhysRevLett.126.120605} {\bibfield  {journal} {\bibinfo  {journal} {Phys. Rev. Lett.}\ }\textbf {\bibinfo {volume} {126}},\ \bibinfo {pages} {120605} (\bibinfo {year} {2021})}\BibitemShut {NoStop}%
\bibitem [{\citenamefont {Chand}\ \emph {et~al.}(2021)\citenamefont {Chand}, \citenamefont {Dasgupta},\ and\ \citenamefont {Biswas}}]{chand2021finite}%
  \BibitemOpen
  \bibfield  {author} {\bibinfo {author} {\bibfnamefont {S.}~\bibnamefont {Chand}}, \bibinfo {author} {\bibfnamefont {S.}~\bibnamefont {Dasgupta}},\ and\ \bibinfo {author} {\bibfnamefont {A.}~\bibnamefont {Biswas}},\ }\bibfield  {title} {\bibinfo {title} {Finite-time performance of a single-ion quantum {Otto} engine},\ }\href {https://doi.org/https://doi.org/10.1103/PhysRevE.103.032144} {\bibfield  {journal} {\bibinfo  {journal} {Phys. Rev. E}\ }\textbf {\bibinfo {volume} {103}},\ \bibinfo {pages} {032144} (\bibinfo {year} {2021})}\BibitemShut {NoStop}%
\bibitem [{\citenamefont {Lin}\ \emph {et~al.}(2021)\citenamefont {Lin}, \citenamefont {Su}, \citenamefont {Chen}, \citenamefont {Chen},\ and\ \citenamefont {Santos}}]{lin2021suppressing}%
  \BibitemOpen
  \bibfield  {author} {\bibinfo {author} {\bibfnamefont {Z.}~\bibnamefont {Lin}}, \bibinfo {author} {\bibfnamefont {S.}~\bibnamefont {Su}}, \bibinfo {author} {\bibfnamefont {J.}~\bibnamefont {Chen}}, \bibinfo {author} {\bibfnamefont {J.}~\bibnamefont {Chen}},\ and\ \bibinfo {author} {\bibfnamefont {J.~F.}\ \bibnamefont {Santos}},\ }\bibfield  {title} {\bibinfo {title} {Suppressing coherence effects in quantum-measurement-based engines},\ }\href {https://doi.org/https://doi.org/10.1103/PhysRevA.104.062210} {\bibfield  {journal} {\bibinfo  {journal} {Phys. Rev. A}\ }\textbf {\bibinfo {volume} {104}},\ \bibinfo {pages} {062210} (\bibinfo {year} {2021})}\BibitemShut {NoStop}%
\bibitem [{\citenamefont {Naghiloo}\ \emph {et~al.}(2020)\citenamefont {Naghiloo}, \citenamefont {Tan}, \citenamefont {Harrington}, \citenamefont {Alonso}, \citenamefont {Lutz}, \citenamefont {Romito},\ and\ \citenamefont {Murch}}]{PhysRevLett.124.110604}%
  \BibitemOpen
  \bibfield  {author} {\bibinfo {author} {\bibfnamefont {M.}~\bibnamefont {Naghiloo}}, \bibinfo {author} {\bibfnamefont {D.}~\bibnamefont {Tan}}, \bibinfo {author} {\bibfnamefont {P.~M.}\ \bibnamefont {Harrington}}, \bibinfo {author} {\bibfnamefont {J.~J.}\ \bibnamefont {Alonso}}, \bibinfo {author} {\bibfnamefont {E.}~\bibnamefont {Lutz}}, \bibinfo {author} {\bibfnamefont {A.}~\bibnamefont {Romito}},\ and\ \bibinfo {author} {\bibfnamefont {K.~W.}\ \bibnamefont {Murch}},\ }\bibfield  {title} {\bibinfo {title} {Heat and work along individual trajectories of a quantum bit},\ }\href {https://doi.org/10.1103/PhysRevLett.124.110604} {\bibfield  {journal} {\bibinfo  {journal} {Phys. Rev. Lett.}\ }\textbf {\bibinfo {volume} {124}},\ \bibinfo {pages} {110604} (\bibinfo {year} {2020})}\BibitemShut {NoStop}%
\bibitem [{\citenamefont {Naghiloo}\ \emph {et~al.}(2018)\citenamefont {Naghiloo}, \citenamefont {Alonso}, \citenamefont {Romito}, \citenamefont {Lutz},\ and\ \citenamefont {Murch}}]{PhysRevLett.121.030604}%
  \BibitemOpen
  \bibfield  {author} {\bibinfo {author} {\bibfnamefont {M.}~\bibnamefont {Naghiloo}}, \bibinfo {author} {\bibfnamefont {J.~J.}\ \bibnamefont {Alonso}}, \bibinfo {author} {\bibfnamefont {A.}~\bibnamefont {Romito}}, \bibinfo {author} {\bibfnamefont {E.}~\bibnamefont {Lutz}},\ and\ \bibinfo {author} {\bibfnamefont {K.~W.}\ \bibnamefont {Murch}},\ }\bibfield  {title} {\bibinfo {title} {Information gain and loss for a quantum {Maxwell's} demon},\ }\href {https://doi.org/10.1103/PhysRevLett.121.030604} {\bibfield  {journal} {\bibinfo  {journal} {Phys. Rev. Lett.}\ }\textbf {\bibinfo {volume} {121}},\ \bibinfo {pages} {030604} (\bibinfo {year} {2018})}\BibitemShut {NoStop}%
\bibitem [{\citenamefont {Hern\'andez-G\'omez}\ \emph {et~al.}(2022)\citenamefont {Hern\'andez-G\'omez}, \citenamefont {Gherardini}, \citenamefont {Staudenmaier}, \citenamefont {Poggiali}, \citenamefont {Campisi}, \citenamefont {Trombettoni}, \citenamefont {Cataliotti}, \citenamefont {Cappellaro},\ and\ \citenamefont {Fabbri}}]{PRXQuantum.3.020329}%
  \BibitemOpen
  \bibfield  {author} {\bibinfo {author} {\bibfnamefont {S.}~\bibnamefont {Hern\'andez-G\'omez}}, \bibinfo {author} {\bibfnamefont {S.}~\bibnamefont {Gherardini}}, \bibinfo {author} {\bibfnamefont {N.}~\bibnamefont {Staudenmaier}}, \bibinfo {author} {\bibfnamefont {F.}~\bibnamefont {Poggiali}}, \bibinfo {author} {\bibfnamefont {M.}~\bibnamefont {Campisi}}, \bibinfo {author} {\bibfnamefont {A.}~\bibnamefont {Trombettoni}}, \bibinfo {author} {\bibfnamefont {F.}~\bibnamefont {Cataliotti}}, \bibinfo {author} {\bibfnamefont {P.}~\bibnamefont {Cappellaro}},\ and\ \bibinfo {author} {\bibfnamefont {N.}~\bibnamefont {Fabbri}},\ }\bibfield  {title} {\bibinfo {title} {Autonomous dissipative {Maxwell's} demon in a diamond spin qutrit},\ }\href {https://doi.org/10.1103/PRXQuantum.3.020329} {\bibfield  {journal} {\bibinfo  {journal} {PRX Quantum}\ }\textbf {\bibinfo {volume} {3}},\ \bibinfo {pages} {020329} (\bibinfo {year} {2022})}\BibitemShut {NoStop}%
\bibitem [{\citenamefont {Camati}\ \emph {et~al.}(2016)\citenamefont {Camati}, \citenamefont {Peterson}, \citenamefont {Batalh\~ao}, \citenamefont {Micadei}, \citenamefont {Souza}, \citenamefont {Sarthour}, \citenamefont {Oliveira},\ and\ \citenamefont {Serra}}]{PhysRevLett.117.240502}%
  \BibitemOpen
  \bibfield  {author} {\bibinfo {author} {\bibfnamefont {P.~A.}\ \bibnamefont {Camati}}, \bibinfo {author} {\bibfnamefont {J.~P.~S.}\ \bibnamefont {Peterson}}, \bibinfo {author} {\bibfnamefont {T.~B.}\ \bibnamefont {Batalh\~ao}}, \bibinfo {author} {\bibfnamefont {K.}~\bibnamefont {Micadei}}, \bibinfo {author} {\bibfnamefont {A.~M.}\ \bibnamefont {Souza}}, \bibinfo {author} {\bibfnamefont {R.~S.}\ \bibnamefont {Sarthour}}, \bibinfo {author} {\bibfnamefont {I.~S.}\ \bibnamefont {Oliveira}},\ and\ \bibinfo {author} {\bibfnamefont {R.~M.}\ \bibnamefont {Serra}},\ }\bibfield  {title} {\bibinfo {title} {Experimental rectification of entropy production by {Maxwell's} demon in a quantum system},\ }\href {https://doi.org/10.1103/PhysRevLett.117.240502} {\bibfield  {journal} {\bibinfo  {journal} {Phys. Rev. Lett.}\ }\textbf {\bibinfo {volume} {117}},\ \bibinfo {pages} {240502} (\bibinfo {year} {2016})}\BibitemShut {NoStop}%
\bibitem [{\citenamefont {Anka}\ \emph {et~al.}(2021)\citenamefont {Anka}, \citenamefont {de~Oliveira},\ and\ \citenamefont {Jonathan}}]{anka2021measurement}%
  \BibitemOpen
  \bibfield  {author} {\bibinfo {author} {\bibfnamefont {M.~F.}\ \bibnamefont {Anka}}, \bibinfo {author} {\bibfnamefont {T.~R.}\ \bibnamefont {de~Oliveira}},\ and\ \bibinfo {author} {\bibfnamefont {D.}~\bibnamefont {Jonathan}},\ }\bibfield  {title} {\bibinfo {title} {Measurement-based quantum heat engine in a multilevel system},\ }\href {https://doi.org/https://doi.org/10.1103/PhysRevE.104.054128} {\bibfield  {journal} {\bibinfo  {journal} {Phys. Rev. E}\ }\textbf {\bibinfo {volume} {104}},\ \bibinfo {pages} {054128} (\bibinfo {year} {2021})}\BibitemShut {NoStop}%
\bibitem [{\citenamefont {Alam}\ and\ \citenamefont {Venkatesh}(2022)}]{alam2022two}%
  \BibitemOpen
  \bibfield  {author} {\bibinfo {author} {\bibfnamefont {M.~S.}\ \bibnamefont {Alam}}\ and\ \bibinfo {author} {\bibfnamefont {B.~P.}\ \bibnamefont {Venkatesh}},\ }\bibfield  {title} {\bibinfo {title} {Two-stroke quantum measurement heat engine},\ }\href {https://doi.org/10.48550/arXiv.2201.06303} {\bibfield  {journal} {\bibinfo  {journal} {arXiv:2201.06303}\ } (\bibinfo {year} {2022})}\BibitemShut {NoStop}%
\bibitem [{\citenamefont {Manikandan}\ \emph {et~al.}(2022)\citenamefont {Manikandan}, \citenamefont {Elouard}, \citenamefont {Murch}, \citenamefont {Auff{\`e}ves},\ and\ \citenamefont {Jordan}}]{manikandan2022efficiently}%
  \BibitemOpen
  \bibfield  {author} {\bibinfo {author} {\bibfnamefont {S.~K.}\ \bibnamefont {Manikandan}}, \bibinfo {author} {\bibfnamefont {C.}~\bibnamefont {Elouard}}, \bibinfo {author} {\bibfnamefont {K.~W.}\ \bibnamefont {Murch}}, \bibinfo {author} {\bibfnamefont {A.}~\bibnamefont {Auff{\`e}ves}},\ and\ \bibinfo {author} {\bibfnamefont {A.~N.}\ \bibnamefont {Jordan}},\ }\bibfield  {title} {\bibinfo {title} {Efficiently fueling a quantum engine with incompatible measurements},\ }\href {https://doi.org/https://doi.org/10.1103/PhysRevE.105.044137} {\bibfield  {journal} {\bibinfo  {journal} {Phys. Rev. E}\ }\textbf {\bibinfo {volume} {105}},\ \bibinfo {pages} {044137} (\bibinfo {year} {2022})}\BibitemShut {NoStop}%
\bibitem [{\citenamefont {Myers}\ \emph {et~al.}(2022)\citenamefont {Myers}, \citenamefont {Abah},\ and\ \citenamefont {Deffner}}]{myers2022quantum}%
  \BibitemOpen
  \bibfield  {author} {\bibinfo {author} {\bibfnamefont {N.~M.}\ \bibnamefont {Myers}}, \bibinfo {author} {\bibfnamefont {O.}~\bibnamefont {Abah}},\ and\ \bibinfo {author} {\bibfnamefont {S.}~\bibnamefont {Deffner}},\ }\bibfield  {title} {\bibinfo {title} {Quantum thermodynamic devices: {From} theoretical proposals to experimental reality},\ }\href {https://doi.org/https://doi.org/10.1116/5.0083192} {\bibfield  {journal} {\bibinfo  {journal} {{AVS} Quantum Sci.}\ }\textbf {\bibinfo {volume} {4}},\ \bibinfo {pages} {027101} (\bibinfo {year} {2022})}\BibitemShut {NoStop}%
\bibitem [{\citenamefont {Lisboa}\ \emph {et~al.}(2022)\citenamefont {Lisboa}, \citenamefont {Dieguez}, \citenamefont {Guimar{\~a}es}, \citenamefont {Santos},\ and\ \citenamefont {Serra}}]{lisboa2022experimental}%
  \BibitemOpen
  \bibfield  {author} {\bibinfo {author} {\bibfnamefont {V.}~\bibnamefont {Lisboa}}, \bibinfo {author} {\bibfnamefont {P.}~\bibnamefont {Dieguez}}, \bibinfo {author} {\bibfnamefont {J.}~\bibnamefont {Guimar{\~a}es}}, \bibinfo {author} {\bibfnamefont {J.}~\bibnamefont {Santos}},\ and\ \bibinfo {author} {\bibfnamefont {R.}~\bibnamefont {Serra}},\ }\bibfield  {title} {\bibinfo {title} {Experimental investigation of a quantum heat engine powered by generalized measurements},\ }\href {https://doi.org/10.1103/PhysRevA.106.022436} {\bibfield  {journal} {\bibinfo  {journal} {Phys. Rev. A}\ }\textbf {\bibinfo {volume} {106}},\ \bibinfo {pages} {022436} (\bibinfo {year} {2022})}\BibitemShut {NoStop}%
\bibitem [{\citenamefont {Dieguez}\ \emph {et~al.}(2023)\citenamefont {Dieguez}, \citenamefont {Lisboa},\ and\ \citenamefont {Serra}}]{dieguez2023thermal}%
  \BibitemOpen
  \bibfield  {author} {\bibinfo {author} {\bibfnamefont {P.~R.}\ \bibnamefont {Dieguez}}, \bibinfo {author} {\bibfnamefont {V.~F.}\ \bibnamefont {Lisboa}},\ and\ \bibinfo {author} {\bibfnamefont {R.~M.}\ \bibnamefont {Serra}},\ }\bibfield  {title} {\bibinfo {title} {Thermal devices powered by generalized measurements with indefinite causal order},\ }\href {https://doi.org/https://doi.org/10.1103/PhysRevA.107.012423} {\bibfield  {journal} {\bibinfo  {journal} {Phys. Rev. A}\ }\textbf {\bibinfo {volume} {107}},\ \bibinfo {pages} {012423} (\bibinfo {year} {2023})}\BibitemShut {NoStop}%
\bibitem [{\citenamefont {Oreshkov}\ and\ \citenamefont {Brun}(2005)}]{oreshkov2005weak}%
  \BibitemOpen
  \bibfield  {author} {\bibinfo {author} {\bibfnamefont {O.}~\bibnamefont {Oreshkov}}\ and\ \bibinfo {author} {\bibfnamefont {T.~A.}\ \bibnamefont {Brun}},\ }\bibfield  {title} {\bibinfo {title} {Weak measurements are universal},\ }\href {https://doi.org/https://doi.org/10.1103/PhysRevLett.95.110409} {\bibfield  {journal} {\bibinfo  {journal} {Phys. Rev. Lett.}\ }\textbf {\bibinfo {volume} {95}},\ \bibinfo {pages} {110409} (\bibinfo {year} {2005})}\BibitemShut {NoStop}%
\bibitem [{\citenamefont {Dieguez}\ and\ \citenamefont {Angelo}(2018{\natexlab{a}})}]{dieguez2018information}%
  \BibitemOpen
  \bibfield  {author} {\bibinfo {author} {\bibfnamefont {P.~R.}\ \bibnamefont {Dieguez}}\ and\ \bibinfo {author} {\bibfnamefont {R.~M.}\ \bibnamefont {Angelo}},\ }\bibfield  {title} {\bibinfo {title} {Information-reality complementarity: {T}he role of measurements and quantum reference frames},\ }\href {https://doi.org/10.1103/PhysRevA.97.022107} {\bibfield  {journal} {\bibinfo  {journal} {Phys. Rev. A}\ }\textbf {\bibinfo {volume} {97}},\ \bibinfo {pages} {022107} (\bibinfo {year} {2018}{\natexlab{a}})}\BibitemShut {NoStop}%
\bibitem [{\citenamefont {Pan}\ \emph {et~al.}(2020)\citenamefont {Pan}, \citenamefont {Zhang}, \citenamefont {Cohen}, \citenamefont {Wu}, \citenamefont {Chen},\ and\ \citenamefont {Davidson}}]{pan2020weak}%
  \BibitemOpen
  \bibfield  {author} {\bibinfo {author} {\bibfnamefont {Y.}~\bibnamefont {Pan}}, \bibinfo {author} {\bibfnamefont {J.}~\bibnamefont {Zhang}}, \bibinfo {author} {\bibfnamefont {E.}~\bibnamefont {Cohen}}, \bibinfo {author} {\bibfnamefont {C.-w.}\ \bibnamefont {Wu}}, \bibinfo {author} {\bibfnamefont {P.-X.}\ \bibnamefont {Chen}},\ and\ \bibinfo {author} {\bibfnamefont {N.}~\bibnamefont {Davidson}},\ }\bibfield  {title} {\bibinfo {title} {Weak-to-strong transition of quantum measurement in a trapped-ion system},\ }\href {https://doi.org/https://doi.org/10.1038/s41567-020-0973-y} {\bibfield  {journal} {\bibinfo  {journal} {Nature Physics}\ }\textbf {\bibinfo {volume} {16}},\ \bibinfo {pages} {1206} (\bibinfo {year} {2020})}\BibitemShut {NoStop}%
\bibitem [{\citenamefont {Mancino}\ \emph {et~al.}(2018)\citenamefont {Mancino}, \citenamefont {Sbroscia}, \citenamefont {Roccia}, \citenamefont {Gianani}, \citenamefont {Somma}, \citenamefont {Mataloni}, \citenamefont {Paternostro},\ and\ \citenamefont {Barbieri}}]{mancino2018entropic}%
  \BibitemOpen
  \bibfield  {author} {\bibinfo {author} {\bibfnamefont {L.}~\bibnamefont {Mancino}}, \bibinfo {author} {\bibfnamefont {M.}~\bibnamefont {Sbroscia}}, \bibinfo {author} {\bibfnamefont {E.}~\bibnamefont {Roccia}}, \bibinfo {author} {\bibfnamefont {I.}~\bibnamefont {Gianani}}, \bibinfo {author} {\bibfnamefont {F.}~\bibnamefont {Somma}}, \bibinfo {author} {\bibfnamefont {P.}~\bibnamefont {Mataloni}}, \bibinfo {author} {\bibfnamefont {M.}~\bibnamefont {Paternostro}},\ and\ \bibinfo {author} {\bibfnamefont {M.}~\bibnamefont {Barbieri}},\ }\bibfield  {title} {\bibinfo {title} {The entropic cost of quantum generalized measurements},\ }\href {https://doi.org/10.1038/s41534-018-0069-z} {\bibfield  {journal} {\bibinfo  {journal} {npj Quan. Inf.}\ }\textbf {\bibinfo {volume} {4}},\ \bibinfo {pages} {1} (\bibinfo {year} {2018})}\BibitemShut {NoStop}%
\bibitem [{\citenamefont {Alonso}\ \emph {et~al.}(2016)\citenamefont {Alonso}, \citenamefont {Lutz},\ and\ \citenamefont {Romito}}]{alonso2016thermodynamics}%
  \BibitemOpen
  \bibfield  {author} {\bibinfo {author} {\bibfnamefont {J.~J.}\ \bibnamefont {Alonso}}, \bibinfo {author} {\bibfnamefont {E.}~\bibnamefont {Lutz}},\ and\ \bibinfo {author} {\bibfnamefont {A.}~\bibnamefont {Romito}},\ }\bibfield  {title} {\bibinfo {title} {Thermodynamics of weakly measured quantum systems},\ }\href {https://doi.org/https://doi.org/10.1103/PhysRevLett.116.080403} {\bibfield  {journal} {\bibinfo  {journal} {Physical review letters}\ }\textbf {\bibinfo {volume} {116}},\ \bibinfo {pages} {080403} (\bibinfo {year} {2016})}\BibitemShut {NoStop}%
\bibitem [{\citenamefont {Vieira}\ \emph {et~al.}(2023)\citenamefont {Vieira}, \citenamefont {{de Oliveira}}, \citenamefont {Santos}, \citenamefont {Dieguez},\ and\ \citenamefont {Serra}}]{VIEIRA2023100105}%
  \BibitemOpen
  \bibfield  {author} {\bibinfo {author} {\bibfnamefont {C.}~\bibnamefont {Vieira}}, \bibinfo {author} {\bibfnamefont {J.}~\bibnamefont {{de Oliveira}}}, \bibinfo {author} {\bibfnamefont {J.}~\bibnamefont {Santos}}, \bibinfo {author} {\bibfnamefont {P.}~\bibnamefont {Dieguez}},\ and\ \bibinfo {author} {\bibfnamefont {R.}~\bibnamefont {Serra}},\ }\bibfield  {title} {\bibinfo {title} {Exploring quantum thermodynamics with nmr},\ }\href {https://doi.org/https://doi.org/10.1016/j.jmro.2023.100105} {\bibfield  {journal} {\bibinfo  {journal} {Journal of Magnetic Resonance Open}\ }\textbf {\bibinfo {volume} {16-17}},\ \bibinfo {pages} {100105} (\bibinfo {year} {2023})}\BibitemShut {NoStop}%
\bibitem [{\citenamefont {Zhang}\ \emph {et~al.}(2024)\citenamefont {Zhang}, \citenamefont {Yang},\ and\ \citenamefont {Fei}}]{PhysRevA.109.042424}%
  \BibitemOpen
  \bibfield  {author} {\bibinfo {author} {\bibfnamefont {T.}~\bibnamefont {Zhang}}, \bibinfo {author} {\bibfnamefont {H.}~\bibnamefont {Yang}},\ and\ \bibinfo {author} {\bibfnamefont {S.-M.}\ \bibnamefont {Fei}},\ }\bibfield  {title} {\bibinfo {title} {Local-projective-measurement-enhanced quantum battery capacity},\ }\href {https://doi.org/10.1103/PhysRevA.109.042424} {\bibfield  {journal} {\bibinfo  {journal} {Phys. Rev. A}\ }\textbf {\bibinfo {volume} {109}},\ \bibinfo {pages} {042424} (\bibinfo {year} {2024})}\BibitemShut {NoStop}%
\bibitem [{\citenamefont {Balkanlu}\ \emph {et~al.}(2024)\citenamefont {Balkanlu}, \citenamefont {Faizi},\ and\ \citenamefont {Ahansaz}}]{balkanlu2024selective}%
  \BibitemOpen
  \bibfield  {author} {\bibinfo {author} {\bibfnamefont {M.~A.}\ \bibnamefont {Balkanlu}}, \bibinfo {author} {\bibfnamefont {E.}~\bibnamefont {Faizi}},\ and\ \bibinfo {author} {\bibfnamefont {B.}~\bibnamefont {Ahansaz}},\ }\bibfield  {title} {\bibinfo {title} {Selective weak measurement reveals super-ergotropy},\ }\href {https://doi.org/10.1016/j.rinp.2023.107308 Get rights and content Under a Creative Commons license} {\bibfield  {journal} {\bibinfo  {journal} {Results in Physics}\ }\textbf {\bibinfo {volume} {57}},\ \bibinfo {pages} {107308} (\bibinfo {year} {2024})}\BibitemShut {NoStop}%
\bibitem [{\citenamefont {Francica}\ \emph {et~al.}(2017)\citenamefont {Francica}, \citenamefont {Goold}, \citenamefont {Plastina},\ and\ \citenamefont {Paternostro}}]{Francica2017}%
  \BibitemOpen
  \bibfield  {author} {\bibinfo {author} {\bibfnamefont {G.}~\bibnamefont {Francica}}, \bibinfo {author} {\bibfnamefont {J.}~\bibnamefont {Goold}}, \bibinfo {author} {\bibfnamefont {F.}~\bibnamefont {Plastina}},\ and\ \bibinfo {author} {\bibfnamefont {M.}~\bibnamefont {Paternostro}},\ }\bibfield  {title} {\bibinfo {title} {Daemonic ergotropy: enhanced work extraction from quantum correlations},\ }\href {https://doi.org/10.1038/s41534-017-0012-8} {\bibfield  {journal} {\bibinfo  {journal} {npj Quan. Inf.}\ }\textbf {\bibinfo {volume} {3}},\ \bibinfo {pages} {12} (\bibinfo {year} {2017})}\BibitemShut {NoStop}%
\bibitem [{\citenamefont {Lu}\ \emph {et~al.}(2014)\citenamefont {Lu}, \citenamefont {Brodutch}, \citenamefont {Li}, \citenamefont {Li},\ and\ \citenamefont {Laflamme}}]{Lu_2014}%
  \BibitemOpen
  \bibfield  {author} {\bibinfo {author} {\bibfnamefont {D.}~\bibnamefont {Lu}}, \bibinfo {author} {\bibfnamefont {A.}~\bibnamefont {Brodutch}}, \bibinfo {author} {\bibfnamefont {J.}~\bibnamefont {Li}}, \bibinfo {author} {\bibfnamefont {H.}~\bibnamefont {Li}},\ and\ \bibinfo {author} {\bibfnamefont {R.}~\bibnamefont {Laflamme}},\ }\bibfield  {title} {\bibinfo {title} {Experimental realization of post-selected weak measurements on an nmr quantum processor},\ }\href {https://doi.org/10.1088/1367-2630/16/5/053015} {\bibfield  {journal} {\bibinfo  {journal} {New Journal of Physics}\ }\textbf {\bibinfo {volume} {16}},\ \bibinfo {pages} {053015} (\bibinfo {year} {2014})}\BibitemShut {NoStop}%
\bibitem [{\citenamefont {Singh}\ \emph {et~al.}(2024)\citenamefont {Singh}, \citenamefont {Gaikwad}, \citenamefont {Arvind},\ and\ \citenamefont {Dorai}}]{protect5}%
  \BibitemOpen
  \bibfield  {author} {\bibinfo {author} {\bibfnamefont {G.}~\bibnamefont {Singh}}, \bibinfo {author} {\bibfnamefont {A.}~\bibnamefont {Gaikwad}}, \bibinfo {author} {\bibnamefont {Arvind}},\ and\ \bibinfo {author} {\bibfnamefont {K.}~\bibnamefont {Dorai}},\ }\bibfield  {title} {\bibinfo {title} {Experimental decoherence mitigation using a weak measurement-based scheme and the duality quantum algorithm},\ }\href {https://arxiv.org/abs/2409.12752} {\bibfield  {journal} {\bibinfo  {journal} {arXiv:2409.12752}\ } (\bibinfo {year} {2024})}\BibitemShut {NoStop}%
\bibitem [{\citenamefont {Cujia}\ \emph {et~al.}(2019)\citenamefont {Cujia}, \citenamefont {Boss}, \citenamefont {Herb}, \citenamefont {Zopes},\ and\ \citenamefont {Degen}}]{Cujia2019}%
  \BibitemOpen
  \bibfield  {author} {\bibinfo {author} {\bibfnamefont {K.~S.}\ \bibnamefont {Cujia}}, \bibinfo {author} {\bibfnamefont {J.~M.}\ \bibnamefont {Boss}}, \bibinfo {author} {\bibfnamefont {K.}~\bibnamefont {Herb}}, \bibinfo {author} {\bibfnamefont {J.}~\bibnamefont {Zopes}},\ and\ \bibinfo {author} {\bibfnamefont {C.~L.}\ \bibnamefont {Degen}},\ }\bibfield  {title} {\bibinfo {title} {Tracking the precession of single nuclear spins by weak measurements},\ }\href {https://doi.org/10.1038/s41586-019-1334-9} {\bibfield  {journal} {\bibinfo  {journal} {Nature}\ }\textbf {\bibinfo {volume} {571}},\ \bibinfo {pages} {230} (\bibinfo {year} {2019})}\BibitemShut {NoStop}%
\bibitem [{\citenamefont {Kim}\ \emph {et~al.}(2009)\citenamefont {Kim}, \citenamefont {Cho}, \citenamefont {Ra},\ and\ \citenamefont {Kim}}]{Kim09}%
  \BibitemOpen
  \bibfield  {author} {\bibinfo {author} {\bibfnamefont {Y.-S.}\ \bibnamefont {Kim}}, \bibinfo {author} {\bibfnamefont {Y.-W.}\ \bibnamefont {Cho}}, \bibinfo {author} {\bibfnamefont {Y.-S.}\ \bibnamefont {Ra}},\ and\ \bibinfo {author} {\bibfnamefont {Y.-H.}\ \bibnamefont {Kim}},\ }\bibfield  {title} {\bibinfo {title} {Reversing the weak quantum measurement for a photonic qubit},\ }\href {https://doi.org/10.1364/OE.17.011978} {\bibfield  {journal} {\bibinfo  {journal} {Opt. Express}\ }\textbf {\bibinfo {volume} {17}},\ \bibinfo {pages} {11978} (\bibinfo {year} {2009})}\BibitemShut {NoStop}%
\bibitem [{\citenamefont {Kim}\ \emph {et~al.}(2012)\citenamefont {Kim}, \citenamefont {Lee}, \citenamefont {Kwon},\ and\ \citenamefont {Kim}}]{kim2012protecting}%
  \BibitemOpen
  \bibfield  {author} {\bibinfo {author} {\bibfnamefont {Y.-S.}\ \bibnamefont {Kim}}, \bibinfo {author} {\bibfnamefont {J.-C.}\ \bibnamefont {Lee}}, \bibinfo {author} {\bibfnamefont {O.}~\bibnamefont {Kwon}},\ and\ \bibinfo {author} {\bibfnamefont {Y.-H.}\ \bibnamefont {Kim}},\ }\bibfield  {title} {\bibinfo {title} {Protecting entanglement from decoherence using weak measurement and quantum measurement reversal},\ }\href {https://www.nature.com/articles/nphys2178} {\bibfield  {journal} {\bibinfo  {journal} {Nature Physics}\ }\textbf {\bibinfo {volume} {8}},\ \bibinfo {pages} {117} (\bibinfo {year} {2012})}\BibitemShut {NoStop}%
\bibitem [{\citenamefont {Zou}\ \emph {et~al.}(2017)\citenamefont {Zou}, \citenamefont {Li}, \citenamefont {Wang}, \citenamefont {Cao}, \citenamefont {Ren}, \citenamefont {Yin}, \citenamefont {Peng}, \citenamefont {Wang},\ and\ \citenamefont {Pan}}]{protect1}%
  \BibitemOpen
  \bibfield  {author} {\bibinfo {author} {\bibfnamefont {W.-J.}\ \bibnamefont {Zou}}, \bibinfo {author} {\bibfnamefont {Y.-H.}\ \bibnamefont {Li}}, \bibinfo {author} {\bibfnamefont {S.-C.}\ \bibnamefont {Wang}}, \bibinfo {author} {\bibfnamefont {Y.}~\bibnamefont {Cao}}, \bibinfo {author} {\bibfnamefont {J.-G.}\ \bibnamefont {Ren}}, \bibinfo {author} {\bibfnamefont {J.}~\bibnamefont {Yin}}, \bibinfo {author} {\bibfnamefont {C.-Z.}\ \bibnamefont {Peng}}, \bibinfo {author} {\bibfnamefont {X.-B.}\ \bibnamefont {Wang}},\ and\ \bibinfo {author} {\bibfnamefont {J.-W.}\ \bibnamefont {Pan}},\ }\bibfield  {title} {\bibinfo {title} {Protecting entanglement from finite-temperature thermal noise via weak measurement and quantum measurement reversal},\ }\href {https://doi.org/10.1103/PhysRevA.95.042342} {\bibfield  {journal} {\bibinfo  {journal} {Phys. Rev. A}\ }\textbf {\bibinfo {volume} {95}},\ \bibinfo {pages} {042342} (\bibinfo {year} {2017})}\BibitemShut {NoStop}%
\bibitem [{\citenamefont {Katz}\ \emph {et~al.}(2006)\citenamefont {Katz}, \citenamefont {Ansmann}, \citenamefont {Bialczak}, \citenamefont {Lucero}, \citenamefont {McDermott}, \citenamefont {Neeley}, \citenamefont {Steffen}, \citenamefont {Weig}, \citenamefont {Cleland}, \citenamefont {Martinis},\ and\ \citenamefont {Korotkov}}]{doi:10.1126/science.1126475}%
  \BibitemOpen
  \bibfield  {author} {\bibinfo {author} {\bibfnamefont {N.}~\bibnamefont {Katz}}, \bibinfo {author} {\bibfnamefont {M.}~\bibnamefont {Ansmann}}, \bibinfo {author} {\bibfnamefont {R.~C.}\ \bibnamefont {Bialczak}}, \bibinfo {author} {\bibfnamefont {E.}~\bibnamefont {Lucero}}, \bibinfo {author} {\bibfnamefont {R.}~\bibnamefont {McDermott}}, \bibinfo {author} {\bibfnamefont {M.}~\bibnamefont {Neeley}}, \bibinfo {author} {\bibfnamefont {M.}~\bibnamefont {Steffen}}, \bibinfo {author} {\bibfnamefont {E.~M.}\ \bibnamefont {Weig}}, \bibinfo {author} {\bibfnamefont {A.~N.}\ \bibnamefont {Cleland}}, \bibinfo {author} {\bibfnamefont {J.~M.}\ \bibnamefont {Martinis}},\ and\ \bibinfo {author} {\bibfnamefont {A.~N.}\ \bibnamefont {Korotkov}},\ }\bibfield  {title} {\bibinfo {title} {Coherent state evolution in a superconducting qubit from partial-collapse measurement},\ }\href {https://www.science.org/doi/abs/10.1126/science.1126475} {\bibfield  {journal} {\bibinfo  {journal} {Science}\ }\textbf {\bibinfo {volume} {312}},\
  \bibinfo {pages} {1498} (\bibinfo {year} {2006})}\BibitemShut {NoStop}%
\bibitem [{\citenamefont {Katz}\ \emph {et~al.}(2008)\citenamefont {Katz}, \citenamefont {Neeley}, \citenamefont {Ansmann}, \citenamefont {Bialczak}, \citenamefont {Hofheinz}, \citenamefont {Lucero}, \citenamefont {O'Connell}, \citenamefont {Wang}, \citenamefont {Cleland}, \citenamefont {Martinis},\ and\ \citenamefont {Korotkov}}]{PhysRevLett.101.200401}%
  \BibitemOpen
  \bibfield  {author} {\bibinfo {author} {\bibfnamefont {N.}~\bibnamefont {Katz}}, \bibinfo {author} {\bibfnamefont {M.}~\bibnamefont {Neeley}}, \bibinfo {author} {\bibfnamefont {M.}~\bibnamefont {Ansmann}}, \bibinfo {author} {\bibfnamefont {R.~C.}\ \bibnamefont {Bialczak}}, \bibinfo {author} {\bibfnamefont {M.}~\bibnamefont {Hofheinz}}, \bibinfo {author} {\bibfnamefont {E.}~\bibnamefont {Lucero}}, \bibinfo {author} {\bibfnamefont {A.}~\bibnamefont {O'Connell}}, \bibinfo {author} {\bibfnamefont {H.}~\bibnamefont {Wang}}, \bibinfo {author} {\bibfnamefont {A.~N.}\ \bibnamefont {Cleland}}, \bibinfo {author} {\bibfnamefont {J.~M.}\ \bibnamefont {Martinis}},\ and\ \bibinfo {author} {\bibfnamefont {A.~N.}\ \bibnamefont {Korotkov}},\ }\bibfield  {title} {\bibinfo {title} {Reversal of the weak measurement of a quantum state in a superconducting phase qubit},\ }\href {https://doi.org/10.1103/PhysRevLett.101.200401} {\bibfield  {journal} {\bibinfo  {journal} {Phys. Rev. Lett.}\ }\textbf {\bibinfo {volume} {101}},\
  \bibinfo {pages} {200401} (\bibinfo {year} {2008})}\BibitemShut {NoStop}%
\bibitem [{\citenamefont {Zhong}\ \emph {et~al.}(2014)\citenamefont {Zhong}, \citenamefont {Wang}, \citenamefont {Martinis}, \citenamefont {Cleland}, \citenamefont {Korotkov},\ and\ \citenamefont {Wang}}]{Zhong2014}%
  \BibitemOpen
  \bibfield  {author} {\bibinfo {author} {\bibfnamefont {Y.~P.}\ \bibnamefont {Zhong}}, \bibinfo {author} {\bibfnamefont {Z.~L.}\ \bibnamefont {Wang}}, \bibinfo {author} {\bibfnamefont {J.~M.}\ \bibnamefont {Martinis}}, \bibinfo {author} {\bibfnamefont {A.~N.}\ \bibnamefont {Cleland}}, \bibinfo {author} {\bibfnamefont {A.~N.}\ \bibnamefont {Korotkov}},\ and\ \bibinfo {author} {\bibfnamefont {H.}~\bibnamefont {Wang}},\ }\bibfield  {title} {\bibinfo {title} {Reducing the impact of intrinsic dissipation in a superconducting circuit by quantum error detection},\ }\href {https://doi.org/10.1038/ncomms4135} {\bibfield  {journal} {\bibinfo  {journal} {Nature Communications}\ }\textbf {\bibinfo {volume} {5}},\ \bibinfo {pages} {3135} (\bibinfo {year} {2014})}\BibitemShut {NoStop}%
\bibitem [{\citenamefont {Weber}\ \emph {et~al.}(2014)\citenamefont {Weber}, \citenamefont {Chantasri}, \citenamefont {Dressel}, \citenamefont {Jordan}, \citenamefont {Murch},\ and\ \citenamefont {Siddiqi}}]{Weber2014}%
  \BibitemOpen
  \bibfield  {author} {\bibinfo {author} {\bibfnamefont {S.~J.}\ \bibnamefont {Weber}}, \bibinfo {author} {\bibfnamefont {A.}~\bibnamefont {Chantasri}}, \bibinfo {author} {\bibfnamefont {J.}~\bibnamefont {Dressel}}, \bibinfo {author} {\bibfnamefont {A.~N.}\ \bibnamefont {Jordan}}, \bibinfo {author} {\bibfnamefont {K.~W.}\ \bibnamefont {Murch}},\ and\ \bibinfo {author} {\bibfnamefont {I.}~\bibnamefont {Siddiqi}},\ }\bibfield  {title} {\bibinfo {title} {Mapping the optimal route between two quantum states},\ }\href {https://doi.org/10.1038/nature13559} {\bibfield  {journal} {\bibinfo  {journal} {Nature}\ }\textbf {\bibinfo {volume} {511}},\ \bibinfo {pages} {570} (\bibinfo {year} {2014})}\BibitemShut {NoStop}%
\bibitem [{\citenamefont {Monroe}\ \emph {et~al.}(2021)\citenamefont {Monroe}, \citenamefont {Yunger~Halpern}, \citenamefont {Lee},\ and\ \citenamefont {Murch}}]{PhysRevLett.126.100403}%
  \BibitemOpen
  \bibfield  {author} {\bibinfo {author} {\bibfnamefont {J.~T.}\ \bibnamefont {Monroe}}, \bibinfo {author} {\bibfnamefont {N.}~\bibnamefont {Yunger~Halpern}}, \bibinfo {author} {\bibfnamefont {T.}~\bibnamefont {Lee}},\ and\ \bibinfo {author} {\bibfnamefont {K.~W.}\ \bibnamefont {Murch}},\ }\bibfield  {title} {\bibinfo {title} {Weak measurement of a superconducting qubit reconciles incompatible operators},\ }\href {https://doi.org/10.1103/PhysRevLett.126.100403} {\bibfield  {journal} {\bibinfo  {journal} {Phys. Rev. Lett.}\ }\textbf {\bibinfo {volume} {126}},\ \bibinfo {pages} {100403} (\bibinfo {year} {2021})}\BibitemShut {NoStop}%
\bibitem [{\citenamefont {Latune}\ and\ \citenamefont {Elouard}(2025)}]{latune2025thermodynamically}%
  \BibitemOpen
  \bibfield  {author} {\bibinfo {author} {\bibfnamefont {C.~L.}\ \bibnamefont {Latune}}\ and\ \bibinfo {author} {\bibfnamefont {C.}~\bibnamefont {Elouard}},\ }\bibfield  {title} {\bibinfo {title} {A thermodynamically consistent approach to the energy costs of quantum measurements},\ }\href {https://doi.org/10.22331/q-2025-01-28-1614} {\bibfield  {journal} {\bibinfo  {journal} {Quantum}\ }\textbf {\bibinfo {volume} {9}},\ \bibinfo {pages} {1614} (\bibinfo {year} {2025})}\BibitemShut {NoStop}%
\bibitem [{\citenamefont {Ferraz}\ and\ \citenamefont {Elouard}(2025)}]{ferraz2025weak}%
  \BibitemOpen
  \bibfield  {author} {\bibinfo {author} {\bibfnamefont {L.~B.}\ \bibnamefont {Ferraz}}\ and\ \bibinfo {author} {\bibfnamefont {C.}~\bibnamefont {Elouard}},\ }\bibfield  {title} {\bibinfo {title} {Weak continuous measurements require more work than strong ones},\ }\href {https://doi.org/10.48550/arXiv.2502.09732} {\bibfield  {journal} {\bibinfo  {journal} {arXiv:2502.09732}\ } (\bibinfo {year} {2025})}\BibitemShut {NoStop}%
\bibitem [{\citenamefont {Xiao}\ and\ \citenamefont {Li}(2013)}]{Xiao2013}%
  \BibitemOpen
  \bibfield  {author} {\bibinfo {author} {\bibfnamefont {X.}~\bibnamefont {Xiao}}\ and\ \bibinfo {author} {\bibfnamefont {Y.-L.}\ \bibnamefont {Li}},\ }\bibfield  {title} {\bibinfo {title} {Protecting qutrit-qutrit entanglement by weak measurement and reversal},\ }\href {https://doi.org/10.1140/epjd/e2013-40036-3} {\bibfield  {journal} {\bibinfo  {journal} {The European Physical Journal D}\ }\textbf {\bibinfo {volume} {67}},\ \bibinfo {pages} {204} (\bibinfo {year} {2013})}\BibitemShut {NoStop}%
\bibitem [{\citenamefont {Korotkov}\ and\ \citenamefont {Keane}(2010)}]{protect2}%
  \BibitemOpen
  \bibfield  {author} {\bibinfo {author} {\bibfnamefont {A.~N.}\ \bibnamefont {Korotkov}}\ and\ \bibinfo {author} {\bibfnamefont {K.}~\bibnamefont {Keane}},\ }\bibfield  {title} {\bibinfo {title} {Decoherence suppression by quantum measurement reversal},\ }\href {https://doi.org/10.1103/PhysRevA.81.040103} {\bibfield  {journal} {\bibinfo  {journal} {Phys. Rev. A}\ }\textbf {\bibinfo {volume} {81}},\ \bibinfo {pages} {040103} (\bibinfo {year} {2010})}\BibitemShut {NoStop}%
\bibitem [{\citenamefont {Wang}\ \emph {et~al.}(2014{\natexlab{a}})\citenamefont {Wang}, \citenamefont {Yu}, \citenamefont {Zou},\ and\ \citenamefont {Wang}}]{protect3}%
  \BibitemOpen
  \bibfield  {author} {\bibinfo {author} {\bibfnamefont {S.-C.}\ \bibnamefont {Wang}}, \bibinfo {author} {\bibfnamefont {Z.-W.}\ \bibnamefont {Yu}}, \bibinfo {author} {\bibfnamefont {W.-J.}\ \bibnamefont {Zou}},\ and\ \bibinfo {author} {\bibfnamefont {X.-B.}\ \bibnamefont {Wang}},\ }\bibfield  {title} {\bibinfo {title} {Protecting quantum states from decoherence of finite temperature using weak measurement},\ }\href {https://doi.org/10.1103/PhysRevA.89.022318} {\bibfield  {journal} {\bibinfo  {journal} {Phys. Rev. A}\ }\textbf {\bibinfo {volume} {89}},\ \bibinfo {pages} {022318} (\bibinfo {year} {2014}{\natexlab{a}})}\BibitemShut {NoStop}%
\bibitem [{\citenamefont {Wang}\ \emph {et~al.}(2014{\natexlab{b}})\citenamefont {Wang}, \citenamefont {Zhao},\ and\ \citenamefont {Yu}}]{protect4}%
  \BibitemOpen
  \bibfield  {author} {\bibinfo {author} {\bibfnamefont {K.}~\bibnamefont {Wang}}, \bibinfo {author} {\bibfnamefont {X.}~\bibnamefont {Zhao}},\ and\ \bibinfo {author} {\bibfnamefont {T.}~\bibnamefont {Yu}},\ }\bibfield  {title} {\bibinfo {title} {Environment-assisted quantum state restoration via weak measurements},\ }\href {https://doi.org/10.1103/PhysRevA.89.042320} {\bibfield  {journal} {\bibinfo  {journal} {Phys. Rev. A}\ }\textbf {\bibinfo {volume} {89}},\ \bibinfo {pages} {042320} (\bibinfo {year} {2014}{\natexlab{b}})}\BibitemShut {NoStop}%
\bibitem [{\citenamefont {Lee}\ \emph {et~al.}(2011)\citenamefont {Lee}, \citenamefont {Jeong}, \citenamefont {Kim},\ and\ \citenamefont {Kim}}]{Lee:11}%
  \BibitemOpen
  \bibfield  {author} {\bibinfo {author} {\bibfnamefont {J.-C.}\ \bibnamefont {Lee}}, \bibinfo {author} {\bibfnamefont {Y.-C.}\ \bibnamefont {Jeong}}, \bibinfo {author} {\bibfnamefont {Y.-S.}\ \bibnamefont {Kim}},\ and\ \bibinfo {author} {\bibfnamefont {Y.-H.}\ \bibnamefont {Kim}},\ }\bibfield  {title} {\bibinfo {title} {Experimental demonstration of decoherence suppression via quantum measurement reversal},\ }\href {https://doi.org/10.1364/OE.19.016309} {\bibfield  {journal} {\bibinfo  {journal} {Opt. Express}\ }\textbf {\bibinfo {volume} {19}},\ \bibinfo {pages} {16309} (\bibinfo {year} {2011})}\BibitemShut {NoStop}%
\bibitem [{\citenamefont {Lalita}\ and\ \citenamefont {Banerjee}(2024)}]{Lalita_2024}%
  \BibitemOpen
  \bibfield  {author} {\bibinfo {author} {\bibfnamefont {J.}~\bibnamefont {Lalita}}\ and\ \bibinfo {author} {\bibfnamefont {S.}~\bibnamefont {Banerjee}},\ }\bibfield  {title} {\bibinfo {title} {Protecting quantum correlations of negative quantum states using weak measurement under non-markovian noise},\ }\href {https://doi.org/10.1088/1402-4896/ad273e} {\bibfield  {journal} {\bibinfo  {journal} {Physica Scripta}\ }\textbf {\bibinfo {volume} {99}},\ \bibinfo {pages} {035116} (\bibinfo {year} {2024})}\BibitemShut {NoStop}%
\bibitem [{\citenamefont {Wang}\ \emph {et~al.}(2025)\citenamefont {Wang}, \citenamefont {Liu}, \citenamefont {Cao}, \citenamefont {Yang}, \citenamefont {Liu}, \citenamefont {Sun},\ and\ \citenamefont {Zhao}}]{e27040350}%
  \BibitemOpen
  \bibfield  {author} {\bibinfo {author} {\bibfnamefont {M.}~\bibnamefont {Wang}}, \bibinfo {author} {\bibfnamefont {H.}~\bibnamefont {Liu}}, \bibinfo {author} {\bibfnamefont {L.}~\bibnamefont {Cao}}, \bibinfo {author} {\bibfnamefont {Y.}~\bibnamefont {Yang}}, \bibinfo {author} {\bibfnamefont {X.}~\bibnamefont {Liu}}, \bibinfo {author} {\bibfnamefont {B.}~\bibnamefont {Sun}},\ and\ \bibinfo {author} {\bibfnamefont {J.}~\bibnamefont {Zhao}},\ }\bibfield  {title} {\bibinfo {title} {Protecting the entanglement of x-type systems via weak measurement and reversal in the generalized amplitude damping channel},\ }\bibfield  {journal} {\bibinfo  {journal} {Entropy}\ }\textbf {\bibinfo {volume} {27}},\ \href {https://doi.org/10.3390/e27040350} {10.3390/e27040350} (\bibinfo {year} {2025})\BibitemShut {NoStop}%
\bibitem [{\citenamefont {Mpemba}\ and\ \citenamefont {Osborne}(1969)}]{mpemba1969cool}%
  \BibitemOpen
  \bibfield  {author} {\bibinfo {author} {\bibfnamefont {E.~B.}\ \bibnamefont {Mpemba}}\ and\ \bibinfo {author} {\bibfnamefont {D.~G.}\ \bibnamefont {Osborne}},\ }\bibfield  {title} {\bibinfo {title} {Cool?},\ }\href@noop {} {\bibfield  {journal} {\bibinfo  {journal} {Physics Education}\ }\textbf {\bibinfo {volume} {4}},\ \bibinfo {pages} {172} (\bibinfo {year} {1969})}\BibitemShut {NoStop}%
\bibitem [{\citenamefont {Medina}\ \emph {et~al.}(2024)\citenamefont {Medina}, \citenamefont {Culhane}, \citenamefont {Binder}, \citenamefont {Landi},\ and\ \citenamefont {Goold}}]{medina2024anomalous}%
  \BibitemOpen
  \bibfield  {author} {\bibinfo {author} {\bibfnamefont {I.}~\bibnamefont {Medina}}, \bibinfo {author} {\bibfnamefont {O.}~\bibnamefont {Culhane}}, \bibinfo {author} {\bibfnamefont {F.~C.}\ \bibnamefont {Binder}}, \bibinfo {author} {\bibfnamefont {G.~T.}\ \bibnamefont {Landi}},\ and\ \bibinfo {author} {\bibfnamefont {J.}~\bibnamefont {Goold}},\ }\bibfield  {title} {\bibinfo {title} {Anomalous discharging of quantum batteries: the ergotropic mpemba effect},\ }\href {https://doi.org/10.48550/arXiv.2412.13259} {\bibfield  {journal} {\bibinfo  {journal} {arXiv:2412.13259}\ } (\bibinfo {year} {2024})}\BibitemShut {NoStop}%
\bibitem [{\citenamefont {Streltsov}\ \emph {et~al.}(2017)\citenamefont {Streltsov}, \citenamefont {Adesso},\ and\ \citenamefont {Plenio}}]{RevModPhys.89.041003}%
  \BibitemOpen
  \bibfield  {author} {\bibinfo {author} {\bibfnamefont {A.}~\bibnamefont {Streltsov}}, \bibinfo {author} {\bibfnamefont {G.}~\bibnamefont {Adesso}},\ and\ \bibinfo {author} {\bibfnamefont {M.~B.}\ \bibnamefont {Plenio}},\ }\bibfield  {title} {\bibinfo {title} {Colloquium: Quantum coherence as a resource},\ }\href {https://doi.org/10.1103/RevModPhys.89.041003} {\bibfield  {journal} {\bibinfo  {journal} {Rev. Mod. Phys.}\ }\textbf {\bibinfo {volume} {89}},\ \bibinfo {pages} {041003} (\bibinfo {year} {2017})}\BibitemShut {NoStop}%
\bibitem [{\citenamefont {Francica}\ \emph {et~al.}(2020)\citenamefont {Francica}, \citenamefont {Binder}, \citenamefont {Guarnieri}, \citenamefont {Mitchison}, \citenamefont {Goold},\ and\ \citenamefont {Plastina}}]{PhysRevLett.125.180603}%
  \BibitemOpen
  \bibfield  {author} {\bibinfo {author} {\bibfnamefont {G.}~\bibnamefont {Francica}}, \bibinfo {author} {\bibfnamefont {F.~C.}\ \bibnamefont {Binder}}, \bibinfo {author} {\bibfnamefont {G.}~\bibnamefont {Guarnieri}}, \bibinfo {author} {\bibfnamefont {M.~T.}\ \bibnamefont {Mitchison}}, \bibinfo {author} {\bibfnamefont {J.}~\bibnamefont {Goold}},\ and\ \bibinfo {author} {\bibfnamefont {F.}~\bibnamefont {Plastina}},\ }\bibfield  {title} {\bibinfo {title} {Quantum coherence and ergotropy},\ }\href {https://doi.org/10.1103/PhysRevLett.125.180603} {\bibfield  {journal} {\bibinfo  {journal} {Phys. Rev. Lett.}\ }\textbf {\bibinfo {volume} {125}},\ \bibinfo {pages} {180603} (\bibinfo {year} {2020})}\BibitemShut {NoStop}%
\bibitem [{\citenamefont {Guha}\ \emph {et~al.}(2022)\citenamefont {Guha}, \citenamefont {Roy}, \citenamefont {Simonov},\ and\ \citenamefont {Zimbor{\'a}s}}]{guha2022activation}%
  \BibitemOpen
  \bibfield  {author} {\bibinfo {author} {\bibfnamefont {T.}~\bibnamefont {Guha}}, \bibinfo {author} {\bibfnamefont {S.}~\bibnamefont {Roy}}, \bibinfo {author} {\bibfnamefont {K.}~\bibnamefont {Simonov}},\ and\ \bibinfo {author} {\bibfnamefont {Z.}~\bibnamefont {Zimbor{\'a}s}},\ }\bibfield  {title} {\bibinfo {title} {Activation of thermal states by quantum switch-driven thermalization and its limits},\ }\href {https://arxiv.org/abs/2208.04034v2} {\bibfield  {journal} {\bibinfo  {journal} {arXiv:2208.04034}\ } (\bibinfo {year} {2022})}\BibitemShut {NoStop}%
\bibitem [{\citenamefont {Niu}\ \emph {et~al.}(2024)\citenamefont {Niu}, \citenamefont {Wu}, \citenamefont {Wang}, \citenamefont {Rong},\ and\ \citenamefont {Du}}]{niu2024experimental}%
  \BibitemOpen
  \bibfield  {author} {\bibinfo {author} {\bibfnamefont {Z.}~\bibnamefont {Niu}}, \bibinfo {author} {\bibfnamefont {Y.}~\bibnamefont {Wu}}, \bibinfo {author} {\bibfnamefont {Y.}~\bibnamefont {Wang}}, \bibinfo {author} {\bibfnamefont {X.}~\bibnamefont {Rong}},\ and\ \bibinfo {author} {\bibfnamefont {J.}~\bibnamefont {Du}},\ }\bibfield  {title} {\bibinfo {title} {Experimental investigation of coherent ergotropy in a single spin system},\ }\href {https://arxiv.org/abs/2409.06249} {\bibfield  {journal} {\bibinfo  {journal} {arXiv:2409.06249}\ } (\bibinfo {year} {2024})}\BibitemShut {NoStop}%
\bibitem [{\citenamefont {Cattaneo}\ \emph {et~al.}(2019)\citenamefont {Cattaneo}, \citenamefont {Giorgi}, \citenamefont {Maniscalco},\ and\ \citenamefont {Zambrini}}]{cattaneo2019local}%
  \BibitemOpen
  \bibfield  {author} {\bibinfo {author} {\bibfnamefont {M.}~\bibnamefont {Cattaneo}}, \bibinfo {author} {\bibfnamefont {G.~L.}\ \bibnamefont {Giorgi}}, \bibinfo {author} {\bibfnamefont {S.}~\bibnamefont {Maniscalco}},\ and\ \bibinfo {author} {\bibfnamefont {R.}~\bibnamefont {Zambrini}},\ }\bibfield  {title} {\bibinfo {title} {Local versus global master equation with common and separate baths: superiority of the global approach in partial secular approximation},\ }\href {https://dx.doi.org/10.1088/1367-2630/ab54ac} {\bibfield  {journal} {\bibinfo  {journal} {New Journal of Physics}\ }\textbf {\bibinfo {volume} {21}},\ \bibinfo {pages} {113045} (\bibinfo {year} {2019})}\BibitemShut {NoStop}%
\bibitem [{\citenamefont {Cidrim}\ \emph {et~al.}(2020)\citenamefont {Cidrim}, \citenamefont {do~Espirito~Santo}, \citenamefont {Schachenmayer}, \citenamefont {Kaiser},\ and\ \citenamefont {Bachelard}}]{PhysRevLett.125.073601}%
  \BibitemOpen
  \bibfield  {author} {\bibinfo {author} {\bibfnamefont {A.}~\bibnamefont {Cidrim}}, \bibinfo {author} {\bibfnamefont {T.~S.}\ \bibnamefont {do~Espirito~Santo}}, \bibinfo {author} {\bibfnamefont {J.}~\bibnamefont {Schachenmayer}}, \bibinfo {author} {\bibfnamefont {R.}~\bibnamefont {Kaiser}},\ and\ \bibinfo {author} {\bibfnamefont {R.}~\bibnamefont {Bachelard}},\ }\bibfield  {title} {\bibinfo {title} {Photon blockade with ground-state neutral atoms},\ }\href {https://doi.org/10.1103/PhysRevLett.125.073601} {\bibfield  {journal} {\bibinfo  {journal} {Phys. Rev. Lett.}\ }\textbf {\bibinfo {volume} {125}},\ \bibinfo {pages} {073601} (\bibinfo {year} {2020})}\BibitemShut {NoStop}%
\bibitem [{\citenamefont {Masson}\ \emph {et~al.}(2020)\citenamefont {Masson}, \citenamefont {Ferrier-Barbut}, \citenamefont {Orozco}, \citenamefont {Browaeys},\ and\ \citenamefont {Asenjo-Garcia}}]{PhysRevLett.125.263601}%
  \BibitemOpen
  \bibfield  {author} {\bibinfo {author} {\bibfnamefont {S.~J.}\ \bibnamefont {Masson}}, \bibinfo {author} {\bibfnamefont {I.}~\bibnamefont {Ferrier-Barbut}}, \bibinfo {author} {\bibfnamefont {L.~A.}\ \bibnamefont {Orozco}}, \bibinfo {author} {\bibfnamefont {A.}~\bibnamefont {Browaeys}},\ and\ \bibinfo {author} {\bibfnamefont {A.}~\bibnamefont {Asenjo-Garcia}},\ }\bibfield  {title} {\bibinfo {title} {Many-body signatures of collective decay in atomic chains},\ }\href {https://doi.org/10.1103/PhysRevLett.125.263601} {\bibfield  {journal} {\bibinfo  {journal} {Phys. Rev. Lett.}\ }\textbf {\bibinfo {volume} {125}},\ \bibinfo {pages} {263601} (\bibinfo {year} {2020})}\BibitemShut {NoStop}%
\bibitem [{\citenamefont {Benedetti}\ \emph {et~al.}(2016)\citenamefont {Benedetti}, \citenamefont {Galve}, \citenamefont {Mandarino}, \citenamefont {Paris},\ and\ \citenamefont {Zambrini}}]{PhysRevA.94.052118}%
  \BibitemOpen
  \bibfield  {author} {\bibinfo {author} {\bibfnamefont {C.}~\bibnamefont {Benedetti}}, \bibinfo {author} {\bibfnamefont {F.}~\bibnamefont {Galve}}, \bibinfo {author} {\bibfnamefont {A.}~\bibnamefont {Mandarino}}, \bibinfo {author} {\bibfnamefont {M.~G.~A.}\ \bibnamefont {Paris}},\ and\ \bibinfo {author} {\bibfnamefont {R.}~\bibnamefont {Zambrini}},\ }\bibfield  {title} {\bibinfo {title} {Minimal model for spontaneous quantum synchronization},\ }\href {https://doi.org/10.1103/PhysRevA.94.052118} {\bibfield  {journal} {\bibinfo  {journal} {Phys. Rev. A}\ }\textbf {\bibinfo {volume} {94}},\ \bibinfo {pages} {052118} (\bibinfo {year} {2016})}\BibitemShut {NoStop}%
\bibitem [{\citenamefont {Cattaneo}\ \emph {et~al.}(2021)\citenamefont {Cattaneo}, \citenamefont {Giorgi}, \citenamefont {Maniscalco}, \citenamefont {Paraoanu},\ and\ \citenamefont {Zambrini}}]{Cattaneo2021}%
  \BibitemOpen
  \bibfield  {author} {\bibinfo {author} {\bibfnamefont {M.}~\bibnamefont {Cattaneo}}, \bibinfo {author} {\bibfnamefont {G.~L.}\ \bibnamefont {Giorgi}}, \bibinfo {author} {\bibfnamefont {S.}~\bibnamefont {Maniscalco}}, \bibinfo {author} {\bibfnamefont {G.~S.}\ \bibnamefont {Paraoanu}},\ and\ \bibinfo {author} {\bibfnamefont {R.}~\bibnamefont {Zambrini}},\ }\bibfield  {title} {\bibinfo {title} {Bath-induced collective phenomena on superconducting qubits: Synchronization, subradiance, and entanglement generation},\ }\href {https://doi.org/10.1002/andp.202100038} {\bibfield  {journal} {\bibinfo  {journal} {Annalen der Physik}\ }\textbf {\bibinfo {volume} {533}},\ \bibinfo {pages} {2100038} (\bibinfo {year} {2021})}\BibitemShut {NoStop}%
\bibitem [{\citenamefont {Ma}\ \emph {et~al.}(2012)\citenamefont {Ma}, \citenamefont {Sun}, \citenamefont {Wang},\ and\ \citenamefont {Nori}}]{PhysRevA.85.062323}%
  \BibitemOpen
  \bibfield  {author} {\bibinfo {author} {\bibfnamefont {J.}~\bibnamefont {Ma}}, \bibinfo {author} {\bibfnamefont {Z.}~\bibnamefont {Sun}}, \bibinfo {author} {\bibfnamefont {X.}~\bibnamefont {Wang}},\ and\ \bibinfo {author} {\bibfnamefont {F.}~\bibnamefont {Nori}},\ }\bibfield  {title} {\bibinfo {title} {Entanglement dynamics of two qubits in a common bath},\ }\href {https://doi.org/10.1103/PhysRevA.85.062323} {\bibfield  {journal} {\bibinfo  {journal} {Phys. Rev. A}\ }\textbf {\bibinfo {volume} {85}},\ \bibinfo {pages} {062323} (\bibinfo {year} {2012})}\BibitemShut {NoStop}%
\bibitem [{\citenamefont {Hu}\ \emph {et~al.}(2018)\citenamefont {Hu}, \citenamefont {Man},\ and\ \citenamefont {Xia}}]{Hu2018}%
  \BibitemOpen
  \bibfield  {author} {\bibinfo {author} {\bibfnamefont {L.-Z.}\ \bibnamefont {Hu}}, \bibinfo {author} {\bibfnamefont {Z.-X.}\ \bibnamefont {Man}},\ and\ \bibinfo {author} {\bibfnamefont {Y.-J.}\ \bibnamefont {Xia}},\ }\bibfield  {title} {\bibinfo {title} {Steady-state entanglement and thermalization of coupled qubits in two common heat baths},\ }\href {https://doi.org/10.1007/s11128-018-1825-x} {\bibfield  {journal} {\bibinfo  {journal} {Quantum Information Processing}\ }\textbf {\bibinfo {volume} {17}},\ \bibinfo {pages} {45} (\bibinfo {year} {2018})}\BibitemShut {NoStop}%
\bibitem [{\citenamefont {Ghasemian}\ and\ \citenamefont {Tavassoly}(2020)}]{Ghasemian2020}%
  \BibitemOpen
  \bibfield  {author} {\bibinfo {author} {\bibfnamefont {E.}~\bibnamefont {Ghasemian}}\ and\ \bibinfo {author} {\bibfnamefont {M.~K.}\ \bibnamefont {Tavassoly}},\ }\bibfield  {title} {\bibinfo {title} {Entanglement dynamics of a dissipative two-qubit system under the influence of a global environment},\ }\href {https://doi.org/10.1007/s10773-020-04440-1} {\bibfield  {journal} {\bibinfo  {journal} {International Journal of Theoretical Physics}\ }\textbf {\bibinfo {volume} {59}},\ \bibinfo {pages} {1742} (\bibinfo {year} {2020})}\BibitemShut {NoStop}%
\bibitem [{\citenamefont {Hewgill}\ \emph {et~al.}(2018)\citenamefont {Hewgill}, \citenamefont {Ferraro},\ and\ \citenamefont {De~Chiara}}]{PhysRevA.98.042102}%
  \BibitemOpen
  \bibfield  {author} {\bibinfo {author} {\bibfnamefont {A.}~\bibnamefont {Hewgill}}, \bibinfo {author} {\bibfnamefont {A.}~\bibnamefont {Ferraro}},\ and\ \bibinfo {author} {\bibfnamefont {G.}~\bibnamefont {De~Chiara}},\ }\bibfield  {title} {\bibinfo {title} {Quantum correlations and thermodynamic performances of two-qubit engines with local and common baths},\ }\href {https://doi.org/10.1103/PhysRevA.98.042102} {\bibfield  {journal} {\bibinfo  {journal} {Phys. Rev. A}\ }\textbf {\bibinfo {volume} {98}},\ \bibinfo {pages} {042102} (\bibinfo {year} {2018})}\BibitemShut {NoStop}%
\bibitem [{\citenamefont {Manzano}\ \emph {et~al.}(2019)\citenamefont {Manzano}, \citenamefont {Giorgi}, \citenamefont {Fazio},\ and\ \citenamefont {Zambrini}}]{Manzano_2019}%
  \BibitemOpen
  \bibfield  {author} {\bibinfo {author} {\bibfnamefont {G.}~\bibnamefont {Manzano}}, \bibinfo {author} {\bibfnamefont {G.-L.}\ \bibnamefont {Giorgi}}, \bibinfo {author} {\bibfnamefont {R.}~\bibnamefont {Fazio}},\ and\ \bibinfo {author} {\bibfnamefont {R.}~\bibnamefont {Zambrini}},\ }\bibfield  {title} {\bibinfo {title} {Boosting the performance of small autonomous refrigerators via common environmental effects},\ }\href {https://doi.org/10.1088/1367-2630/ab5c58} {\bibfield  {journal} {\bibinfo  {journal} {New Journal of Physics}\ }\textbf {\bibinfo {volume} {21}},\ \bibinfo {pages} {123026} (\bibinfo {year} {2019})}\BibitemShut {NoStop}%
\bibitem [{\citenamefont {Liu}\ \emph {et~al.}(2022)\citenamefont {Liu}, \citenamefont {Yu},\ and\ \citenamefont {Yu}}]{e24010032}%
  \BibitemOpen
  \bibfield  {author} {\bibinfo {author} {\bibfnamefont {Y.-Q.}\ \bibnamefont {Liu}}, \bibinfo {author} {\bibfnamefont {D.-H.}\ \bibnamefont {Yu}},\ and\ \bibinfo {author} {\bibfnamefont {C.-S.}\ \bibnamefont {Yu}},\ }\bibfield  {title} {\bibinfo {title} {Common environmental effects on quantum thermal transistor},\ }\href {https://www.mdpi.com/1099-4300/24/1/32} {\bibfield  {journal} {\bibinfo  {journal} {Entropy}\ }\textbf {\bibinfo {volume} {24}},\ \bibinfo {pages} {32} (\bibinfo {year} {2022})}\BibitemShut {NoStop}%
\bibitem [{\citenamefont {Ekanayake}\ \emph {et~al.}(2023)\citenamefont {Ekanayake}, \citenamefont {Gunapala},\ and\ \citenamefont {Premaratne}}]{PhysRevB.107.075440}%
  \BibitemOpen
  \bibfield  {author} {\bibinfo {author} {\bibfnamefont {U.~N.}\ \bibnamefont {Ekanayake}}, \bibinfo {author} {\bibfnamefont {S.~D.}\ \bibnamefont {Gunapala}},\ and\ \bibinfo {author} {\bibfnamefont {M.}~\bibnamefont {Premaratne}},\ }\bibfield  {title} {\bibinfo {title} {Engineered common environmental effects on multitransistor systems},\ }\href {https://doi.org/10.1103/PhysRevB.107.075440} {\bibfield  {journal} {\bibinfo  {journal} {Phys. Rev. B}\ }\textbf {\bibinfo {volume} {107}},\ \bibinfo {pages} {075440} (\bibinfo {year} {2023})}\BibitemShut {NoStop}%
\bibitem [{\citenamefont {Rajapaksha}\ \emph {et~al.}(2024)\citenamefont {Rajapaksha}, \citenamefont {Gunapala},\ and\ \citenamefont {Premaratne}}]{10.1063/5.0237842}%
  \BibitemOpen
  \bibfield  {author} {\bibinfo {author} {\bibfnamefont {A.}~\bibnamefont {Rajapaksha}}, \bibinfo {author} {\bibfnamefont {S.~D.}\ \bibnamefont {Gunapala}},\ and\ \bibinfo {author} {\bibfnamefont {M.}~\bibnamefont {Premaratne}},\ }\bibfield  {title} {\bibinfo {title} {Enhanced thermal rectification in coupled qutrit–qubit quantum thermal diode},\ }\href {https://doi.org/10.1063/5.0237842} {\bibfield  {journal} {\bibinfo  {journal} {APL Quantum}\ }\textbf {\bibinfo {volume} {1}},\ \bibinfo {pages} {046123} (\bibinfo {year} {2024})}\BibitemShut {NoStop}%
\bibitem [{\citenamefont {Behzadi}\ \emph {et~al.}(2018)\citenamefont {Behzadi}, \citenamefont {Soltani},\ and\ \citenamefont {Faizi}}]{behzadi2018thermodynamic}%
  \BibitemOpen
  \bibfield  {author} {\bibinfo {author} {\bibfnamefont {N.}~\bibnamefont {Behzadi}}, \bibinfo {author} {\bibfnamefont {E.}~\bibnamefont {Soltani}},\ and\ \bibinfo {author} {\bibfnamefont {E.}~\bibnamefont {Faizi}},\ }\bibfield  {title} {\bibinfo {title} {Thermodynamic cost of creating global quantum discord and local quantum uncertainty},\ }\href {https://doi.org/10.1007/s10773-018-3838-8} {\bibfield  {journal} {\bibinfo  {journal} {International Journal of Theoretical Physics}\ }\textbf {\bibinfo {volume} {57}},\ \bibinfo {pages} {3207} (\bibinfo {year} {2018})}\BibitemShut {NoStop}%
\bibitem [{\citenamefont {Gomes}\ \emph {et~al.}(2022)\citenamefont {Gomes}, \citenamefont {Dieguez},\ and\ \citenamefont {Vasconcelos}}]{gomes2022realism}%
  \BibitemOpen
  \bibfield  {author} {\bibinfo {author} {\bibfnamefont {V.}~\bibnamefont {Gomes}}, \bibinfo {author} {\bibfnamefont {P.}~\bibnamefont {Dieguez}},\ and\ \bibinfo {author} {\bibfnamefont {H.}~\bibnamefont {Vasconcelos}},\ }\bibfield  {title} {\bibinfo {title} {Realism-based nonlocality: Invariance under local unitary operations and asymptotic decay for thermal correlated states},\ }\href {https://doi.org/10.1016/j.physa.2022.127568} {\bibfield  {journal} {\bibinfo  {journal} {Physica A: Statistical Mechanics and its Applications}\ }\textbf {\bibinfo {volume} {601}},\ \bibinfo {pages} {127568} (\bibinfo {year} {2022})}\BibitemShut {NoStop}%
\bibitem [{\citenamefont {Vinjanampathy}\ and\ \citenamefont {Anders}(2016)}]{Vinjanampathy2016}%
  \BibitemOpen
  \bibfield  {author} {\bibinfo {author} {\bibfnamefont {S.}~\bibnamefont {Vinjanampathy}}\ and\ \bibinfo {author} {\bibfnamefont {J.}~\bibnamefont {Anders}},\ }\bibfield  {title} {\bibinfo {title} {Quantum thermodynamics},\ }\href {https://doi.org/10.1080/00107514.2016.1201896} {\bibfield  {journal} {\bibinfo  {journal} {Contemporary Physics}\ }\textbf {\bibinfo {volume} {57}},\ \bibinfo {pages} {545} (\bibinfo {year} {2016})}\BibitemShut {NoStop}%
\bibitem [{\citenamefont {Werlang}\ \emph {et~al.}(2014)\citenamefont {Werlang}, \citenamefont {Marchiori}, \citenamefont {Cornelio},\ and\ \citenamefont {Valente}}]{werlang2014optimal}%
  \BibitemOpen
  \bibfield  {author} {\bibinfo {author} {\bibfnamefont {T.}~\bibnamefont {Werlang}}, \bibinfo {author} {\bibfnamefont {M.~A.}\ \bibnamefont {Marchiori}}, \bibinfo {author} {\bibfnamefont {M.~F.}\ \bibnamefont {Cornelio}},\ and\ \bibinfo {author} {\bibfnamefont {D.}~\bibnamefont {Valente}},\ }\bibfield  {title} {\bibinfo {title} {Optimal rectification in the ultrastrong coupling regime},\ }\href {https://doi.org/10.1103/PhysRevE.89.062109} {\bibfield  {journal} {\bibinfo  {journal} {Phys. Rev. E}\ }\textbf {\bibinfo {volume} {89}},\ \bibinfo {pages} {062109} (\bibinfo {year} {2014})}\BibitemShut {NoStop}%
\bibitem [{\citenamefont {{\L{}}obejko}\ \emph {et~al.}(2020)\citenamefont {{\L{}}obejko}, \citenamefont {Mazurek},\ and\ \citenamefont {Horodecki}}]{lobejko2020thermodynamics}%
  \BibitemOpen
  \bibfield  {author} {\bibinfo {author} {\bibfnamefont {M.}~\bibnamefont {{\L{}}obejko}}, \bibinfo {author} {\bibfnamefont {P.}~\bibnamefont {Mazurek}},\ and\ \bibinfo {author} {\bibfnamefont {M.}~\bibnamefont {Horodecki}},\ }\bibfield  {title} {\bibinfo {title} {Thermodynamics of {M}inimal {C}oupling {Q}uantum {H}eat {E}ngines},\ }\href {https://doi.org/10.22331/q-2020-12-23-375} {\bibfield  {journal} {\bibinfo  {journal} {{Quantum}}\ }\textbf {\bibinfo {volume} {4}},\ \bibinfo {pages} {375} (\bibinfo {year} {2020})}\BibitemShut {NoStop}%
\bibitem [{\citenamefont {Bhattacharjee}\ and\ \citenamefont {Dutta}(2021)}]{bhattacharjee2021quantum}%
  \BibitemOpen
  \bibfield  {author} {\bibinfo {author} {\bibfnamefont {S.}~\bibnamefont {Bhattacharjee}}\ and\ \bibinfo {author} {\bibfnamefont {A.}~\bibnamefont {Dutta}},\ }\bibfield  {title} {\bibinfo {title} {Quantum thermal machines and batteries},\ }\href {https://doi.org/10.1140/epjb/s10051-021-00235-3} {\bibfield  {journal} {\bibinfo  {journal} {The European Physical Journal B}\ }\textbf {\bibinfo {volume} {94}},\ \bibinfo {pages} {239} (\bibinfo {year} {2021})}\BibitemShut {NoStop}%
\bibitem [{\citenamefont {Malavazi}\ \emph {et~al.}(2024)\citenamefont {Malavazi}, \citenamefont {Ahmadi}, \citenamefont {Mazurek},\ and\ \citenamefont {Mandarino}}]{PhysRevE.109.064146}%
  \BibitemOpen
  \bibfield  {author} {\bibinfo {author} {\bibfnamefont {A.~H.~A.}\ \bibnamefont {Malavazi}}, \bibinfo {author} {\bibfnamefont {B.}~\bibnamefont {Ahmadi}}, \bibinfo {author} {\bibfnamefont {P.}~\bibnamefont {Mazurek}},\ and\ \bibinfo {author} {\bibfnamefont {A.}~\bibnamefont {Mandarino}},\ }\bibfield  {title} {\bibinfo {title} {Detuning effects for heat-current control in quantum thermal devices},\ }\href {https://doi.org/10.1103/PhysRevE.109.064146} {\bibfield  {journal} {\bibinfo  {journal} {Phys. Rev. E}\ }\textbf {\bibinfo {volume} {109}},\ \bibinfo {pages} {064146} (\bibinfo {year} {2024})}\BibitemShut {NoStop}%
\bibitem [{\citenamefont {Ollivier}\ and\ \citenamefont {Zurek}(2001)}]{ollivier2001quantum}%
  \BibitemOpen
  \bibfield  {author} {\bibinfo {author} {\bibfnamefont {H.}~\bibnamefont {Ollivier}}\ and\ \bibinfo {author} {\bibfnamefont {W.~H.}\ \bibnamefont {Zurek}},\ }\bibfield  {title} {\bibinfo {title} {Quantum discord: A measure of the quantumness of correlations},\ }\href {https://doi.org/10.1103/PhysRevLett.88.017901} {\bibfield  {journal} {\bibinfo  {journal} {Phys. Rev. Lett.}\ }\textbf {\bibinfo {volume} {88}},\ \bibinfo {pages} {017901} (\bibinfo {year} {2001})}\BibitemShut {NoStop}%
\bibitem [{\citenamefont {Dieguez}\ and\ \citenamefont {Angelo}(2018{\natexlab{b}})}]{dieguez2018weak}%
  \BibitemOpen
  \bibfield  {author} {\bibinfo {author} {\bibfnamefont {P.~R.}\ \bibnamefont {Dieguez}}\ and\ \bibinfo {author} {\bibfnamefont {R.~M.}\ \bibnamefont {Angelo}},\ }\bibfield  {title} {\bibinfo {title} {Weak quantum discord},\ }\href {https://doi.org/10.1007/s11128-018-1963-1} {\bibfield  {journal} {\bibinfo  {journal} {Quantum Inf. Process.}\ }\textbf {\bibinfo {volume} {17}},\ \bibinfo {pages} {1} (\bibinfo {year} {2018}{\natexlab{b}})}\BibitemShut {NoStop}%
\bibitem [{\citenamefont {Castellano}\ \emph {et~al.}(2024{\natexlab{a}})\citenamefont {Castellano}, \citenamefont {Nery}, \citenamefont {Simonov},\ and\ \citenamefont {Farina}}]{castellano2024parallel}%
  \BibitemOpen
  \bibfield  {author} {\bibinfo {author} {\bibfnamefont {R.}~\bibnamefont {Castellano}}, \bibinfo {author} {\bibfnamefont {R.}~\bibnamefont {Nery}}, \bibinfo {author} {\bibfnamefont {K.}~\bibnamefont {Simonov}},\ and\ \bibinfo {author} {\bibfnamefont {D.}~\bibnamefont {Farina}},\ }\bibfield  {title} {\bibinfo {title} {Parallel ergotropy: Maximum work extraction via parallel local unitary operations},\ }\href {https://doi.org/10.48550/arXiv.2407.20916} {\bibfield  {journal} {\bibinfo  {journal} {arXiv:2407.20916}\ } (\bibinfo {year} {2024}{\natexlab{a}})}\BibitemShut {NoStop}%
\bibitem [{\citenamefont {Salvia}\ \emph {et~al.}(2023{\natexlab{b}})\citenamefont {Salvia}, \citenamefont {De~Palma},\ and\ \citenamefont {Giovannetti}}]{PhysRevA.107.012405}%
  \BibitemOpen
  \bibfield  {author} {\bibinfo {author} {\bibfnamefont {R.}~\bibnamefont {Salvia}}, \bibinfo {author} {\bibfnamefont {G.}~\bibnamefont {De~Palma}},\ and\ \bibinfo {author} {\bibfnamefont {V.}~\bibnamefont {Giovannetti}},\ }\bibfield  {title} {\bibinfo {title} {Optimal local work extraction from bipartite quantum systems in the presence of hamiltonian couplings},\ }\href {https://doi.org/10.1103/PhysRevA.107.012405} {\bibfield  {journal} {\bibinfo  {journal} {Phys. Rev. A}\ }\textbf {\bibinfo {volume} {107}},\ \bibinfo {pages} {012405} (\bibinfo {year} {2023}{\natexlab{b}})}\BibitemShut {NoStop}%
\bibitem [{\citenamefont {Di~Bello}\ \emph {et~al.}(2024)\citenamefont {Di~Bello}, \citenamefont {Farina}, \citenamefont {Jansen}, \citenamefont {Perroni}, \citenamefont {Cataudella},\ and\ \citenamefont {De~Filippis}}]{di2024local}%
  \BibitemOpen
  \bibfield  {author} {\bibinfo {author} {\bibfnamefont {G.}~\bibnamefont {Di~Bello}}, \bibinfo {author} {\bibfnamefont {D.}~\bibnamefont {Farina}}, \bibinfo {author} {\bibfnamefont {D.}~\bibnamefont {Jansen}}, \bibinfo {author} {\bibfnamefont {C.}~\bibnamefont {Perroni}}, \bibinfo {author} {\bibfnamefont {V.}~\bibnamefont {Cataudella}},\ and\ \bibinfo {author} {\bibfnamefont {G.}~\bibnamefont {De~Filippis}},\ }\bibfield  {title} {\bibinfo {title} {Local ergotropy and its fluctuations across a dissipative quantum phase transition},\ }\href {https://doi.org/10.48550/arXiv.2408.02655} {\bibfield  {journal} {\bibinfo  {journal} {arXiv:2408.02655}\ } (\bibinfo {year} {2024})}\BibitemShut {NoStop}%
\bibitem [{\citenamefont {Castellano}\ \emph {et~al.}(2024{\natexlab{b}})\citenamefont {Castellano}, \citenamefont {Farina}, \citenamefont {Giovannetti},\ and\ \citenamefont {Acin}}]{PhysRevLett.133.150402}%
  \BibitemOpen
  \bibfield  {author} {\bibinfo {author} {\bibfnamefont {R.}~\bibnamefont {Castellano}}, \bibinfo {author} {\bibfnamefont {D.}~\bibnamefont {Farina}}, \bibinfo {author} {\bibfnamefont {V.}~\bibnamefont {Giovannetti}},\ and\ \bibinfo {author} {\bibfnamefont {A.}~\bibnamefont {Acin}},\ }\bibfield  {title} {\bibinfo {title} {Extended local ergotropy},\ }\href {https://doi.org/10.1103/PhysRevLett.133.150402} {\bibfield  {journal} {\bibinfo  {journal} {Phys. Rev. Lett.}\ }\textbf {\bibinfo {volume} {133}},\ \bibinfo {pages} {150402} (\bibinfo {year} {2024}{\natexlab{b}})}\BibitemShut {NoStop}%
\bibitem [{\citenamefont {Georgescu}(2021)}]{Georgescu2021}%
  \BibitemOpen
  \bibfield  {author} {\bibinfo {author} {\bibfnamefont {I.}~\bibnamefont {Georgescu}},\ }\bibfield  {title} {\bibinfo {title} {60 years of landauer's principle},\ }\href {https://doi.org/10.1038/s42254-021-00400-8} {\bibfield  {journal} {\bibinfo  {journal} {Nature Reviews Physics}\ }\textbf {\bibinfo {volume} {3}},\ \bibinfo {pages} {770} (\bibinfo {year} {2021})}\BibitemShut {NoStop}%
\end{thebibliography}%
\end{document}